\documentclass[aps,prd,reprint,superscriptaddress,showpacs,floatfix]{revtex4-1}
\RequirePackage{lineno}

\usepackage{graphicx}
\usepackage{amsmath, amsthm, amssymb}
\usepackage{xspace}

\newcommand{\CP}{\textit{CP}\xspace}
\newcommand{\Jpsi}{\ensuremath{J/\psi}\xspace}

\newcommand{\Bs}{\ensuremath{B_s^0}\xspace}
\newcommand{\bBs}{\ensuremath{\bar{B}_s^0}\xspace}

\newcommand{\BsJpsiPhi}{\ensuremath{\Bs\rightarrow\Jpsi\phi}\xspace}
\newcommand{\betas}{\ensuremath{\beta_s^{\Jpsi\phi}}\xspace}

\newcommand{\babar}{\mbox{\sl B\hspace{-0.4em} 
                     {\small\sl A}\hspace{-0.37em} \sl B\hspace{-0.4em} 
                     {\small\sl A\hspace{-0.02em}R}}\xspace}

\newcommand{\dedx}{\ensuremath{\mathrm{d}E/\mathrm{d}x}\xspace}

\newcommand{\deltaG}{\ensuremath{\Delta\Gamma_s}\xspace}
\newcommand{\deltaM}{\ensuremath{\Delta m_s}\xspace}

\newcommand{\ps}{\ensuremath{\mathrm{ps}^{-1}}\xspace}
\newcommand{\fb}{\ensuremath{\mathrm{fb}^{-1}}\xspace}
\newcommand{\mevc}{\ensuremath{\mathrm{MeV}/c}\xspace}
\newcommand{\mevcc}{\ensuremath{\mathrm{MeV}/c^2}\xspace}
\newcommand{\gevc}{\ensuremath{\mathrm{GeV}/c}\xspace}
\newcommand{\gevcc}{\ensuremath{\mathrm{GeV}/c^2}\xspace}

\newcommand{\degree}{\ensuremath{^\circ}}

\begin{document}
\title{Measurement of the \CP-Violating Phase \boldmath{\betas} in 
\boldmath{\BsJpsiPhi} Decays \\ with the CDF\,II Detector}

\affiliation{Institute of Physics, Academia Sinica, Taipei, Taiwan 11529, Republic of China}
\affiliation{Argonne National Laboratory, Argonne, Illinois 60439, USA}
\affiliation{University of Athens, 157 71 Athens, Greece}
\affiliation{Institut de Fisica d'Altes Energies, ICREA, Universitat Autonoma de Barcelona, E-08193, Bellaterra (Barcelona), Spain}
\affiliation{Baylor University, Waco, Texas 76798, USA}
\affiliation{Istituto Nazionale di Fisica Nucleare Bologna, $^{ee}$University of Bologna, I-40127 Bologna, Italy}
\affiliation{University of California, Davis, Davis, California 95616, USA}
\affiliation{University of California, Los Angeles, Los Angeles, California 90024, USA}
\affiliation{Instituto de Fisica de Cantabria, CSIC-University of Cantabria, 39005 Santander, Spain}
\affiliation{Carnegie Mellon University, Pittsburgh, Pennsylvania 15213, USA}
\affiliation{Enrico Fermi Institute, University of Chicago, Chicago, Illinois 60637, USA}
\affiliation{Comenius University, 842 48 Bratislava, Slovakia; Institute of Experimental Physics, 040 01 Kosice, Slovakia}
\affiliation{Joint Institute for Nuclear Research, RU-141980 Dubna, Russia}
\affiliation{Duke University, Durham, North Carolina 27708, USA}
\affiliation{Fermi National Accelerator Laboratory, Batavia, Illinois 60510, USA}
\affiliation{University of Florida, Gainesville, Florida 32611, USA}
\affiliation{Laboratori Nazionali di Frascati, Istituto Nazionale di Fisica Nucleare, I-00044 Frascati, Italy}
\affiliation{University of Geneva, CH-1211 Geneva 4, Switzerland}
\affiliation{Glasgow University, Glasgow G12 8QQ, United Kingdom}
\affiliation{Harvard University, Cambridge, Massachusetts 02138, USA}
\affiliation{Division of High Energy Physics, Department of Physics, University of Helsinki and Helsinki Institute of Physics, FIN-00014, Helsinki, Finland}
\affiliation{University of Illinois, Urbana, Illinois 61801, USA}
\affiliation{The Johns Hopkins University, Baltimore, Maryland 21218, USA}
\affiliation{Institut f\"{u}r Experimentelle Kernphysik, Karlsruhe Institute of Technology, D-76131 Karlsruhe, Germany}
\affiliation{Center for High Energy Physics: Kyungpook National University, Daegu 702-701, Korea; Seoul National University, Seoul 151-742, Korea; Sungkyunkwan University, Suwon 440-746, Korea; Korea Institute of Science and Technology Information, Daejeon 305-806, Korea; Chonnam National University, Gwangju 500-757, Korea; Chonbuk National University, Jeonju 561-756, Korea}
\affiliation{Ernest Orlando Lawrence Berkeley National Laboratory, Berkeley, California 94720, USA}
\affiliation{University of Liverpool, Liverpool L69 7ZE, United Kingdom}
\affiliation{University College London, London WC1E 6BT, United Kingdom}
\affiliation{Centro de Investigaciones Energeticas Medioambientales y Tecnologicas, E-28040 Madrid, Spain}
\affiliation{Massachusetts Institute of Technology, Cambridge, Massachusetts 02139, USA}
\affiliation{Institute of Particle Physics: McGill University, Montr\'{e}al, Qu\'{e}bec, Canada H3A~2T8; Simon Fraser University, Burnaby, British Columbia, Canada V5A~1S6; University of Toronto, Toronto, Ontario, Canada M5S~1A7; and TRIUMF, Vancouver, British Columbia, Canada V6T~2A3}
\affiliation{University of Michigan, Ann Arbor, Michigan 48109, USA}
\affiliation{Michigan State University, East Lansing, Michigan 48824, USA}
\affiliation{Institution for Theoretical and Experimental Physics, ITEP, Moscow 117259, Russia}
\affiliation{University of New Mexico, Albuquerque, New Mexico 87131, USA}
\affiliation{The Ohio State University, Columbus, Ohio 43210, USA}
\affiliation{Okayama University, Okayama 700-8530, Japan}
\affiliation{Osaka City University, Osaka 588, Japan}
\affiliation{University of Oxford, Oxford OX1 3RH, United Kingdom}
\affiliation{Istituto Nazionale di Fisica Nucleare, Sezione di Padova-Trento, $^{ff}$University of Padova, I-35131 Padova, Italy}
\affiliation{University of Pennsylvania, Philadelphia, Pennsylvania 19104, USA}
\affiliation{Istituto Nazionale di Fisica Nucleare Pisa, $^{gg}$University of Pisa, $^{hh}$University of Siena and $^{ii}$Scuola Normale Superiore, I-56127 Pisa, Italy}
\affiliation{University of Pittsburgh, Pittsburgh, Pennsylvania 15260, USA}
\affiliation{Purdue University, West Lafayette, Indiana 47907, USA}
\affiliation{University of Rochester, Rochester, New York 14627, USA}
\affiliation{The Rockefeller University, New York, New York 10065, USA}
\affiliation{Istituto Nazionale di Fisica Nucleare, Sezione di Roma 1, $^{jj}$Sapienza Universit\`{a} di Roma, I-00185 Roma, Italy}
\affiliation{Rutgers University, Piscataway, New Jersey 08855, USA}
\affiliation{Texas A\&M University, College Station, Texas 77843, USA}
\affiliation{Istituto Nazionale di Fisica Nucleare Trieste/Udine, I-34100 Trieste, $^{kk}$University of Udine, I-33100 Udine, Italy}
\affiliation{University of Tsukuba, Tsukuba, Ibaraki 305, Japan}
\affiliation{Tufts University, Medford, Massachusetts 02155, USA}
\affiliation{University of Virginia, Charlottesville, Virginia 22906, USA}
\affiliation{Waseda University, Tokyo 169, Japan}
\affiliation{Wayne State University, Detroit, Michigan 48201, USA}
\affiliation{University of Wisconsin, Madison, Wisconsin 53706, USA}
\affiliation{Yale University, New Haven, Connecticut 06520, USA}

\author{T.~Aaltonen}
\affiliation{Division of High Energy Physics, Department of Physics, University of Helsinki and Helsinki Institute of Physics, FIN-00014, Helsinki, Finland}
\author{B.~\'{A}lvarez~Gonz\'{a}lez$^z$}
\affiliation{Instituto de Fisica de Cantabria, CSIC-University of Cantabria, 39005 Santander, Spain}
\author{S.~Amerio}
\affiliation{Istituto Nazionale di Fisica Nucleare, Sezione di Padova-Trento, $^{ff}$University of Padova, I-35131 Padova, Italy}
\author{D.~Amidei}
\affiliation{University of Michigan, Ann Arbor, Michigan 48109, USA}
\author{A.~Anastassov$^x$}
\affiliation{Fermi National Accelerator Laboratory, Batavia, Illinois 60510, USA}
\author{A.~Annovi}
\affiliation{Laboratori Nazionali di Frascati, Istituto Nazionale di Fisica Nucleare, I-00044 Frascati, Italy}
\author{J.~Antos}
\affiliation{Comenius University, 842 48 Bratislava, Slovakia; Institute of Experimental Physics, 040 01 Kosice, Slovakia}
\author{G.~Apollinari}
\affiliation{Fermi National Accelerator Laboratory, Batavia, Illinois 60510, USA}
\author{J.A.~Appel}
\affiliation{Fermi National Accelerator Laboratory, Batavia, Illinois 60510, USA}
\author{T.~Arisawa}
\affiliation{Waseda University, Tokyo 169, Japan}
% ...

\collaboration{CDF Collaboration}\thanks{With visitors from
$^a$Istituto Nazionale di Fisica Nucleare, Sezione di Cagliari, 09042 Monserrato (Cagliari), Italy,
$^b$University of CA Irvine, Irvine, CA 92697, USA,
$^c$University of CA Santa Barbara, Santa Barbara, CA 93106, USA,
$^d$University of CA Santa Cruz, Santa Cruz, CA 95064, USA,
$^e$Institute of Physics, Academy of Sciences of the Czech Republic, Czech Republic,
$^f$CERN, CH-1211 Geneva, Switzerland,
$^g$Cornell University, Ithaca, NY 14853, USA,
$^h$University of Cyprus, Nicosia CY-1678, Cyprus,
$^i$Office of Science, U.S. Department of Energy, Washington, DC 20585, USA,
$^j$University College Dublin, Dublin 4, Ireland,
$^k$ETH, 8092 Zurich, Switzerland,
$^l$University of Fukui, Fukui City, Fukui Prefecture, Japan 910-0017,
$^m$Universidad Iberoamericana, Mexico D.F., Mexico,
$^n$University of Iowa, Iowa City, IA 52242, USA,
$^o$Kinki University, Higashi-Osaka City, Japan 577-8502,
$^p$Kansas State University, Manhattan, KS 66506, USA,
$^q$Korea University, Seoul, 136-713, Korea,
$^r$University of Manchester, Manchester M13 9PL, United Kingdom,
$^s$Queen Mary, University of London, London, E1 4NS, United Kingdom,
$^t$University of Melbourne, Victoria 3010, Australia,
$^u$Muons, Inc., Batavia, IL 60510, USA,
$^v$Nagasaki Institute of Applied Science, Nagasaki, Japan,
$^w$National Research Nuclear University, Moscow, Russia,
$^x$Northwestern University, Evanston, IL 60208, USA,
$^y$University of Notre Dame, Notre Dame, IN 46556, USA,
$^z$Universidad de Oviedo, E-33007 Oviedo, Spain,
$^{aa}$CNRS-IN2P3, Paris, F-75205 France,
$^{bb}$Texas Tech University, Lubbock, TX 79609, USA,
$^{cc}$Universidad Tecnica Federico Santa Maria, 110v Valparaiso, Chile,
$^{dd}$Yarmouk University, Irbid 211-63, Jordan,
$^{ll}$University of Warwick, Coventry CV4 7AL, United Kingdom, 
$^{mm}$CPPM, Aix-Marseille Universit´e, CNRS/IN2P3, Marseille, France. 
}
\noaffiliation

\pacs{13.25.Hw, 14.40.Nd, 14.65.Fy}
\date{\today}
\begin{abstract}

  We present a measurement of the \CP-violating parameter \betas 
  using approximately 6500 $\BsJpsiPhi$ decays reconstructed with the
  CDF\,II detector in a sample of $p\bar p$~collisions at
  $\sqrt{s}=1.96$~TeV corresponding to 5.2~fb$^{-1}$ integrated
  luminosity produced by the Tevatron Collider at Fermilab.
  We find the \CP-violating phase to be within the range $\betas \in
  [0.02, 0.52] \cup [1.08, 1.55]$ at 68\%~confidence level where the
  coverage property of the quoted interval is guaranteed using a
  frequentist statistical analysis.  This result is in agreement with
  the standard model expectation at the level of about one Gaussian
  standard deviation. We consider the inclusion of a potential $S$-wave
  contribution to the $\Bs\to J/\psi K^+K^-$ final state which is found
  to be negligible over the mass interval $1.009 <
  m(K^+K^-)<1.028~\gevcc$.
  Assuming the standard model prediction for the \CP-violating
  phase~\betas, we find the \Bs~decay width difference to be $\deltaG =
  0.075 \pm 0.035\,\textrm{(stat)} \pm 0.006\,\textrm{(syst)}~\ps$.  We
  also present the most precise measurements of the \Bs~mean lifetime
  $\tau(\Bs) = 1.529 \pm 0.025\,\textrm{(stat)} \pm
  0.012\,\textrm{(syst)}$~ps, the polarization fractions $|A_0(0)|^2 =
  0.524 \pm 0.013\,\textrm{(stat)} \pm 0.015\,\textrm{(syst)}$ and
  $|A_{\parallel}(0)|^2 = 0.231 \pm 0.014\,\textrm{(stat)} \pm
  0.015\,\textrm{(syst)}$, as well as the strong phase $\delta_{\perp}=
  2.95 \pm 0.64\,\textrm{(stat)} \pm
  0.07\,\textrm{(syst)}~\textrm{rad}$.  In addition, we report an
  alternative Bayesian analysis
  that gives results consistent with the frequentist approach.

\end{abstract}

\maketitle
\section{Introduction}
\label{sec:introduction}

Since the discovery of the simultaneous violation of charge and parity
quantum numbers (\CP violation) in 1964 in the neutral kaon
system~\cite{Christenson:1964fg}, \CP violation has played a crucial
role in the development of the standard model (SM) of particle physics
and in searches for ``new'' physics (NP) beyond the SM. In 1973, before
the discovery of the fourth (charm)
quark~\cite{Aubert:1974js,*Augustin:1974xw} and the third generation
(bottom~\cite{Herb:1977ek} and top~\cite{Abe:1995hr,*Abachi:1995iq}
quarks), Kobayashi and Maskawa proposed
an extension to a six-quark model~\cite{Kobayashi:1973fv} in which
\CP~violation was explained through the quark mixing parametrized by
the Cabibbo-Kobayashi-Maskawa (CKM) matrix.  A single, irreducible
complex phase in the CKM matrix is responsible for all \CP-violating
effects in the standard model.
\begin{figure}
\includegraphics[width=0.4\textwidth]{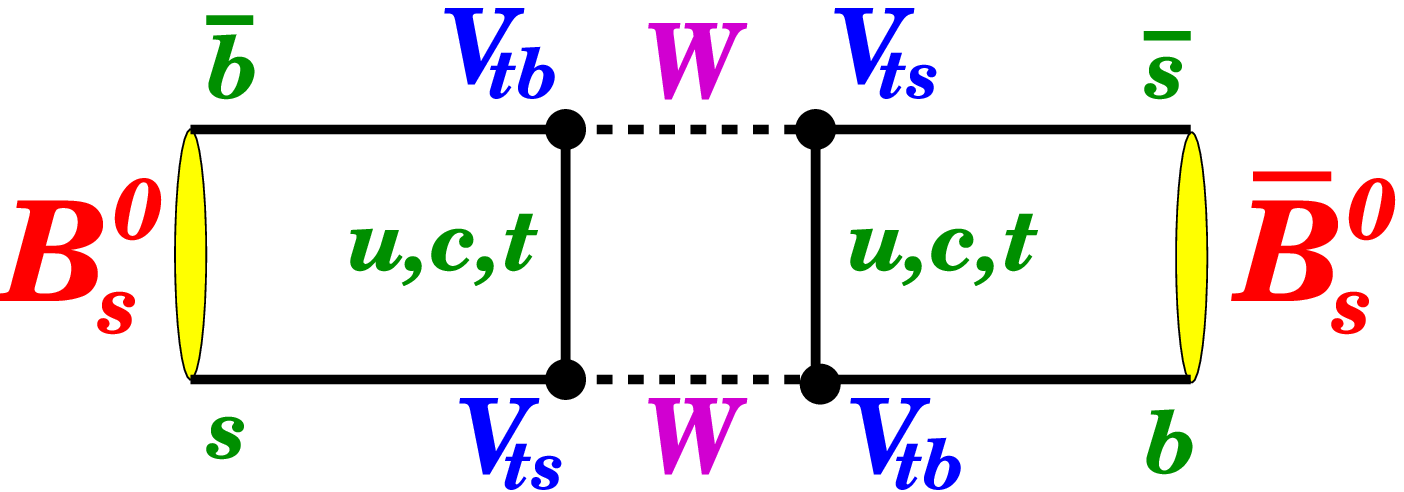}\\
\vspace*{0.4cm}
\includegraphics[width=0.4\textwidth]{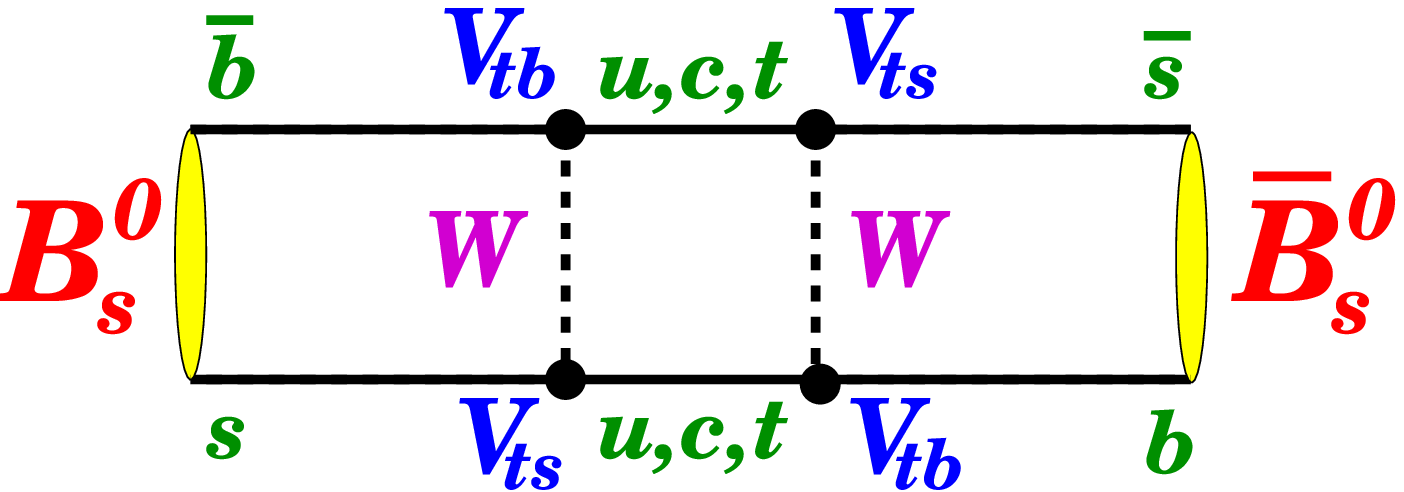}
\caption{Lowest order Feynman diagrams that induce \Bs-\bBs oscillations.
\label{fig:osc_fig}}
\end{figure}

One of the most promising processes for the search for physics beyond the
standard model is through oscillations of \Bs and \bBs mesons. The time
evolution of the \Bs and \bBs mesons can be described by the
Schr\"odinger equation
\begin{eqnarray}
&& i\frac{d}{dt}
\left(\begin{array}{c}
|\Bs(t)\,\rangle \\ |\bBs(t)\,\rangle \\ 
\end{array}\right)
=\left[\hat{M}_s - \frac{i}{2}\,\hat{\Gamma}_s\,\right]
\left(\begin{array}{c}
|\Bs(t)\,\rangle \\ |\bBs(t)\,\rangle \\ 
\end{array}\right) \\
&& {\textrm{with}} \ 
\hat{M}_s =
\left(\begin{array}{cc}
M^s_0 & M^s_{12} \\ M_{12}^{s*} & M^s_0 \\
\end{array}\right) 
{\textrm{and}}\ 
\hat{\Gamma}_s=
\left(\begin{array}{cc}
\Gamma^s_0 & \Gamma^s_{12} \\ \Gamma_{12}^{s*} & \Gamma^s_0 \\
\end{array}\right), \nonumber
\end{eqnarray}
where $\hat{M}_s$ and $\hat{\Gamma}_s$ are the mass and decay rate
$2\times2$ matrices. The $\Bs$ mixing diagrams shown in
Fig.~\ref{fig:osc_fig} give rise to non-zero off-diagonal elements $M^s_{12}$
and $\Gamma^s_{12}$. 
The diagonalization of $\hat{M}_s-i/2\,\hat{\Gamma}_s$ leads to the
heavy and light mass eigenstates $B_{s}^H$ and $B_{s}^L$ which are
admixtures of the flavor eigenstates \Bs\ and \bBs:
\begin{eqnarray}
|B_{s}^H\,\rangle &=& p\,|\Bs\,\rangle - q\,|\bBs\,\rangle 
\nonumber \\
\textrm{and}\quad
|B_{s}^L\,\rangle &=& p\,|\Bs\,\rangle + q\,|\bBs\,\rangle,
\end{eqnarray}
where $p$ and $q$ are complex quantities which are related to the
respective CKM matrix elements (see Fig.~\ref{fig:osc_fig}) through
$q/p=(V_{tb}^*V_{ts})/(V_{tb}V_{ts}^*)$ within the SM, and satisfy
$|p|^{2}+|q|^{2} = 1$.

The off-diagonal elements of the mass and decay matrices can be
related to the mass difference
between the $B_{s}^H$ and $B_{s}^L$ mass eigenstates~\cite{Kreps:2011cb}
\begin{eqnarray}
 \deltaM &=&
m_{s}^H - m_{s}^L \nonumber \\ & =& 
        2 { |M^s_{12}|} \left( 1 +
        \frac{1}{8}
\frac{|\Gamma^s_{12}|^2}{|M^s_{12}|^2} \sin^2 \phi_s +
...\right), 
\end{eqnarray}
and the corresponding decay width difference~\cite{Kreps:2011cb}
\begin{eqnarray}
\deltaG &=& \Gamma_s^L - \Gamma_s^H \nonumber \\ &=& 
        2  |\Gamma^s_{12}| \cos  \phi_s 
        \left( 1 -
         \frac{1}{8}
\frac{|\Gamma^s_{12}|^2}{|M^s_{12}|^2} \sin^2 \phi_s 
        + ...\right), 
\end{eqnarray} 
where $\phi_s=\arg( -M^s_{12}/\Gamma^s_{12})$.  Typically in the
\Bs~system corrections of order $\left({\Gamma^s_{12}}/{M^s_{12}}
\right)^2$ can be neglected.  The total \Bs~decay width $\Gamma_s =
(\Gamma_{s}^L + \Gamma_{s}^H)/2 = \hbar/\tau(\Bs)$ is related to the
mean \Bs~lifetime $\tau(\Bs)$, while the mass difference \deltaM is
proportional to the frequency of \Bs-\bBs~oscillations, first observed by
CDF~\cite{Ref:bsmixing1}, where the current world average is
$\deltaM=17.77\pm 0.10\pm0.07~\ps$~\cite{Ref:PDG}.
Assuming no $CP$~violation in the \Bs~system, which is justified in the
SM where the \CP-violating phase is expected to be small
($\phi_s^{\textit{SM}}\approx 0.004$~\cite{Ref:lenz}), the \Bs~mass eigenstates
are also $CP$~eigenstates where $\Gamma_L$ is the width of the $CP$-even
state corresponding to the short lived state in analogy to the kaon
system where the short-lived state ($K^0_S$) is $CP$~even.  $\Gamma_H$
is the width of the \CP-odd state corresponding to the long lived
\Bs~state.

A broad class of theoretically well-founded extensions of the SM
predicts new sources of \CP-violating phases~\cite{ref:multipleHiggs,
Nandi:2010zx,
*Drobnak:2011wj, *Wang:2011ax, *Shelton:2011hq, *Girrbach:2011an, 
*Buras:2010pm, *Ajaltouni:2010iv, *Bai:2010kf, 
*Everett:2009cn, *Saha:2010vw, *Isidori:2010kg, *Botella:2009pq, *Chiang:2009ev,
Ref:ligeti}.  In the presence of physics beyond the standard model,
the quantities describing the \Bs~system can be modified by a phase
$\phi_s^{\textit{NP}}$ as follows~\cite{Ref:lenz,Ref:ligeti}:
\begin{eqnarray}
\Gamma^s_{12} &=&  {\Gamma^s_{12}}^{\textit{SM}}, \nonumber \\
M^s_{12}  &=&  {M^s_{12}}^{\textit{SM}} \times { \Delta_s},\quad \textrm{where}\
{\Delta_s} = { |\Delta_s|}\, e^{i\,\phi_s^{\textit{NP}}}.
\end{eqnarray}
In this parameterization it is assumed that new physics has a negligible
effect on $\Gamma^s_{12}$, which is the case for a large class of new
physics models and confirmed by experimental data~\cite{Ref:PDG}, and
only $M^s_{12}$ is changed by the factor $\Delta_s$. As the precise
determination of the $\Bs$ oscillation frequency~\cite{Ref:PDG} is well
within the standard model expectation, contributions of new physics to
the magnitude $|\Delta_s|$ on the level of greater than about 10-20\%
are unlikely~\cite{Ref:ligeti}. A currently preferred place to search
for new physics is through the phase $\phi_s$ which is unconstrained by
measurements of the \Bs-\bBs oscillation frequency.  Since
$\phi_s^{\textit{SM}}$ is small, in a new-physics scenario with a large
contribution to $\phi_s$, the approximation 
$\phi_s=\phi_s^{\textit{SM}}+\phi_s^{\textit{NP}}\approx\phi_s^{\textit{NP}}$ 
can be made.

\begin{figure}
\includegraphics[width=0.4\textwidth]{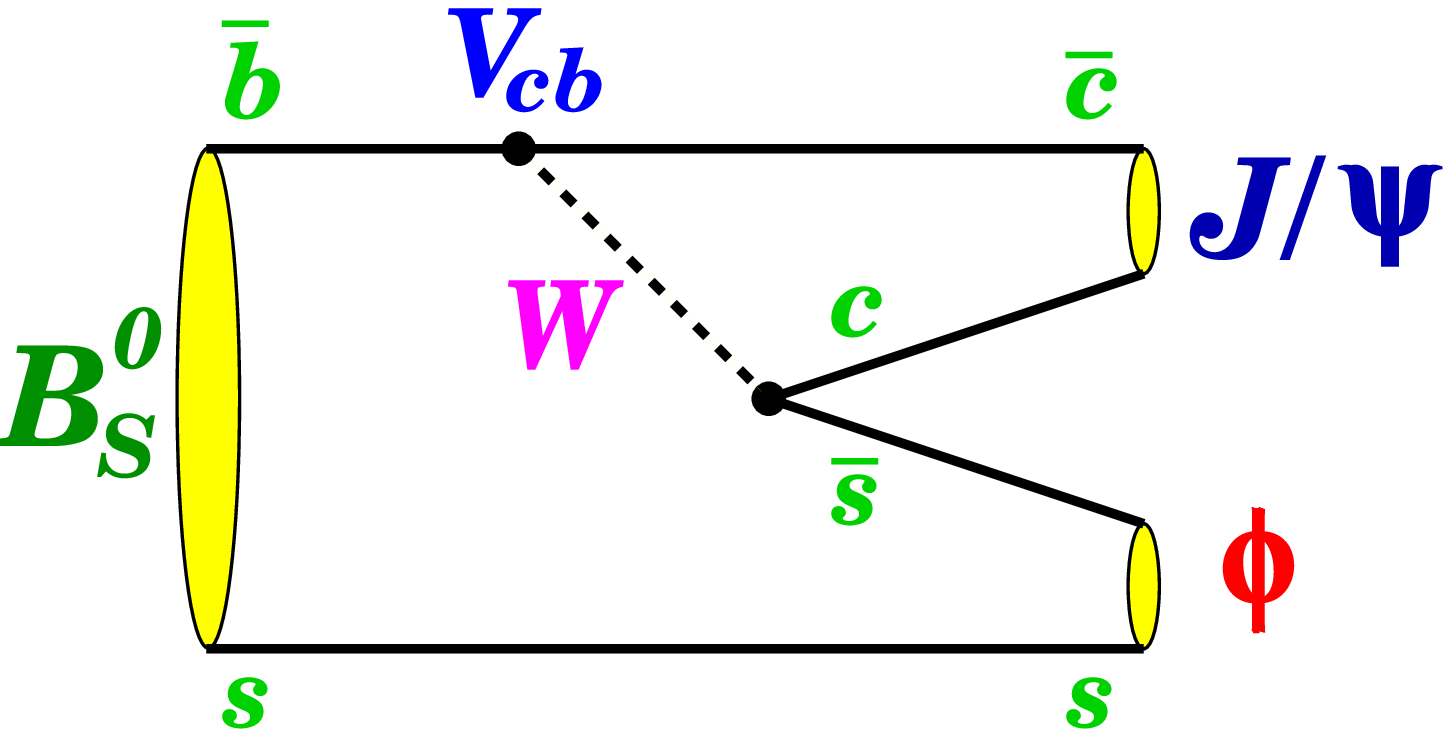}\\
\vspace*{0.2cm}
\includegraphics[width=0.4\textwidth]{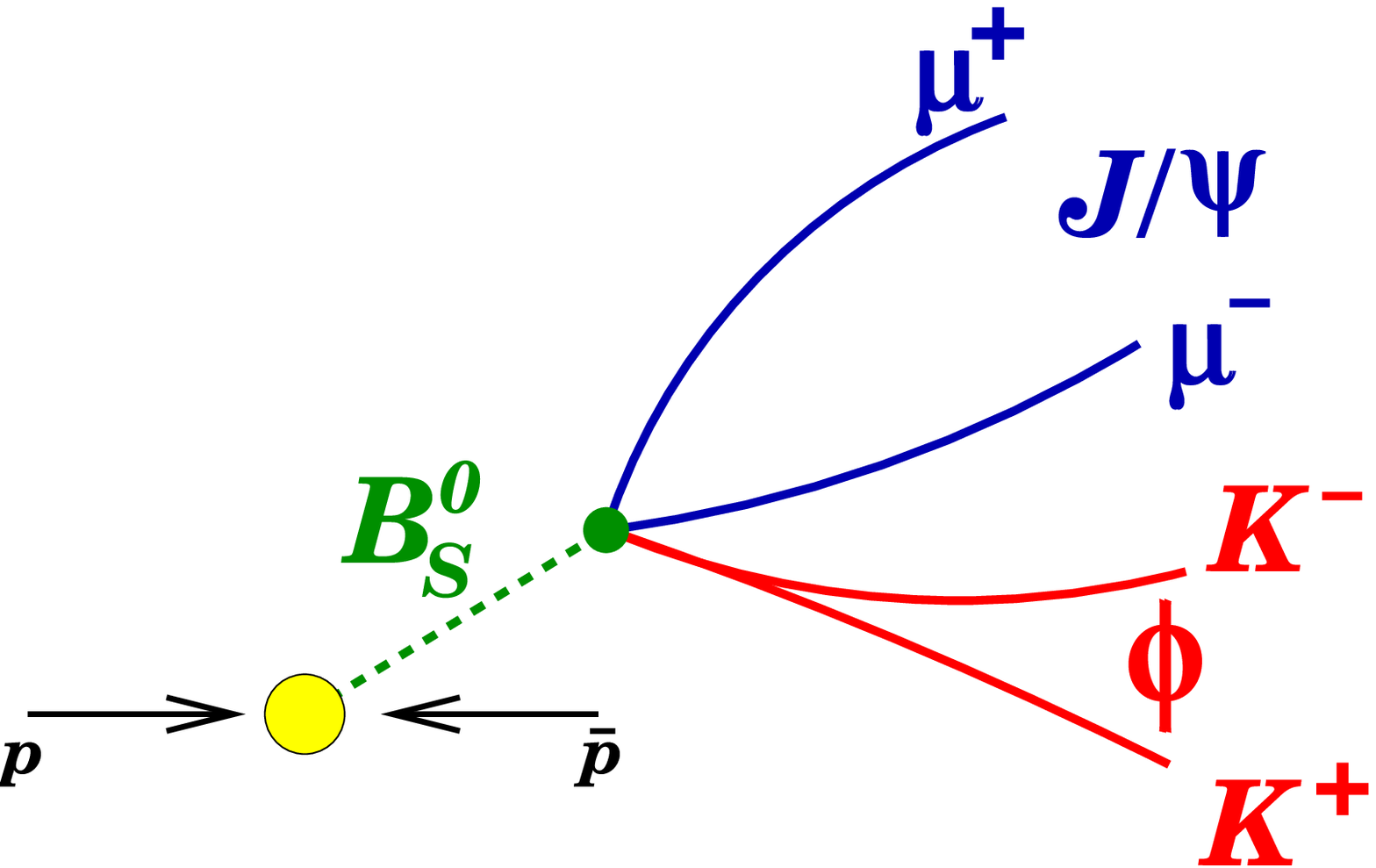}
\caption{Leading \BsJpsiPhi decay diagram (top) and decay topology (bottom).
\label{fig:BsJPsiPhi_decay}}
\end{figure}

An excellent probe of this new-physics phase~\cite{theory} is through
the decay mode $\Bs\rightarrow\Jpsi\phi(1020)$, 
with $\Jpsi \rightarrow \mu^+ \mu^-$ and
$\phi(1020) \rightarrow K^+ K^-$. Note that throughout this paper we
refer to $\phi(1020)$ just as $\phi$ for brevity.  
Figure~\ref{fig:BsJPsiPhi_decay} shows the
leading \BsJpsiPhi decay diagram on the left-hand side while the decay
topology is indicated on the right.  The relative phase between the
decay amplitudes with and without mixing is $2\betas$, which is responsible
for \CP~violation in \BsJpsiPhi decays. Neglecting higher order loop
corrections (penguin contributions) and assuming that there is no
\CP~violation present in the decay amplitude, $2\betas$ can be
associated with $e^{i\,2\betas} = (q/p)\cdot (A_f/\bar{A_f})$, where
$A_f$ and $\bar A_f$ are the decay amplitudes in \BsJpsiPhi and
$\bBs\rightarrow\Jpsi\phi$, respectively.  In the standard model this
phase is
$\beta_s^{\textit{SM}}=\arg\left[-(V_{ts}V_{tb}^{*})/(V_{cs}V_{cb}^{*})\right]$,
where $V_{ij}$ are again the corresponding elements of the CKM quark
mixing matrix. Global fits of experimental data tightly constrain the
\CP-violating phase to small values in the context of the standard
model, $\beta_s^{\textit{SM}}\approx 0.02$~\cite{Ref:lenz, globalfit1,
 *globalfit2}. The presence of new physics could modify this phase by 
the same quantity $\phi_s^{\textit{NP}}$ that affects the phase $\phi_s$
allowing $\betas$ to be expressed as $2\betas = 2\beta_s^{\textit{SM}} -
\phi_s^{\textit{NP}}$~\cite{Ref:lenz,Ref:ligeti}. Assuming that new
physics effects are much larger than the SM phase, we can again
approximate $2\betas \approx - \phi_s^{\textit{NP}} \approx -\phi_s$.

The first measurements of the \CP-violating phase \betas by the CDF and
D0 experiments~\cite{Ref:tagged_cdf_old,Ref:tagged_d0} each showed a
mild inconsistency with the SM prediction where interestingly both
results deviated in the same direction. A preliminary combination of the
CDF and D0 analyses with samples corresponding to 2.8~fb$^{-1}$
integrated luminosity was inconsistent with the SM expectation at the
level of about two standard deviations~\cite{Ref:ICHEP2008}.  In
addition, recent dimuon asymmetry results from the
D0~collaboration~\cite{ref:d0_likesign_dimuon,*Abazov:2011yk} suggest
additional indication for effects of physics beyond the standard model
in \Bs~mixing. During the preparation of this manuscript the
D0~collaboration released an updated measurement of the \CP-violating
phase \betas using a data sample based on 8~fb$^{-1}$ of integrated
luminosity~\cite{Abazov:2011ry}, while the LHCb collaboration presented
a first preliminary measurement of the \Bs~mixing phase showing
confidence regions in agreement with the SM~prediction within one
standard deviation~\cite{Ref:LHCb_first_betas}.

This paper presents a measurement of the \CP-violating phase $\betas$
using about four times the integrated luminosity of our previously
published analysis~\cite{Ref:tagged_cdf_old}, as well as
additional improvements in flavor tagging and the inclusion of
potential $S$-wave contributions to the \BsJpsiPhi signal. 
This article is organized as follows. In Sec.~\ref{sec:overview} we give
an overview of the work flow of the analysis, while we describe the CDF
experiment in Sec.~\ref{sec:detector}. The data selection is summarized
in Sec.~\ref{sec:reconstruction}. The applied flavor tagging is
discussed in Sec.~\ref{sec:tagging} and the likelihood fit function is
detailed in Sec.~\ref{sec:fit}. The measurements of the \Bs~mean
lifetime, \deltaG, the polarization fraction and the respective
systematic uncertainties are described in Sec.~\ref{sec:smresult}
and Sec.~\ref{sec:systematics}, respectively. The results on \betas and
\deltaG using a frequentist analysis are summarized in
Sec.~\ref{sec:statistics} while Sec.~\ref{06_CDF} describes an
alternative Bayesian approach. A summary is given in
Sec.~\ref{sec:conclusions}.

\section{Measurement Overview}
\label{sec:overview}

The measurement of the phase \betas relies on an analysis of the
time-evolution and kinematics of the \BsJpsiPhi decay, which features a
pseudoscalar meson decaying to two vector mesons.  Consequently, the
total spin in the final $\Jpsi\phi$ state is either 0, 1 or 2.  To
conserve the total angular momentum, the orbital angular momentum $L$
between the final state decay products must be either 0, 1 or 2. While
the \Jpsi and $\phi$ are \CP-even eigenstates, the $\Jpsi\phi$ final
state has a \CP eigenvalue given as $(-1)^L$.  Consequently, the states
with orbital angular momentum 0 and 2 are \CP-even while the state with
angular momentum 1 is \CP-odd.  We use both the decay time of the \Bs
and the decay angles of the $\Jpsi\rightarrow\mu^+\mu^-$ and
$\phi\rightarrow K^+K^-$~mesons to statistically separate the \CP-odd
and \CP-even components of the $\Jpsi\phi$ final state.

\begin{figure}
\includegraphics[width=0.48\textwidth]{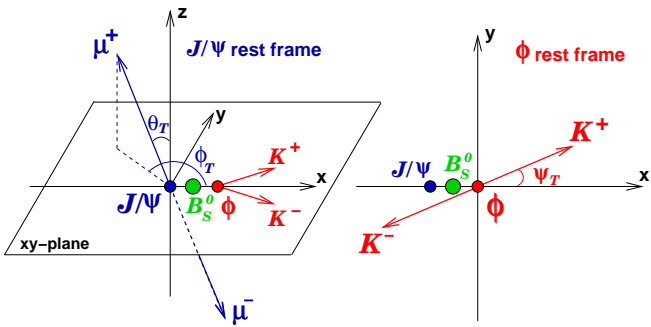}
\caption{Illustration of definition of transversity angles $\theta_T$, 
$\phi_T$, and $\psi_T$.
\label{Fig:sketch_transversity_angles}}
\end{figure}

There are three angles that completely define the directions of the
four particles in the final state. We use the angular variables $\vec\rho =
\{\cos\theta_T,\phi_T, \cos\psi_T\}$ as defined in the transversity
basis~\cite{Ref:Dighe}.   
In the following relations we use a notation where $\vec{p}\,(A)_{B}$
denotes the three-momentum of particle $A$ in the rest frame of
particle~$B$.
With this notation, the helicity angle $\psi_T$ of the $K^+$ is defined 
in the $\phi$~rest frame as the angle between $\vec{p}\,(K^+)$ and the
negative $J/\psi$~direction: 
\begin{eqnarray}
\cos\psi_T = -\frac{\vec{p}\,(K^+)_{\phi}\cdot\vec{p}\,(\Jpsi)_{\phi}}
{\left|\vec{p}\,(K^+)_{\phi}\right|\cdot
\left|\vec{p}\,(\Jpsi)_{\phi}\right|}.
\end{eqnarray}
To calculate the other two angles, we first define a coordinate
system through the directions
\begin{eqnarray}
\hat{x}&=&\frac{\vec{p}\,(\phi)_{\Jpsi}}{\left|\vec{p}\,(\phi)_{\Jpsi}\right|}, \nonumber \\
\hat{y}&=&
\frac{\vec{p}\,(K^+)_{\Jpsi}-\left[\vec{p}\,(K^+)_{\Jpsi}\cdot\hat{x}\right]\hat{x}}
{\left|\vec{p}\,(K^+)_{\Jpsi}-\left[\vec{p}\,(K^+)_{\Jpsi}\cdot\hat{x}\right]\hat{x}\right|}, \nonumber \\
\hat{z}&=&\hat{x}\times\hat{y}.
\end{eqnarray}
With this coordinate system the following angles of the direction of the
$\mu^+$ in the $J/\psi$~rest frame are calculated as
\begin{eqnarray}
\cos\theta_T&=&\frac{\vec{p}\,(\mu^+)_{\Jpsi}}
{\left|\vec{p}\,(\mu^+)_{\Jpsi}\right|}\cdot\hat{z}, \\
\phi_T&=&\tan^{-1}\left(\frac{
\frac{\vec{p}\,(\mu^+)_{\Jpsi}}
{\left|\vec{p}\,(\mu^+)_{\Jpsi}\right|}\cdot\hat{y}}
{\frac{\vec{p}\,(\mu^+)_{\Jpsi}}
{\left|\vec{p}\,(\mu^+)_{\Jpsi}\right|}\cdot\hat{x}} \right),
\end{eqnarray}
where the ambiguity of the angle $\phi_T$ is resolved using signs of
$\vec{p}\,(\mu^+)_{\Jpsi}\cdot\hat{x}$ and
$\vec{p}\,(\mu^+)_{\Jpsi}\cdot\hat{y}$.  The definitions of the
transversity angles are illustrated in
Fig.~\ref{Fig:sketch_transversity_angles}.

The decay is further described
in terms of the polarization states of the vector mesons, either
longitudinal ($0$), or transverse to their directions of motion, and in
the latter case, parallel ($\parallel$) or perpendicular ($\perp$) to
each other. The corresponding amplitudes, which depend on time $t$, are
called $A_0$, $A_{\parallel}$ and $A_{\perp}$, respectively.  The
transverse linear polarization amplitudes $A_{\parallel}$ and
$A_{\perp}$ correspond to \CP-even and \CP-odd final states at decay
time $t=0$, respectively. The longitudinal polarization amplitude $A_0$
corresponds to a \CP-even final state.
The three states in the transversity basis are easily
expressed as linear combinations of states in either the helicity basis
($++,\, 00,\, --$) or the orbital angular momentum basis ($S$, $P$,
$D$). In the helicity basis, $A_{\parallel}$ and $A_{\perp}$ are linear
combinations of the states with helicities $++$ and $--$, while the
state corresponding to $A_0$ is the same in both transversity and
helicity bases. In terms of the $S$, $P$ and $D$-waves, the states
described by $A_0$ and $A_{\parallel}$ are linear combinations of $S$
and $D$ waves, while $A_{\perp}$ corresponds to the $P$-wave state.
Since only differences between the strong phases of these amplitudes are
observable, we define the strong phases relative to $A_{0}(0)$ at 
time $t=0$: $\delta_{0} = 0$, $\delta_{\parallel} =
\arg[A_{\parallel}(0)A^*_{0}(0)]$ and $\delta_{\perp} =
\arg[A_{\perp}(0)A^*_{0}(0)]$.  We note that the strong phases
$\delta_{\parallel}$ and $\delta_{\perp}$ are either $0$ or $\pi$ in the
absence of final state $\Jpsi\phi$ interactions. Deviations of these
phases from $0$ or $\pi$ indicate breaking of the factorization
hypothesis which assumes no interaction between the $\Jpsi$ and $\phi$
in the final state~\cite{Ref:lenz,theory}.

If the decay width difference between the $\Bs$ mass eigenstates
$\deltaG$ is different from zero, a time-dependent angular analysis
without flavor tagging is sensitive to \betas because of the
interference between \CP-odd and \CP-even components
\cite{Aaltonen:2007gf}.  The sensitivity to \betas can be improved by
separating mesons produced as \Bs from those produced as \bBs in order
to detect \CP~asymmetries in the fast \Bs-\bBs~flavor oscillations given
sufficient decay time resolution. The process of separating \Bs mesons
from \bBs mesons at production is called flavor~tagging.

The angular-dependence and flavor~tagged (see Sec.~\ref{sec:tagging})
time-dependence are combined in an unbinned maximum likelihood fit.
The fit is used to extract \betas, the
\Bs~decay width difference \deltaG, the average \Bs lifetime, the
transversity amplitudes and the strong phases.
Since a contamination from $K^+K^-$ final states that do not originate
from a $\phi$~decay can contribute to the $K^+K^-$~mass window used to
identify $\phi$~candidates in this analysis, we consider potential
contributions from other $\Bs\rightarrow \Jpsi K^{+}K^{-}$ decays in our
\BsJpsiPhi candidate sample. In such decays the relative angular
momentum of the two kaons is assumed to be zero ($S$-wave) as expected,
for example, from $f_0(980)\rightarrow K^+K^-$ decays. Continuum
$\Bs\rightarrow \Jpsi K^{+}K^{-}$ decays with angular momentum higher
than zero are expected to be suppressed.  In all such cases the
$K^{+}K^{-}$~system is assumed to be in a partial $S$-wave whose angular
momentum combined with that of the $\Jpsi$ leads to a \CP-odd final
state~\cite{Stone:2008ak}.
The $S$-wave contribution is included in the time-dependent angular
analysis and the $S$-wave fraction together with its corresponding phase
$\delta_{SW}$ are determined as parameters in the maximum likelihood fit.
The inclusion of the $S$-wave in the likelihood function 
constitutes a significant improvement with respect to earlier 
measurements~\cite{Ref:tagged_cdf_old,Ref:tagged_d0}. 

Due to the non-Gaussian behavior of the likelihood function with respect
to the parameters $\betas$ and
$\deltaG$~\cite{Ref:tagged_cdf_old,Ref:tagged_d0}, we use a
frequentist analysis to obtain
confidence regions for both parameters.  
We also determine point estimates for other parameters of interest, like
the polarization fractions and the $\Bs$~lifetime. 
In addition, we perform an alternative 
Bayesian approach, through the use of priors,
applied to probability densities determined with Markov chain Monte
Carlo.

\section{CDF\,II detector and trigger}
\label{sec:detector}

The CDF\,II detector employs a cylindrical geometry around the
$p\bar{p}$ interaction region with the proton direction defining the
positive $z$-direction.  Most of the quantities used for candidate
selection are measured in the plane transverse to the $z$-axis.  In
the CDF coordinate system, $\varphi$ is the azimuthal angle, $\theta$
is the polar angle measured from the proton direction, and $r$ is the
radius perpendicular to the beam axis. The pseudorapidity $\eta$ is
defined as $\eta=-\ln[\,\tan(\theta/2)\,]$.  The transverse
momentum,~$p_T$, is the component of the particle momentum,~$p$,
transverse to the $z$-axis ($p_T = p\cdot \sin\theta$),
while $E_T = E\cdot\sin\theta$, with $E$ being the energy measured in
the calorimeter.
 
The CDF\,II detector features excellent lepton identification and
charged particle tracking and is described in detail
elsewhere~\cite{Ref:CDFdet,Ref:CDF_TDR}. The parts of the detector
relevant to the reconstruction of the $\BsJpsiPhi$ decays used in this
measurement are briefly summarized below.  The detector nearest to the
$p\bar{p}$ interaction region is a silicon vertex detector
(SVX\,II)~\cite{Ref:SVXII}, which consists of five concentric layers of
double-sided sensors located at radii between 2.5 and 10.6~cm plus one
additional single-sided layer of silicon (L00)~\cite{Ref:L00} mounted
directly onto the beam pipe at radius $r$\,$\sim$\,1.5~cm.
In addition, two forward layers plus one central layer of double-sided
silicon located outside the SVX\,II at radii of 20-29~cm make up the
intermediate silicon layers (ISL)~\cite{Ref:ISL}. Together with the
SVX\,II, the ISL detector extends the sensitive region of the CDF\,II
tracking detector to $|\eta|\le2.0$. The CDF silicon system has a
typical hit resolution of $\sim\!11~\mu$m and provides three-dimensional
track reconstruction. It is used to identify displaced vertices
associated with bottom hadron decays which are reconstructed with a
typical transverse resolution of 10-20~$\mu$m. The central outer tracker
(COT)~\cite{Ref:COT}, an open-cell drift chamber with 30\,240 sense
wires arranged in 96 layers combined into four axial and four stereo
super-layers (SL), provides tracking from a radius of $\sim$\,40~cm out
to a radius of 132~cm covering $|z|<155$~cm, as well as the main
measurement of track momentum with a resolution of
$\sigma(p_T)/p_T^2\sim0.15\%\,[\gevc]^{-1}$ for high momentum tracks.
The COT also provides specific ionization energy loss, \dedx, information for
charged particle identification with approximately 1.5\,$\sigma$
separation between pions and kaons with momenta greater than
2~\gevc~\cite{Ref:CDF_dEdx1,*Ref:CDF_dEdx2}.  The central tracking system
is immersed in a superconducting solenoid that provides a 1.4~T axial
magnetic field.  Right outside the solenoid, the time-of-flight (TOF)
detector provides additional particle identification for low-momentum
particles.
 
Central electromagnetic (CEM)~\cite{Ref:CEM} and hadronic
(CHA)~\cite{Ref:CHA} calorimeters ($|\eta|<1.1$) are located outside the
COT and the solenoid, where they are arranged in a projective-tower
geometry.  The electromagnetic and hadronic calorimeters are
lead-scintillator and iron-scintillator sampling devices, respectively.
The energy resolution for the CDF central calorimeter is $\sigma(E_T) /
E_T = [(13.5\% / \sqrt{E_T})^2 + (1.5\%)^2]^{1/2}$ for electromagnetic
showers~\cite{Ref:CEM,Ref:CEM2} and $\sigma(E_T) / E_T = [(75\% /
\sqrt{E_T})^2 + (3\%)^2]^{1/2}$ for hadrons~\cite{Ref:CDF_TDR,Ref:CHA},
where $E_T$ is measured in GeV.  A layer of proportional chambers (CES),
with wire and strip readout, is located six radiation lengths deep in
the CEM calorimeters, near the electromagnetic shower maximum.  The CES
provides a measurement of electromagnetic shower profiles in both the
$\varphi$ and $z$~directions for use in electron identification.

Muon candidates are identified by multi-layer drift chambers and
scintillator counters~\cite{Ref:CMU,*Ref:mu_upgrade}.  Four layers of
planar drift chambers (CMU) are located outside the central calorimeter
at a radius of $347$~cm from the beam line.  The CMU system covers
$|\eta|\leq0.6$ and can be reached by muons with $p_T$ in excess of
$\sim\!1.4~\gevc$.  To reduce the probability of misidentifying
penetrating hadrons as muon candidates in the central detector region,
four additional layers of drift chambers (CMP) are located behind 0.6~m
of steel outside the CMU system. Approximately 84\% of the solid angle
for $|\eta|\leq0.6$ is covered by the CMU detector, 63\% by the CMP,
and 53\% by both. To reach these two detectors, particles produced at
the primary interaction vertex, with a polar angle of $90^{\circ}$, must
traverse material totaling 5.5 and 8.8 pion interaction lengths,
respectively.  Muons with hits in both the CMU and CMP detectors are
called CMUP muons.  An additional set of muon chambers (CMX) is located
in the pseudorapidity interval $0.6<|\eta|<1.0$ to extend the
polar acceptance of the muon system to the forward region. Approximately
71\% of the solid angle for $0.6<|\eta|<1.0$ is covered by the
free-standing conical arches of the CMX. The calorimeter, magnet yoke
of the detector, and the steel support structure provide shielding of
about 6.2 pion interaction lengths.

The data used in this measurement are collected with dimuon
triggers~\cite{Ref:CDFdet}.  Muons are reconstructed as track stubs in
the CMU, CMP and CMX chambers. Muon stubs are matched to tracks
reconstructed using COT axial information from the extremely fast
trigger (XFT)~\cite{Thomson:2002xp}. The dimuon trigger requires at
least one central muon matching the CMU or CMUP chambers, while the
second muon can be either central or forward, matching to the CMU or CMX
detectors, respectively.  The CMU, CMUP, or CMX muons must satisfy
$p_T>1.5~\gevc$, $p_T>3.0~\gevc$ and $p_T>2.0~\gevc$, respectively. The
two trigger muon candidates
are required to be oppositely charged, have an opening angle
inconsistent with a cosmic ray event, and the invariant mass of the muon
pair must satisfy $2.7<m(\mu^+\mu^-)<4~\gevcc$.

\section{Data reconstruction and selection}
\label{sec:reconstruction}

We use a data sample corresponding to an integrated luminosity of
5.2~\fb collected with all CDF\,II detector subsystems functioning. In
addition, all analyzed data passed the dimuon trigger requirements
given above. We begin our offline reconstruction of the $\Bs \rightarrow
\Jpsi (\rightarrow \mu^{+}\mu^{-}) \phi (\rightarrow K^{+}K^{-})$ decay
mode by requiring two muon candidate tracks that extrapolate to a track
segment in the muon detectors reapplying the appropriate transverse
momentum requirements for the respective trigger muons using
offline-reconstructed quantities. To reconstruct the \Jpsi candidate, a
kinematic fit constraining the two oppositely charged muon candidate
tracks to a common interaction point (vertex) is applied.  All other
charged particles in the event are assumed to be kaon candidates and
combined as opposite-charge pairs to reconstruct $\phi$ meson candidates.
Finally, all four candidate tracks are combined in a kinematic
fit that constrains the muon candidates to the \Jpsi world average
mass~\cite{Ref:PDG} and requires the four tracks to originate from a
common three-dimensional vertex point.

For each event a primary interaction point is constructed from all
reconstructed tracks in an event, excluding the $\Jpsi$ and $\phi$
candidate tracks.  This interaction point is used in the calculation of
the \Bs proper decay time, $ct = m(\Bs) L_{xy}(\Bs)/p_T(\Bs)$, where
$L_{xy}(\Bs)$ is the distance from the primary vertex point to the
$\BsJpsiPhi$ decay vertex projected onto the momentum of the $\Bs$ in
the plane transverse to the proton beam direction, $m(\Bs)$ is the
nominal mass of the $\Bs$~meson~\cite{Ref:PDG}, and $p_T(\Bs)$ is its
measured transverse momentum.

\subsection{Basic selection criteria}

For the final selection of \BsJpsiPhi candidates, we use the
Neurobayes~\cite{Ref:Neurobayes} artificial neural network (ANN) to
distinguish signal events from background. Prior to the training of the
ANN, we apply basic selection requirements 
in order to ensure track and vertex quality, as well as the
reconstructed particle candidates to have kinematic properties
appropriate for \Bs, \Jpsi, and $\phi$~mesons. These basic selection
criteria are listed in Table~\ref{tab:preselec} summarizing the
standard quality requirements that were imposed on
track candidates. 
We require kaon candidates to have transverse momentum greater than
400~\mevc and all kinematic fits are required to have $\chi_{r\phi}^{2}<
50$, where $\chi_{r\phi}^{2}$ is the $\chi^{2}$ of the two-dimensional
$r\phi$-vertex fit for four degrees of freedom (dof). To select
\BsJpsiPhi candidates, we require that the invariant mass of the muon
pair lies within the mass region $3.04 < m(\mu^+\mu^-) < 3.14~\gevcc$
corresponding to an interval around the world average $\Jpsi$
mass~\cite{Ref:PDG} of about $\pm2.5\sigma$ where $\sigma$ is the
$J/\psi$~mass resolution. The invariant mass of the kaon pair is
required to be within $1.009 < m(K^+K^-) <1.028~\gevcc$ corresponding to
a $\pm2.5\sigma$ interval around the nominal $\phi$~mass~\cite{Ref:PDG}
where $\sigma$ corresponds to the $\phi$~mass resolution. The mass of
the reconstructed $\Jpsi\phi$ candidate has to be in the mass window
$5.1 < m(\Jpsi K^+K^-) < 5.6~\gevcc$ corresponding to a $\pm 250~\mevcc$
interval around the nominal \Bs~mass~\cite{Ref:PDG}. Additionally we
require that the transverse momentum of the $\phi$~candidate is greater
than 1.0~\gevc and the \Bs~candidate has a transverse momentum of more
than 4.0~\gevc.

\begin{table}[tbp]
\caption{Basic selection requirements applied to the \BsJpsiPhi
  four-track system as used in training the artificial neural network.
\label{tab:preselec}}
\begin{ruledtabular} 
\begin{tabular}{ll}
Quantity                 & Selection requirement            \\
\hline		           				                    
COT hits                 & $\ge2$ stereo and $\ge2$ axial super-layers \\
                         & with $\ge5$ hits each \\
$r$-$\phi$ silicon hits  & $\ge 3$                                        \\
Kaon track $p_T$         & $> 0.4~\gevc$                                  \\
Vertex $\chi_{r\phi}^{2}$ (4 dof)  & $< 50$                            \\
\Jpsi  mass region       & $3.04 < m(\mu^+\mu^-) < 3.14~\gevcc$           \\
$\phi$ mass region       & $1.009 < m(K^+K^-) <1.028~\gevcc$              \\
\Bs mass region          & $5.1 < m(\Jpsi K^+K^-) < 5.6~\gevcc$             \\
$p_T(\phi)$              & $> 1.0~\gevc$                                  \\
$p_T(\Bs)$               & $> 4.0~\gevc$                                  \\
\end{tabular}
\end{ruledtabular} 
\end{table}

\subsection{Monte Carlo simulation}

We use simulated \BsJpsiPhi Monte Carlo (MC) event samples to describe
the signal in the training of the artificial neural network. These MC
samples are also employed in the determination of the transversity angle
efficiencies due to the non-hermeticity of the CDF\,II detector (see
Sec.~\ref{sec:fit}). We simulate the generation and fragmentation of
\textit{b}~quarks using the \textsc{bgenerator} program~\cite{Ref:BGen}. It
is based on next-to-leading-order QCD calculations and the Peterson
fragmentation function~\cite{Ref:Peterson} tuned to the 
$b$-quark momentum spectrum measured at CDF~\cite{Ref:CDFdet}.
The decay of the \Bs~meson is simulated with the \textsc{evtgen} decay
package~\cite{Ref:Evtgen}. The interaction of the generated particles
with the CDF\,II detector is simulated with the full 
\textsc{geant}~\cite{Ref:Geant1,*Ref:Geant2} based CDF\,II detector simulation
package~\cite{Ref:CDF_simulation}. We subject the simulated events to
the same trigger requirements and reconstruction process as our data
events.  The $\Bs$ decays are simulated according to the phase space
available to the decay averaging over the spin states of the decay
daughters. This procedure ensures that all transversity angles are
generated flat for \Bs decays.

\subsection{Selection using artificial neural network}

The information from several kinematic variables is combined into a single
discriminant by the artificial neural network. Based on the
discriminant, an event is classified as background-like or signal-like
on a scale of $-1$ to $+1$. Correlations between variables are taken
into account by the ANN, and the weight of each variable in the overall
discriminant depends on its correlation with other variables. 
Our artificial neural network
is trained on a signal sample based on 350\,000 \BsJpsiPhi Monte Carlo
events. The background sample used in training the ANN consists of
$\sim300\,000$ data events taken from the $\Bs$~invariant mass sideband
regions $(5.2, 5.3) \cup (5.45, 5.55)$~\gevcc.

We use the following variables as input to the ANN listed in order of
discriminating power and relevance
to the final discriminant: the transverse momentum $p_{T}$ of
the $\phi$ meson, the kaon likelihood~\cite{Aaltonen:2007gf} based
on TOF and \dedx~information, the muon
likelihood~\cite{Ref:Thesis_Giurgiu} for the \Jpsi muon daughters,
$\chi^{2}_{r\phi}$ for the \Bs~decay vertex reconstruction, the
transverse momentum $p_{T}$ of the \Bs~meson, and the probabilities to
reconstruct vertices of the \Bs, $\phi$, and \Jpsi candidates. These
vertex probabilities are $\chi^{2}$ probabilities for the three-dimensional
vertex fit, while $\chi^{2}_{r\phi}$ is the goodness of fit for the
two-dimensional vertex fit. The muon and kaon likelihoods are quantities
used for particle identification. The algorithm determining the muon
likelihood is described in Ref.~\cite{Ref:Thesis_Giurgiu}. The kaon
likelihood~\cite{Aaltonen:2007gf} is a combined
discriminant constructed from the kaon track specific energy loss,
\dedx, and its time-of-flight information. Both likelihood variables
have been calibrated on large control samples.

\subsection{Optimization of selection}

Once the artificial neural network is trained, we choose a discriminant
cut value that provides the best expected average resolution
(sensitivity) to the \CP-violating phase $\betas$. This optimization
differs from our previous analysis~\cite{Ref:tagged_cdf_old} where the
significance of \Bs~signal yield was maximized using $S/\sqrt{S+B}$ as
figure of merit, in which the signal $S$ was obtained from Monte Carlo
simulation of \Bs events and the background $B$ taken from the
\Bs~sideband regions. In the current optimization we study the
sensitivity to $\betas$ as a function of ANN discriminant $C_{NN}$ using
pseudoexperiments that are generated to mimic our data with a specific
signal-to-background ratio for a given cut value on $C_{NN}$.
Using the likelihood fit function described in Sec.~\ref{sec:fit}, we 
repeat the entire analysis procedure for each pseudoexperiment and
evaluate the distributions of the estimated variance on \betas to find
the ANN discriminant cut value that gives the best expected average
resolution for \betas~\cite{Ref:Thesis_Elisa}.

In detail, the pseudoexperiments are created by randomly sampling the
probability density functions (PDF) for variables used in the fit to
describe the data as outlined in Sec.~\ref{sec:fit}.
We simulate the effect of varying the cut on the ANN output variable by
generating pseudoexperiments at different values of $S/N_{TOT}$, where
$N_{TOT}$ is the total number of events in the $\Jpsi\phi$ invariant
mass window, and \textit{S} is the number of $\Bs$ signal events.
$N_{TOT}$ and \textit{S} are determined from mass fits to the data for
different cut values of $C_{NN}$.  The input values of all other
parameters in the PDF are kept the same for all pseudoexperiments
corresponding to the same ANN cut; only $N_{TOT}$ and $S/N_{TOT}$ are
varied. We take the parameter input values from the results of the
unbinned maximum likelihood fit of \BsJpsiPhi~decays using data
corresponding to an integrated luminosity of
2.8~\fb~\cite{Ref:ICHEP2008}. The only exceptions are the parameters
describing the tagging power, which correspond to the total tagging
effectiveness of both the opposite and same side tagging algorithms (see
Sec.~\ref{sec:tagging}) valid for the full data set of 5.2~\fb
integrated luminosity. To verify in our optimization that the
expected average resolution to $\betas$ is independent of the true values of our
parameters of interest, \betas and \deltaG, we generate
pseudoexperiments at a few points in (\betas, \deltaG) parameter space:
($\betas=0.5$, $\deltaG=0.12$~\ps), ($\betas=0.02$, $\deltaG=0.1$~\ps), and
($\betas=0.3$, $\deltaG=0.09$~\ps).  We also verify that the widths of the
uncertainty distributions do not vary as a function of artificial neural
network discriminant~\cite{Ref:Thesis_Elisa}.

\begin{figure}[tbp]
\includegraphics[width=0.48\textwidth]{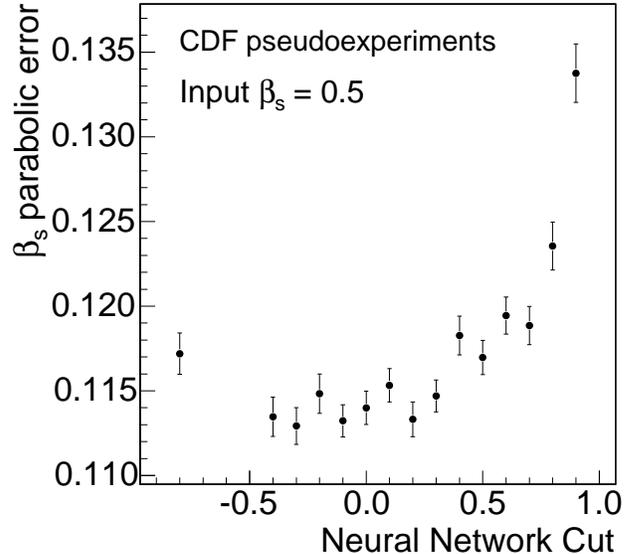}
\caption{Magnitude of expected uncertainty on \betas as a function of 
cut value on the artificial neural
network discriminant for pseudoexperiments generated with
input values $\betas=0.5$ and $\deltaG=0.12$~\ps.
\label{sensitivity_v_nn}}
\end{figure}

The most probable values of the \betas~uncertainty for pseudoexperiments
generated at ($\betas=0.5$, $\deltaG=0.12$~\ps) are shown as a function
of the cut value on the ANN output variable in
Fig.~\ref{sensitivity_v_nn}. It is apparent that tight cuts on $C_{NN}$
correspond to larger \betas expected uncertainties. It should be
emphasized that this technique is not intended to guarantee a particular
uncertainty on \betas, but is merely used to identify the trend in
\betas uncertainty size as a function of cut value on the ANN output
variable.

The trend in size of the \betas uncertainty distributions is similar for
the other sets of (\betas, \deltaG) at which pseudoexperiments are
generated. The expected statistical uncertainty on \betas shows a shallow
minimum around $C_{NN}\sim0$. We adopt a cut on the ANN output
discriminant at $>0.2$, where the uncertainty on \betas is small,
allowing us to avoid adding unnecessary amounts of background which we
would include by going to a lower cut value on $C_{NN}$. For comparison,
an optimization using $S/\sqrt{S+B}$ as figure of merit yields a cut
value of the ANN discriminant of about 0.9 resulting in an
almost 20\% larger expected uncertainty on \betas.
 
\begin{figure*}[tbp]
\centerline{
\includegraphics[width=0.49\textwidth]{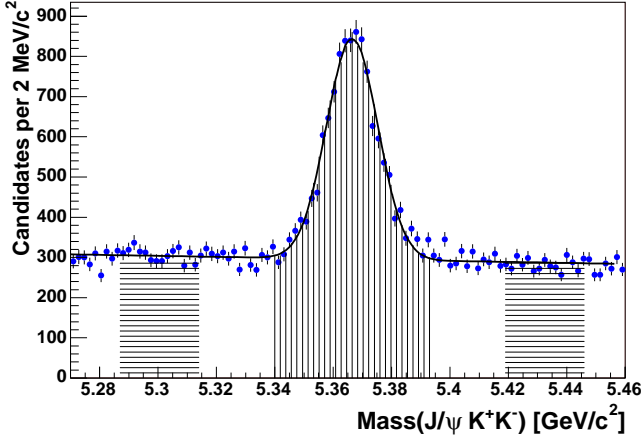}
\ \ \ 
\includegraphics[width=0.48\textwidth]{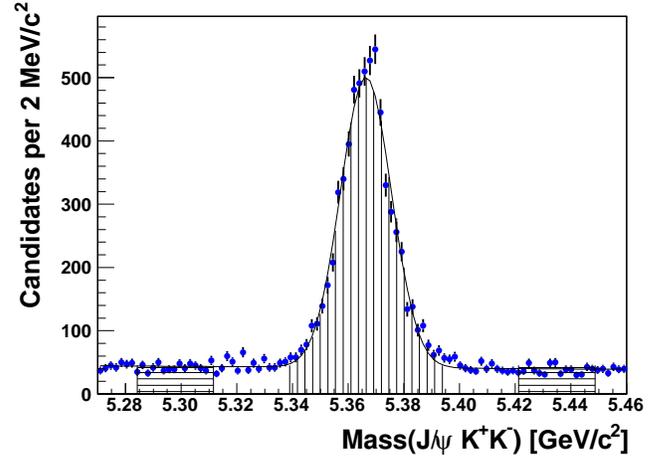}
}
\caption{$\Jpsi K^+K^-$ invariant mass distribution with a cut of 0.2 on
  the ANN discriminant (left) and in addition with a lifetime
  requirement $ct>60~\mu$m on the \Bs~candidate (right). The areas with
  vertical (horizontal) lines indicate the signal (sideband) regions
  used in Sec.~\ref{sec:fit}.
\label{bsmass}}
\end{figure*}

A cut on the ANN output discriminant of 0.2 yields $6504 \pm 85$
\BsJpsiPhi signal events, as extracted by a fit to the invariant mass
with a single Gaussian with flat background as shown in
Fig.~\ref{bsmass} on the left-hand side. The signal and sideband regions
used to describe 
signal and background events in the PDF (see Sec.~\ref{sec:fit}) are
indicated by the areas with vertical and horizontal lines,
respectively. They explicitly correspond to
$5.340<m(J/\psi K^+K^-)<5.393~\gevcc$ for the signal region and
$(5.287,5.314) \cup (5.419,5.446)~\gevcc$ for the sidebands.
The right-hand side of Fig.~\ref{bsmass} shows the $\Jpsi K^+K^-$
invariant mass distribution with an additional lifetime requirement  
$ct>60~\mu$m on the \Bs~candidate indicating the combinatorial
background to be mainly prompt. 

\subsection{Initial study of \textit{S}-wave contribution}

\begin{figure}[tbp]
\includegraphics[width=0.48\textwidth]{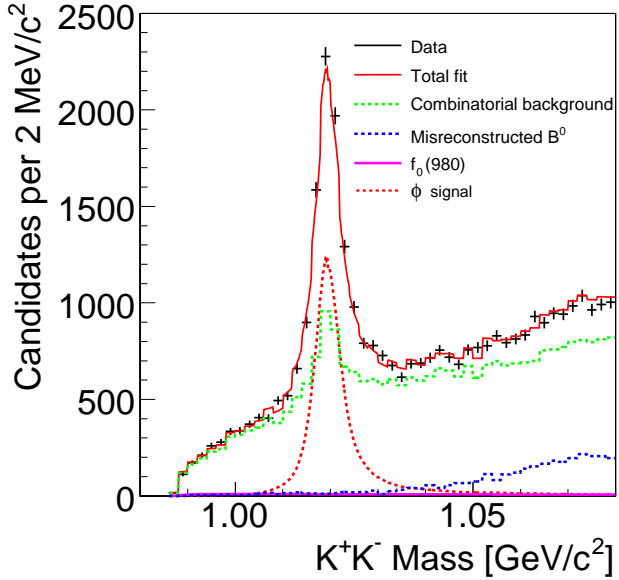}
\caption{(color online). $K^+K^-$ invariant mass distribution with combinatorial
  background, $B^0$~reflection, and potential $S$-wave contribution.
\label{KKmass}}
\end{figure}

Since this analysis considers a potential $S$-wave contribution, which
is not coming from $\phi\rightarrow K^+K^-$ decays, to the
$\Bs\rightarrow \Jpsi K^{+}K^{-}$ signal in the description of the
likelihood function (see Sec.~\ref{sec:fit}), we perform an initial
study to obtain an estimate of the size of the $S$-wave contribution
using a data sample corresponding to integrated luminosity of
3.8~fb$^{-1}$. We select $\phi$~candidates in a larger $K^+K^-$~mass
window $0.98<m(K^+K^-)<1.08~\gevcc$ and form \Bs~candidates.  Taking
\Bs~candidates from the signal region defined above, we obtain the
$K^+K^-$~invariant mass distribution shown in Fig.~\ref{KKmass}.  Using
a binned likelihood method, we fit the $K^+K^-$~invariant mass
distribution with a $\phi$~signal component modeled by a template from
\BsJpsiPhi~MC simulation allowing for a mass-dependent width consistent
with the parameterization in Eq.~(\ref{eqn:phi_prop}) (see
Sec.~\ref{sec:fit}).  The combinatoric background is modeled by a
histogram taken from the \Bs~sidebands.  An additional component, which
takes into account $B^0$~reflections where the pion from a $B^0$~decay
is misidentified as a kaon, is obtained from an inclusive $B^0\to J/\psi
X$ MC simulation.  The fractions of combinatorial background and
$B^0$~reflection are fixed from a fit to the \Bs~invariant mass, which
prevents these components from absorbing a non-resonant component in the
$K^+K^-$~mass distribution if present.  Together with the $\phi$~signal
and the fixed components, an additional contribution is included in the
fit allowing for a possible non-resonant $S$-wave contribution from
$\Bs\to J/\psi K^+K^-$ or $\Bs\to J/\psi f_0(980)$. This component is
modeled either flat in $K^+K^-$~mass or following a mass
parameterization suggested in Ref.~\cite{Flatte:1976xu}. In either case
the $S$-wave fraction is found to be compatible with zero as indicated
in Fig.~\ref{KKmass}. From this study we do not expect a significant
$S$-wave contribution across the $\phi$~mass region.

\section{Flavor Tagging}
\label{sec:tagging}

To maximize the sensitivity to the \CP-violating phase $\betas$, we
employ flavor tagging algorithms to determine whether the reconstructed
$\Jpsi\phi$~candidate was a \Bs~meson or its anti-particle \bBs at the
time of production.
Flavor tagging algorithms assign each \Bs~meson candidate a tagging
decision and a tagging dilution.  The tagging decision can be $\xi = +1$
or $-1$, corresponding to a \Bs~meson at production or a \bBs~meson,
respectively. A value of $\xi=0$ means the tagging algorithm failed and
no tag is assigned.  The tagging dilution $\mathcal{D}$ is related to
the probability that the tagging decision is correct, $\mathcal{D} = 1 -
2p_{W}$, where $p_{W}$ is the probability of an incorrect tag or
mis-tag.  In general, the dilution is obtained by counting the number
of correctly and incorrectly assigned tags $\mathcal{D} = (N_{R} -
N_{W})/(N_{R} + N_{W})$, where $N_{R}$ is the number of correct tags and
$N_{W}$ is the number of incorrect tags.  The flavor tagging performance
is quantified by the product between the tagging efficiency and the
squared dilution $\varepsilon\mathcal{D}^2$.  The tagging efficiency
$\varepsilon$ is defined as the number of events that receive a tag,
divided by the total number of events considered.
We use two types of flavor tagging algorithms: a same side (SST) and an
opposite side (OST) tagger. Due to the use of information from different
event hemispheres, there is no overlap between both taggers by
construction. In particular, the SST algorithm considers only tracks
within a cone of $\sqrt{(\Delta\phi)^2+(\Delta\eta)^2} < 0.7$ around the
\Bs~candidate while the OST tagger only includes tracks outside that
cone, allowing us to treat SST and OST as uncorrelated tagging methods.

\subsection{Same side tagging}
\label{sec:same_side_tagging}

\begin{figure}
\includegraphics[width=0.2\textwidth]{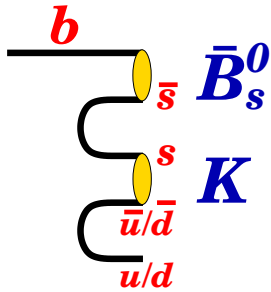}
\caption{
Illustration of $b$~quark fragmentation into \bBs~meson.
\label{fig:Bs_fragmentation}}
\end{figure}

In this analysis we employ a same side kaon
tagging (SSKT) algorithm which uses the charge of the kaon produced in
association with the $b~(\bar{b})$ quark in the fragmentation process
forming the \bBs (\Bs)~meson as illustrated in
Fig.~\ref{fig:Bs_fragmentation}.  To determine the
$b~(\bar{b})$~production flavor we attempt to find kaon tracks produced
in the hadronization of the \Bs~meson.
The strangeness of the \Bs meson preferentially produces associated
kaons in the fragmentation process. The charges of these nearby kaons are
correlated to the $b$~quark content of the \Bs~meson and provide an
opportunity to identify the initial flavor of the \Bs~meson. However,
the \Bs~meson can also be accompanied by a neutral kaon
which cannot be used to tag the \Bs~flavor and therefore lowers the tagging
power.  Misidentification of the associated charged kaon leads to a
further decrease of the tagging dilution.  The SSKT algorithm was
developed on a simulated high statistics Monte Carlo \Bs sample, using
the $B^{0}_{s}\rightarrow J/\psi\phi$ and $\Bs\to
D^{-}_{s}\pi^{+}$~decay modes. We use particle identification (\dedx and
time-of-flight) to help identify the associated track as a
kaon~\cite{Ref:bsmixing1,Ref:bsmixing2,Ref:Thesis_Belloni}.

\begin{figure*}
\centerline{
\includegraphics[width=0.47\textwidth]{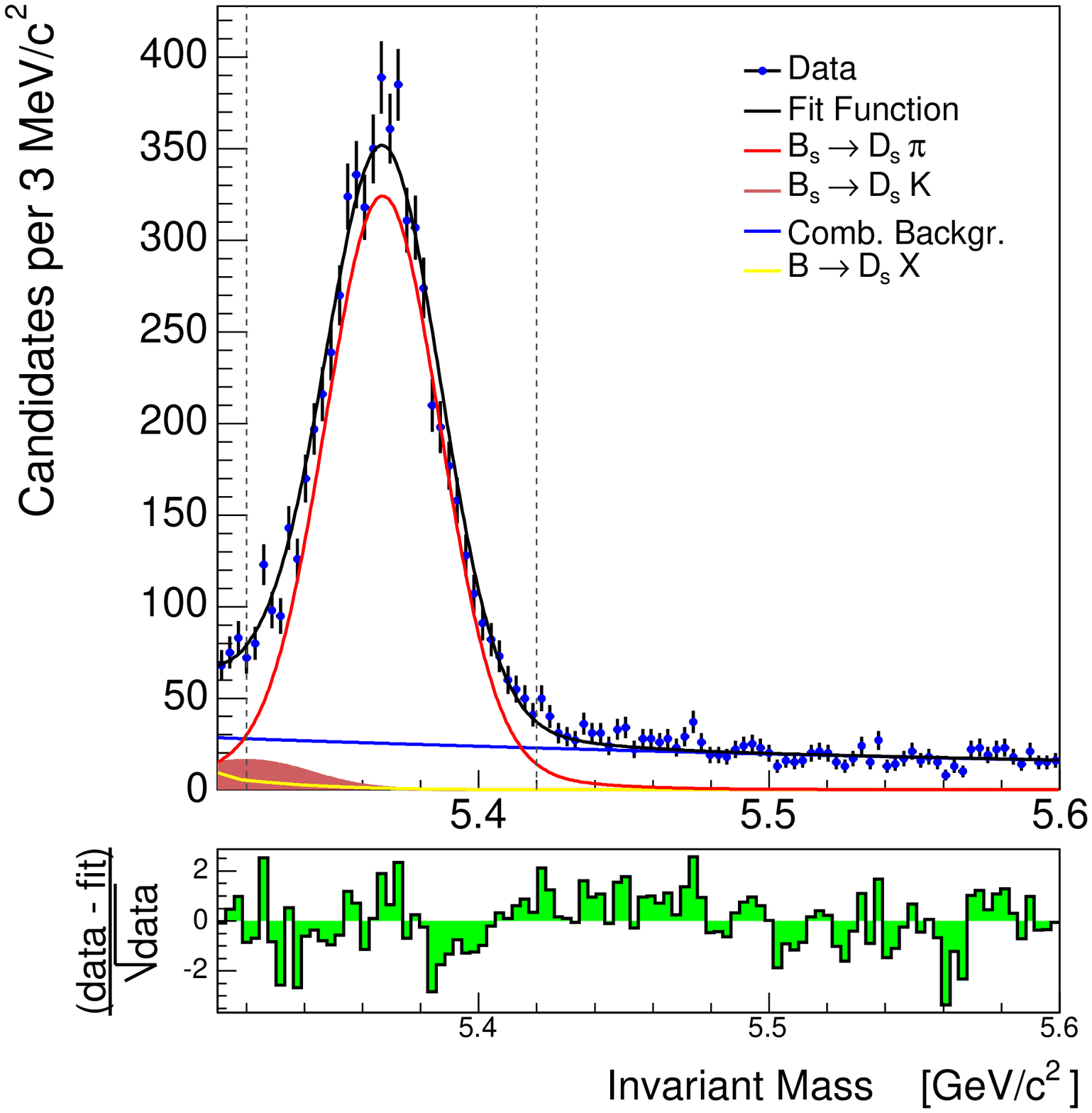}
\includegraphics[width=0.52\textwidth]{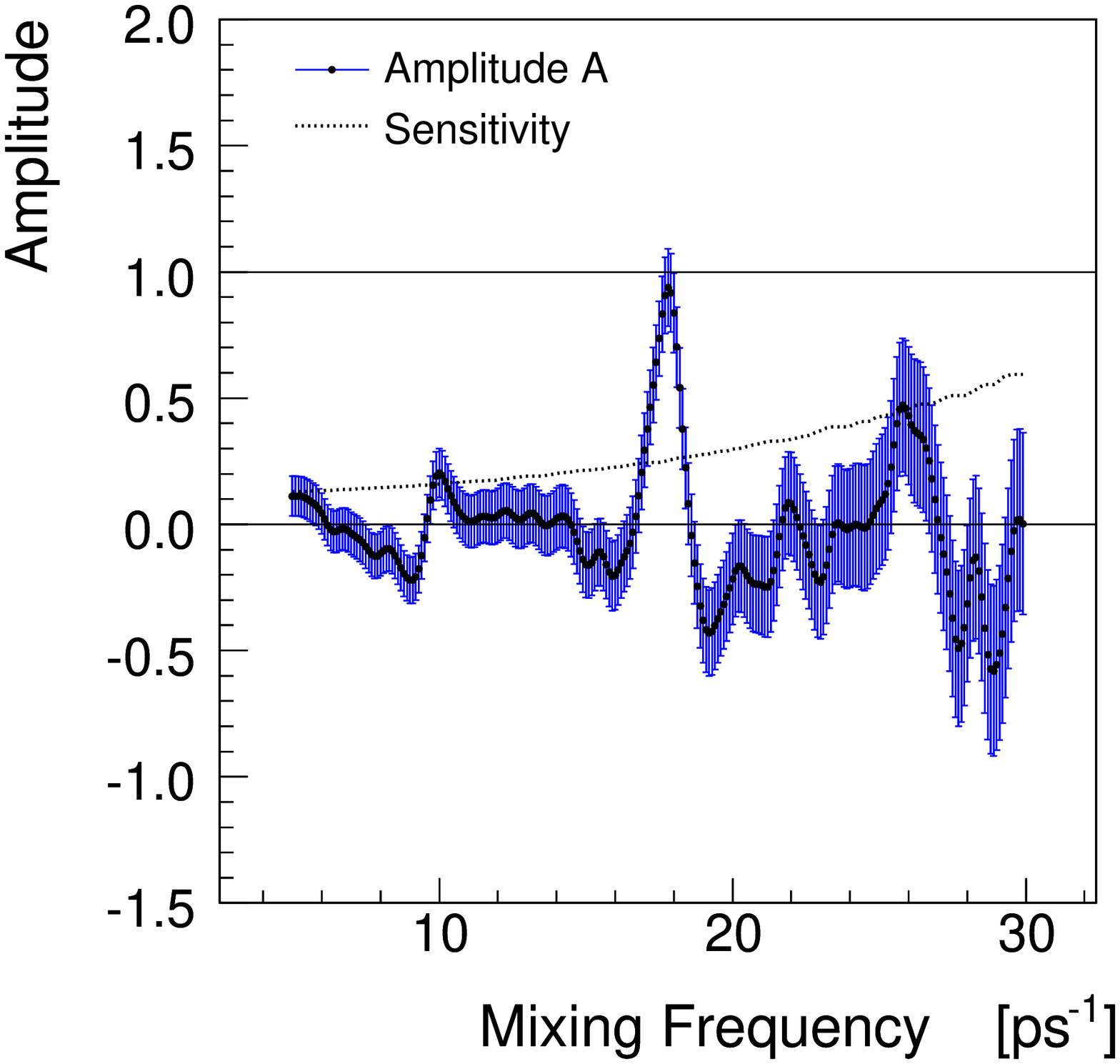}}
\caption{(color online).
Left: Invariant mass of $\Bs \rightarrow D_s^- \pi^+$ candidates with
$D_s^- \rightarrow \phi \pi^-$ and $\phi \rightarrow K^+K^-$ including 
background contributions. The pull distribution at the bottom shows the
difference between data and fit value normalized to the data uncertainty.
Right: The measured amplitude $\mathcal{A}$ and uncertainties versus
the \Bs~oscillation frequency \deltaM for the combined set of
$\Bs\to D_s^-\pi^+$ and $\Bs\to D_s^-\pi^+\pi^+\pi^-$ candidates. 
The sensitivity curve represents $1.645\,\sigma_{\mathcal{A}}$, where 
$\sigma_{\mathcal{A}}\propto e^{(\deltaM\sigma_t)^2/2}$ is the expected
uncertainty on $\mathcal{A}$ for a given value of \deltaM.
Note that the uncertainties shown are correlated between data points.
\label{All_AMP_SSKT}}
\end{figure*}

We calibrate the SSKT algorithm~\cite{Ref:Thesis_Morlock} by measuring
the \Bs mixing frequency on a data set corresponding to 5.2~\fb
integrated luminosity. Using CDF's Silicon Vertex
Trigger~\cite{Ref:SVT}, we select events that contain $\Bs \rightarrow
D_s^- \pi^+$ candidates. The trigger configuration used to collect this
heavy flavor data sample is described in Ref.~\cite{Ref:TTT}.
$\Bs \rightarrow D_s^- \pi^+$ events are fully reconstructed in three
$D_s^-$ decay modes:
$D_s^- \rightarrow \phi \pi^-$ with
$\phi \rightarrow K^+K^-$ (5600 events), 
$D_s^- \rightarrow K^{*0} K^-$ with 
$K^{*0} \rightarrow K^+ \pi^-$ (2760 events), and
$D_s^- \rightarrow \pi^- \pi^- \pi^+$ (2650 events). 
We also include the decay mode
$\Bs \rightarrow D_s^- \pi^+\pi^+\pi^-$, 
with $D_s^- \rightarrow \phi \pi^-$ and 
$\phi \rightarrow K^+K^-$ (1850 events). 
To illustrate this sample of \Bs~candidates, the left-hand side of
Fig.~\ref{All_AMP_SSKT} shows the invariant mass of 
$\Bs \rightarrow D_s^- \pi^+$
candidates with $D_s^- \rightarrow \phi \pi^-$ including background
contributions.

The calibration of the SSKT is achieved via an amplitude scan of the mixing
frequency~\deltaM. 
The probability for observing a \Bs~meson in a \Bs or \bBs~flavor
eigenstate as a function of time is
\begin{eqnarray}
P(t)_{\Bs,\bBs} \propto 1 \pm \mathcal{A\,D}_{p} \cos \deltaM t,
\end{eqnarray} 
where $\mathcal{D}_{p}$ is the event by event predicted dilution and
$\mathcal A$ is a Fourier-like coefficient called ``amplitude''.  The
amplitude scan consists of a series of steps in which the mixing
frequency \deltaM is fixed at values between zero and 30~ps$^{-1}$.  At
each step, the likelihood function based on the above probability
density function, is maximized and the best fit value of the amplitude
parameter is determined.  Whenever the mixing frequency is fixed to
values far from the true mixing frequency, the best fit value of the
amplitude parameter is consistent with zero. On the contrary, when
values of \deltaM close to the true \Bs~mixing frequency are probed, the
best fit value of the amplitude parameter is inconsistent with zero. If
the dilution~$\mathcal{D}_{p}$, which is predicted on an event-by-event
basis by the tagging algorithms, is correct, the amplitude $\mathcal A$
will be close to unity at the true value of \deltaM. Deviations from
unity indicate that the predicted dilution has to be re-scaled by the
actual value of the amplitude parameter at the amplitude maximum. This
value of $\mathcal A$ is also called the dilution scale factor
$\mathcal{S_{D}}$. If the dilution scale factor is larger (smaller) than
unity, the tagging algorithm under (over) estimates the predicted
dilution.  Multiplying the predicted SSKT dilution by $\mathcal{S_{D}}$
will then provide on average the correct event-by-event dilution.

The result of the \deltaM amplitude scan is shown in
Fig.~\ref{All_AMP_SSKT}.  Maximizing the likelihood as a function of
\deltaM measures the frequency of \Bs~oscillations at $\deltaM = 17.79
\pm 0.07\,\textrm{(stat)}~\ps$ in agreement with the world average value of
\deltaM~\cite{Ref:PDG}.  At this point the maximum amplitude is
consistent with one and the measured dilution scale factor for the SSKT
algorithm is thus consistent with unity, indicating that the initial
calibration based on simulated events was accurate. We find
$\mathcal{S_{D}} = 0.94 \pm 0.15\,\textrm{(stat)} \pm 0.13\,\textrm{(syst)}$.
This is the first time the SSKT dilution was calibrated using data only.
We measure a tagging efficiency of $\varepsilon = (52.2 \pm 0.7)\%$ and
an average predicted dilution of $\sqrt{<\mathcal{D}_p^2>} = (27.5 \pm
0.3)\%$.  The total tagging power is found to be $\varepsilon
\mathcal{S_{D}}^2 <\mathcal{D}_p^2>\ = (3.5 \pm 1.4)\%$.


\subsection{Opposite side tagging}
\label{sec:opposite_side_tagging}

The opposite side tagger capitalizes on the fact that most $b$ quarks
produced in $p\bar{p}$ collisions originate from $b\bar{b}$ pairs. The
$b$ or $\bar{b}$ quark on the opposite side of the \Bs~or \bBs~candidate
hadronizes into a $B$ hadron, whose flavor can be inferred using its
decay products.

The OST is a combination of several algorithms: the soft muon tagger
(SMT)~\cite{Ref:Thesis_Giurgiu}, the soft electron tagger
(SET)~\cite{Ref:Thesis_Tiwari}, and the jet charge tagger
(JQT)~\cite{Ref:Thesis_Lecci}. 
The JQT combines all tracks from the fragmentation of the
opposite-side $b$~quark into a single jet charge measurement. The
charge of the jet is determined by the momentum-weighted sum over the
momenta $p_T^i$ of all tracks in the jet $Q_{jet}
= \sum_{i}Q^{i}p^{i}_{T}(1 + P^{i}_{trk})/\sum_{i}p^{i}_{T}(1 +
P^{i}_{trk})$, where $Q_i=\pm1$ is the electric charge of track and
$P^{i}_{trk}$ is the probability of the track 
being part of the $b$~jet. Jets are reconstructed by a cone-clustering
algorithm and separated into three classes, based on their probability
of containing a $b$~quark. A class 1 jet has a vertex displaced 
with respect to the primary $p\bar p$~interaction vertex, while
a class 2 jet contains at least one track displaced with respect to the
primary vertex.  If there are no class 1 and 2 jets found, the jet with
the highest transverse momentum in the event is used. These jets
constitute class 3 jets which can be identified for nearly 100\% of the
events.

The lepton taggers, SMT and SET, utilize the charge of a muon or
electron from a $b\rightarrow c\,\ell^-\bar{\nu}_{\ell}$ transition to determine
the production flavor of the parent $B$~meson. Several variables used to
identify electrons and muons are combined with a multivariate technique
into a global likelihood to select the lepton candidates used in the
tagging algorithms~\cite{Ref:Thesis_Giurgiu,Ref:Thesis_Tiwari}.
 
The outcomes of the three OST algorithms are combined to give a single
tag decision and predicted dilution~\cite{Ref:bsmixing1}. We optimize the
total dilution by combining the taggers using an artificial neural
network trained on data from semileptonic $B\to\ell\nu X$ decays. The
neural network handles correlations between the jet charge tagger and
the lepton taggers and improves the tagging performance by 15\%
relative to other combinations that do not account for correlations.

\begin{figure}
\includegraphics[width=0.49\columnwidth]{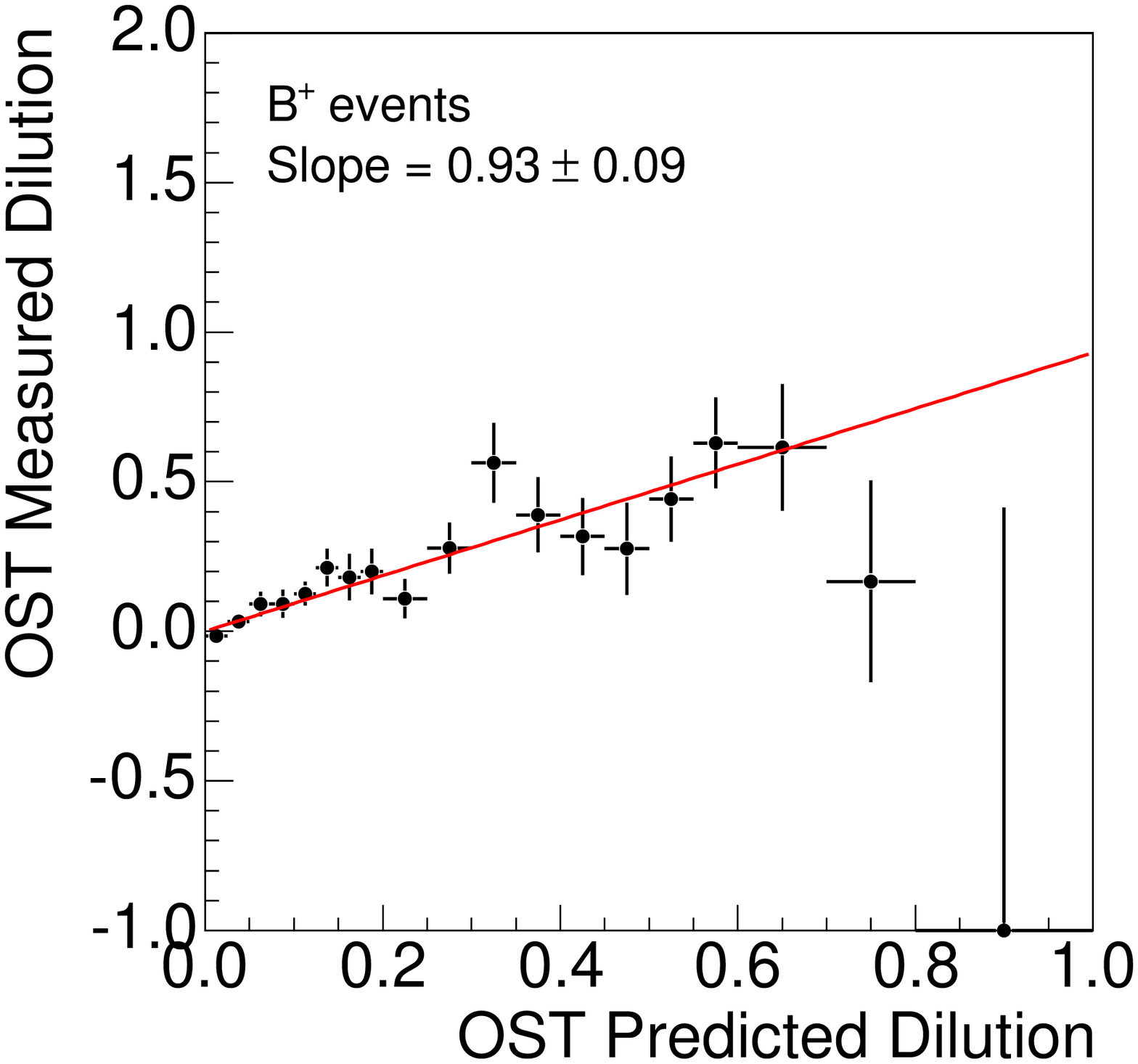}
\includegraphics[width=0.49\columnwidth]{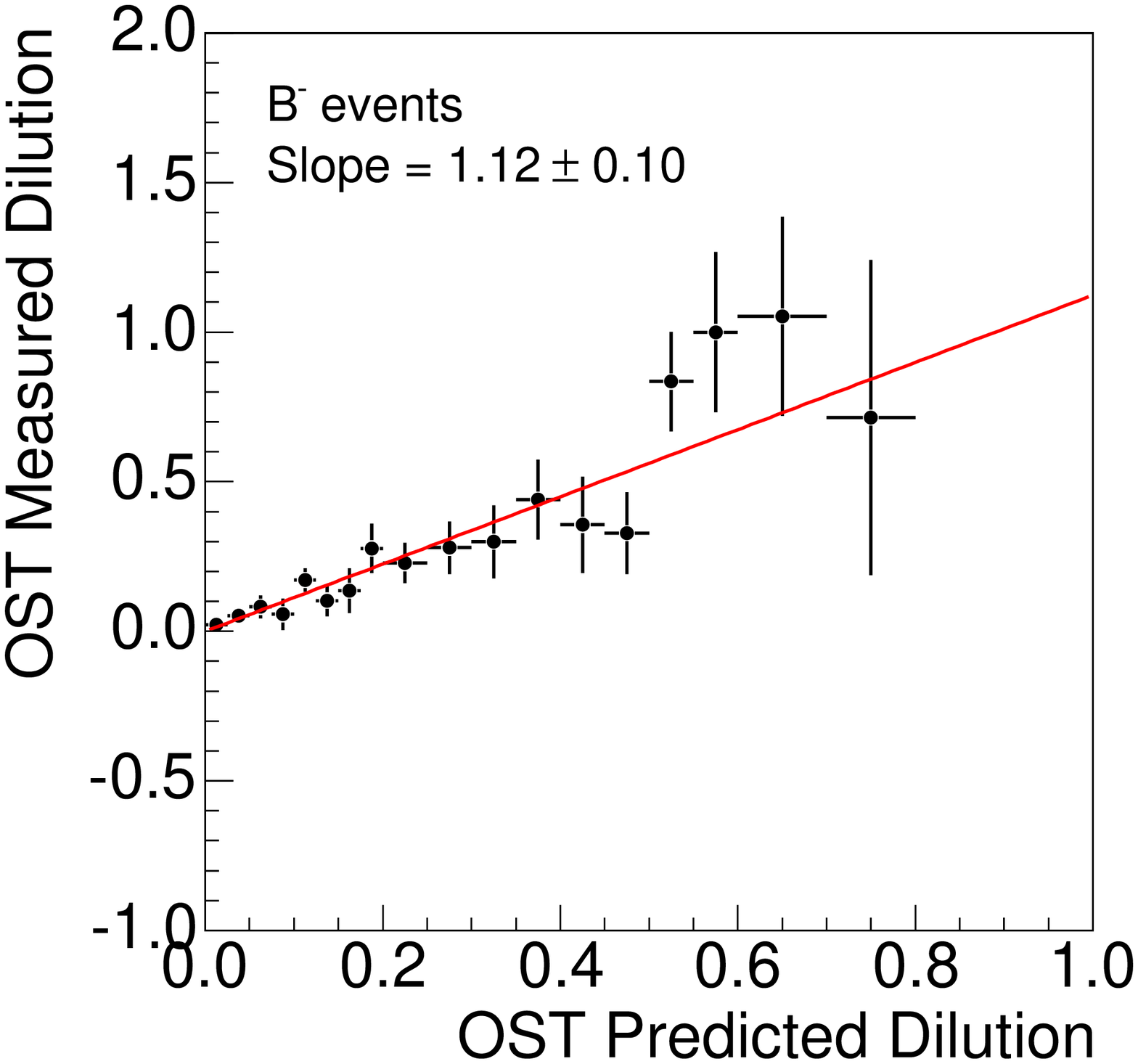}
\caption{Measured versus predicted dilution for $B^{+}$ (left) and
$B^{-}$~candidates (right) from $B^\pm \rightarrow J/\psi K^\pm$ events.
\label{B_all}}
\end{figure}

We calibrate the OST performance on $B^{+} \rightarrow J/\psi K^{+}$
decays. Since charged $B$~mesons do not oscillate, the production flavor
of the $B^{+}$ is identified by the charge of the 
kaon daughter track. Knowing the true flavor of the $B^+$ meson as well
as the flavor predicted by the OST algorithm, we can easily identify the
dilution measured from the number of correct and false tags as a
function of the predicted dilution. This dependence is shown for $B^{+}$
and $B^{-}$~mesons in Fig.~\ref{B_all} and fitted with a linear function
whose slope is expected to be one for a perfectly functioning OST
algorithm. The actual measured slope of the linear fitting function
provides the OST scale factor $\mathcal{S_{D}}$ which is determined to
be $0.93\pm0.09$ ($1.12\pm0.10$) for $B^{+}$ ($B^{-}$)~mesons.

Although the dilution scale factors determined separately from $B^+$ and
$B^-$ decays are both within uncertainties consistent with unity and with
each other, we use two scale factors for the opposite side tagger, one
for $B^{+}$ mesons and one for $B^{-}$ mesons, in order to allow for any
potential asymmetry in the tagging algorithms.  As a cross-check we
determine the scale factors in different data taking periods and find
that the scale factors are stable throughout all parts of the data.
We measure a tagging efficiency of $\varepsilon=(94.2 \pm 0.4)\%$ and an
average predicted dilution from the $B^\pm \rightarrow J/\psi K^\pm$
signal events of $\sqrt{<\mathcal{D}_p^2>} = (11.04 \pm 0.18)\%$. The average
OST dilution scale factor is $\mathcal{S_{D}} = 1.03\pm0.06$. The
total OST tagging power is $\varepsilon \mathcal{S_{D}}^2
<\mathcal{D}_p^2>\ = (1.2\pm0.2)\%$.

\section{The Likelihood Function} 
\label{sec:fit}

A simultaneous unbinned maximum likelihood fit to our data including
information on the $J/\psi K^+K^-$~invariant mass, \Bs~candidate decay
time, and transversity angular variable $\vec{\rho}$ is performed to
extract the main parameters of interest, \betas and \deltaG, plus
additional physics parameters, which include the $\Bs$~meson mass, the
mean $\Bs$ width $\Gamma_s=\hbar/\tau(\Bs)$, the polarization amplitudes
in the transversity basis $|A_0(0)|^2$, $|A_{\parallel}(0)|^2$, and
$|A_{\perp}(0)|^2 \equiv 1 -|A_0(0)|^2 - |A_{\parallel}(0)|^2$, the corresponding
\CP-conserving strong phases
$\delta_{\parallel}$ and $\delta_{\perp}$, the fraction of the
$S$-wave component and its corresponding phase $\delta_{SW}$.  The
likelihood function also includes other technical parameters of less interest
referred to as ``nuisance parameters'' such as the \Bs~signal fraction
$f_s$, parameters describing the $J/\psi\phi$~mass distribution, 
the \Bs~decay time plus angular distributions of background events,
parameters used to describe the estimated decay time uncertainty
distributions for signal and background events, scale factors between
the estimated decay time and mass uncertainties and their true
uncertainties, as well as tagging dilution scale factors and efficiency
and asymmetry parameters. There are a total of 35 fit parameters in the
likelihood function, 11 of which we consider parameters of physical
interest with \betas and \deltaG being the main physics parameters.  All
the physics parameters are listed in Table~\ref{tab:physicsParameters}.

\begin{table}[tbp]
\begin{ruledtabular} 
\caption{Physics parameters of interest in the likelihood fit with
    \betas and \deltaG as main physics parameters.
    \label{tab:physicsParameters}}
\begin{tabular} {ll} 
Parameter            & Description                          \\ 
\hline
\betas               & \CP-violating phase [rad]                 \\
\deltaG              & $\Gamma_{s}^L - \Gamma_{s}^H$ [\ps]  \\
\hline
\deltaM              & \Bs-\bBs oscillation frequency [\ps] \\
$\tau(\Bs)$          & \Bs mean lifetime [ps]               \\
$m(\Bs)$             & \Bs mass [\mevcc]                    \\
$|A_{0}(0)|^2$        & Longitudinally-polarized transition  \\
                     & probability at $t=0$                 \\
$|A_{\parallel}(0)|^2$ & Parallel component of transversely-polarized \\
                       & transition probability at $t=0$      \\
$\delta_{\perp}$        & $\arg[A_{\perp}(0)A_0^{*}(0)]$             \\
$\delta_{\parallel}$     & $\arg(A_{\parallel}(0)A_0^{*}(0)]$         \\ 
$f_{SW}$                & Fraction of $S$-wave in $J/\psi K^+K^-$~sample  \\ 
$\delta_{SW}$           & Relative phase of $S$-wave contribution      \\ 
\end {tabular}
\end{ruledtabular} 
\end{table}

The distributions of the transversity angular variables 
$\vec\rho = (\cos{\theta_T}, \phi_T, \cos{\psi_T})$
observed with the CDF\,II detector
are different from the true distributions 
because the efficiency of detecting the final state muons and kaons
from \Jpsi and $\phi$ decays is non-uniform.  The angular efficiency
function is parametrized in three dimensions using a set of real spherical
harmonics and Legendre polynomials as basis functions with ranges 
$0<\psi_T<\pi$, $0<\theta_T<\pi$ and $0<\phi_T<2\pi$~\cite{Ref:beta_s_JHEP}:
\begin{eqnarray}
  \varepsilon(\psi_T, \theta_T, \phi_T) =\displaystyle \sum_{lmk} a_{lm}^{k}P_k(\cos{\psi_T}) Y_{lm}(\theta_T, \phi_T).
\label{eqn:epsilon}
\end{eqnarray}
The parameters $a_{lm}^{k}$ are obtained from simulated events
where all transversity angles are
generated flat for \Bs decays. These MC events,
which have been passed through the full CDF\,II detector simulation,
enable us to examine how the
initially flat distributions are sculpted by the detector acceptance, thus
allowing us to determine the angle-dependent efficiencies of the
reconstructed particle candidates.
There are no predictions for the distribution of the background
transversity angles, but we find that they can be represented as
the product of three independent functions of $\cos\theta_T$, 
$\phi_T$, and $\cos\psi_T$ that are constant in time:
\begin{eqnarray}
 f(\cos\theta_T) = \frac{a_{0} - a_{1}\cos^{2}(\theta_T)}{2a_{0} - 2a_{1}/3}, \nonumber \\
 f(\phi_T) = \frac{1 + b_{1}\cos(2\phi_T + b_{0})}{2\pi},   \nonumber \\
 f(\cos\psi_T) = \frac{c_{0} + c_{1}\cos^{2}(\psi_T)}{2c_{0} + 2c_{1}/3}.
\label{eqn:bkgang}
\end{eqnarray}
The parameters $a_{0,1}$, $b_{0,1}$ and $c_{0,1}$ are determined
as best fit estimates from the maximum likelihood optimization.  The
above functions follow closely the shapes of the angular efficiencies,
which suggests that the underlying transversity angle distributions of
the background events are flat.

To set up the full unbinned maximum likelihood fit, we
define a set of probability density functions (PDFs), 
$P(\vec x\,|\,\vec{\mu})$, which give the probability density of observing 
the measured variables $\vec x_i$ for an event $i$, given a set of unknown
parameters $\vec{\mu}$. In our likelihood function the measured
variables $\vec x_i$ for each event $i$ are the $J/\psi K^+K^-$ 
invariant mass value $m$ and its uncertainty $\sigma_{m}$, 
the \Bs~candidate proper decay 
length $ct$ and uncertainty $\sigma_{ct}$, the angular
distributions of $\vec\rho = (\cos{\theta_T},
\phi_T, \cos{\psi_T})$ in the transversity basis,  
and the predicted dilution ${\cal D}_p$ and tag decision $\xi$ for
the SSKT and OST method as described in Sec.~\ref{sec:tagging}.
Among the unknown fit parameters $\vec{\mu}=(\vec\theta,\vec\nu)$ are
the physics parameters $\vec\theta$
described in Table~\ref{tab:physicsParameters} as well as the nuisance
parameters $\vec\nu$ discussed above.

The likelihood function for our dataset of $N$ events is given as
\begin{eqnarray}
{\cal L}(\vec{x}\,|\,\vec\theta,\vec\nu) = 
\prod_{i=1}^N P(\vec x_i\,|\,\vec\theta,\vec\nu).  
\end{eqnarray}
We minimize 
\begin{eqnarray}
-\log {\cal L}(\vec{x}\,|\,\vec\theta,\vec\nu) 
= - \displaystyle \sum_{i=1}^N \log P(\vec x_i\,|\,\vec\theta,\vec\nu)
\end{eqnarray}
using the \textsc{minuit} program package~\cite{Ref:Minuit}. 
The likelihood function is composed of separate probability density
functions for signal events, $P_{s}$, and for background
events, $P_{b}$. Both the signal and background components
contain PDFs describing the measured
variables $\vec x_i$ of the \Bs~candidate described above.


The full likelihood function, including flavor tagging, can be
expressed for signal and background events as
\begin{widetext}
\begin{eqnarray}
{\cal L} &=& \prod_{i=1}^N \left[f_s \cdot P_s(m|\sigma_{m}) \cdot P_s(\xi)
  \cdot P_s(\theta_T,\phi_T,\psi_T, ct|\sigma_{ct}, \xi, {\cal D}_p)
  \cdot P_s(\sigma_{ct})\cdot P_s({\cal D}_p) \right. \nonumber \\
&+& \left. (1-f_s)\cdot P_b(m)\cdot P_b(\xi)\cdot P_b(ct|\sigma_{ct})\cdot P_b(\theta_T)\cdot P_b(\phi_T) \cdot P_b(\psi_T)\cdot P_b(\sigma_{ct}) \cdot P_b({\cal D}_p) \right],
\label{lf}
\end{eqnarray}
\end{widetext}
where the product runs over all $N$ events in the data sample and $f_s$
and $(1-f_s)$ are the fraction 
of signal and background events, respectively.
For the case
of the fit without flavor tagging, the likelihood function reduces to
\begin{widetext}
\begin{eqnarray}
{\cal L}^{\mathrm{notag}} &=& \prod_{i=1}^N [f_s \cdot P_s(m|\sigma_{m}) \cdot P_s(\theta_T,\phi_T,\psi_T,
ct|\sigma_{ct}) \cdot P_s(\sigma_{ct}) \nonumber \\ &+& (1-f_s)\cdot
P_b(m)\cdot P_b(ct|\sigma_{ct})\cdot P_b(\theta_T)\cdot P_b(\phi_T) \cdot
P_b(\psi_T)\cdot P_b(\sigma_{ct})]
\label{lfnotag}
\end{eqnarray}
\end{widetext}
which simply corresponds to the flavor tagged case with no tag
decision ($\xi=0$) or a tagging dilution of zero (${\cal D}_p=0$).

In the following, we describe the individual elements of the full
likelihood function in more detail starting with the signal mass PDF 
$P_s(m|\sigma_{m})$. We model the signal mass distribution with a
Gaussian distribution of variable width. To form the probability density
function, $P_s(m|\sigma_{m})$, we use the candidate-by-candidate
observed mass uncertainty $\sigma_{m}$ multiplied by a scale factor
$s_m$, which is a fit parameter and accounts for a collective
mis-estimation of the mass uncertainties. The PDF is normalized over the range
$5.2 < m(J/\psi K^+K^-) < 5.6~\gevcc$. The background mass PDF, $P_b(m)$, is
parametrized as a first order polynomial. 
Since the distributions of the decay time
uncertainty $\sigma_{ct}$ and the event-specific dilution ${\cal D}_p$
are observed to be different in signal and background, we include their PDFs
explicitly in the likelihood. The signal PDFs $P_s(\sigma_{ct})$ and
$P_s({\cal D}_p)$ are determined from sideband-subtracted data
distributions, while the background PDFs $P_b(\sigma_{ct})$ and
$P_b({\cal D}_p)$ are determined from the $\Jpsi K^+K^-$ invariant mass
sidebands.  The PDFs of the decay time uncertainties,
$P_s(\sigma_{t})$ and $P_b(\sigma_{t})$, are described with a sum of
normalized Gamma function distributions as described below, while the
dilution PDFs $P_s({\cal D})$ and $P_b({\cal D})$ are included in the
likelihood as histograms that have been extracted from data.

For the time and angular dependence of the signal PDF
$P_s(\theta_T,\phi_T,\psi_T, ct|\sigma_{ct}, \xi, {\cal D}_p)$,
we follow the method derived in Ref.~\cite{Ref:beta_s_JHEP}. This PDF
includes the additional contribution from $S$-wave 
$\Bs \rightarrow J/\psi K^+K^-$ decays, with fraction
$f_{SW}$ and relative phase $\delta_{SW}$
between the $S$-wave and 
$P$-wave amplitude. 
The major difference between our
treatment and that of Ref.~\cite{Ref:beta_s_JHEP} is a refinement of
the model used to describe the line shape of the $K^+K^-$ invariant mass
$\mu\equiv m(K^+K^-)$~\cite{Ref:Breit-Wigner}.  As in
Ref.~\cite{Ref:beta_s_JHEP} we use a flat model for the $S$-wave
component, whereas for the $P$-wave \BsJpsiPhi component we instead
use an asymmetric relativistic Breit-Wigner distribution with
mass-dependent width 
\begin{widetext}
\begin{eqnarray}
|BW(\mu)|^2&=&\frac{\mu}{m_\phi}\cdot\Gamma_1\cdot
   \frac{E(K^+K^-)}{E(\phi)} 
   \cdot\frac{1}{(m_\phi^2-\mu^2)^2+m_\phi^2\cdot\Gamma_{\textrm{tot}}^2},
\label{eqn:phi_prop}
\end{eqnarray}
\end{widetext}
where $E(\phi)$ $(K^+K^-)$ is the energy of the $\phi$ $(K^+K^-)$ in the
decay of \BsJpsiPhi $(\Bs\rightarrow\Jpsi K^+K^-)$.
This treatment assumes a two-body decay, where the other daughter
particle is the \Jpsi, and the total decay width 
$\Gamma_{\textrm{tot}}=\Gamma_1+\Gamma_2+\Gamma_3$, 
where $\Gamma_{1,2,3}$ are the partial decay widths for the decays $\phi
\rightarrow K^+K^-$ ($48.8\pm0.5\%$), $\phi \rightarrow K_L^0 K_S^0$
($34.2\pm 0.4\%$), and $\phi \rightarrow \rho\pi$ plus 
$\phi\rightarrow\pi^+\pi^-\pi^0$ ($15.32\pm 0.32\%$),
respectively~\cite{Ref:PDG}.
Following Ref.~\cite{Ref:Dighe_cross_prod} we describe the $\Bs$ decay
rate as a function of the transversity angles, decay time and, in
addition, the $K^+ K^-$ invariant mass. When both a $\phi$~component
with kaons in a relative $P$-wave and a $S$-wave
component are present, the amplitudes must be summed and then
squared. The $P$-wave amplitude has a resonant structure due to the
$\phi$~propagator, while the $S$-wave amplitude is flat, but can have an  
arbitrary phase $\delta_{SW}$ with respect to the $P$-wave 
\begin{widetext}
\begin{eqnarray} 
\rho_{B}(\theta_T, \phi_T, \psi_T, t, \mu) &=& \frac{9}{16\pi} \left |\left      [ \sqrt{1-f_{SW}}BW(\mu){\bf A}(t) + e^{i\delta_{SW}}\sqrt{f_{SW}}\frac {h(\mu)}{\sqrt{3}}{\bf B}(t)\right ]\times \hat{n} \right |^2, \nonumber \\
\rho_{\bar{B}}(\theta_T, \phi_T, \psi_T, t, \mu) &=& \frac{9}{16\pi}
\left |\left[ \sqrt{1-f_{SW}}BW(\mu) {\bf {\bar A}} (t) +
    e^{i\delta_{SW}}\sqrt{f_{SW}}\frac {h(\mu)}{\sqrt{3}}{\bf {\bar
        B}(t)} \right ]\times \hat{n}\right |^2.\
\label{eqn:PAndS}
\end{eqnarray}
\end{widetext}
In our analysis we accept events for which the reconstructed $K^+ K^-$
mass $\mu$ lies within a window $\mu_{\rm lo}=1.009 < \mu < \mu_{\rm
  hi}=1.028~\gevcc$.  The $\phi$ mass distribution is described by the
Breit-Wigner function given in Eq.~(\ref{eqn:phi_prop}).  The $S$-wave
mass distribution is given by the flat function $h(\mu) = \frac
{1}{\sqrt{\Delta \mu}}$ between $\mu_{\textrm{lo}}$ and $\mu_{\rm hi}$,
where $\Delta \mu = \mu_{\rm hi} - \mu_{\rm lo}$.  The likelihood
function used in the maximum likelihood fit is obtained by numerically
integrating Eq.~(\ref{eqn:PAndS}) over the $K^+ K^-$ invariant mass
$\mu$.  ${\bf A}(t)$ and ${\bf {\bar A}(t)}$ are time-dependent complex
vector functions describing the $P$-wave component in the transversity
basis. They are defined as
\begin{eqnarray}
{\bf A}(t)&=&\left({\mathcal A}_0(t)\cos{\psi_T}, -\frac{{\mathcal A}_\parallel(t)\sin{\psi_T}}{\sqrt{2}}, i\frac{{\mathcal A}_\perp(t)\sin{\psi_T}}{\sqrt{2}}\right),  \nonumber \\
{\bf {\bar A}}(t)&=&\left({\bar {\mathcal A}}_0(t)\cos{\psi_T},
  -\frac{{\bar {\mathcal A}}_\parallel(t)\sin{\psi_T}}{\sqrt{2}},
  i\frac{{\bar {\mathcal A}}_\perp(t)\sin{\psi_T}}{\sqrt{2}}\right), \nonumber
\label{eqn:fixedAngle}
\end{eqnarray}
with 
\begin{eqnarray}
{\cal A}_i(t) &=& \frac{e^{-\Gamma_s t /2}}{\sqrt{\tau_H + \tau_L \pm
    \cos{2\beta_s}\left(\tau_L-\tau_H\right)}} \nonumber \\
&\times&\left[E_+(t) \pm e^{2i\beta_s} E_-(t)\right] a_i\,,   \nonumber \\
{\bar {\cal A}_i}(t) &=& \frac{e^{-\Gamma_s t /2}}{\sqrt{\tau_H + \tau_L
    \pm \cos{2\beta_s}\left(\tau_L-\tau_H\right)}} \nonumber \\
&\times&\left[\pm E_+(t) + e^{-2i\beta_s} E_-(t)\right] a_i\,,   
\label{eqn:finalAmp}
\end{eqnarray}
where $i \in \{0, \parallel, \perp\}$ and 
the upper sign applies to the \CP-even final states (0 and
$\parallel$), while the lower sign applies to the \CP-odd final
state ($\perp$). Furthermore,
\begin{eqnarray}
E_{\pm}(t) \equiv  \frac{1}{2}\left[e^{+\left(\frac{-\deltaG}{4} + i\frac{\deltaM}{2}\right)t} \pm e^{-\left(\frac{-\deltaG}{4} + i\frac{\deltaM}{2}\right)t}\right],
\label{eqn:functionDef}
\end{eqnarray}
and the $a_i$ are complex amplitude parameters satisfying
\begin{eqnarray}
\sum_i {|a_i|^2} = 1 \,.
\label{eqn:aNorm}
\end{eqnarray}
The $S$-wave component is described in the transversity basis as
\begin{eqnarray}
{\bf B}(t)& = &\left({\mathcal B}(t),0,0\right) \,, \nonumber \\
{\bf{\bar {B}}}(t) & =&\left({\bar {\mathcal B}}(t),0,0\right),
\label{eqn:fixedAngleS}
\end{eqnarray}
where the time-dependent amplitudes are
\begin{eqnarray}
{\cal B}(t) &=& \frac{e^{-\Gamma_s t /2}}{\sqrt{\tau_H + \tau_L -
    \cos{2\beta_s}\left(\tau_L-\tau_H\right)}} \nonumber \\
&\times&\left[E_+(t) - e^{2i\beta_s} E_-(t)\right],   \nonumber \\
{\bar {\cal B}}(t) &=& \frac{e^{-\Gamma_s t /2}}{\sqrt{\tau_H + \tau_L -
    \cos{2\beta_s}\left(\tau_L-\tau_H\right)}} \nonumber \\
&\times&\left[-E_+(t) + e^{-2i\beta_s} E_-(t)\right].  
\label{eqn:finalAmpS}
\end{eqnarray}
In Eq.~(\ref{eqn:PAndS}) the unit vector $\hat n$ is defined as $\hat{n} =
(\sin\theta_T\cos\phi_T, \sin\theta_T\sin\phi_T, \cos\theta_T)$ 
in the transversity basis and the
strong phases $\delta_{\parallel}$ and $\delta_{\perp}$
appear from terms of the form $A_i A^*_0 = |A_i||A_0|e^{\arg(A_i A_0^*)}$, 
where $i = '\parallel'$ or $i = '\perp'$. The decay width difference 
$\deltaG$ and the \Bs~oscillation frequency $\deltaM$ are encoded 
in~Eq.~(\ref{eqn:functionDef}).

%

%

The time-dependent functions in the PDFs above are convolved with a
resolution function composed of two independent Gaussians with
candidate-by-candidate expected width $\sigma_{ct}$.  The widths of each
Gaussian function are multiplied by independent scale factors $s_{ct1}$ and
$s_{ct2}$ which are freely floated in the maximum likelihood fit
to account for an overall mis-estimation of the decay-time resolution. The
PDF describing the decay time distributions for signal events as part of
$P_s(\theta_T,\phi_T,\psi_T, ct|\sigma_{ct}, \xi, {\cal D}_p)$ in
Eq.~(\ref{lf}) is of the form
\begin{eqnarray}
P_s(ct|\sigma_{ct})&=&P_s(ct, \sigma_{ct} | c\tau, s_{ct1,2}) \nonumber \\ 
&=&F(ct,c\tau) \otimes G(ct,\sigma_{ct} | f_{s_{ct1}}, s_{ct1},
s_{ct2}),\quad
\label{eqn:sigctsmear}
\end{eqnarray}
where $\tau=\tau(\Bs)$ and $F(ct,c\tau)$ represents the time dependence
of the signal events which, e.g., for an exponential decay is given as
$e^{-\frac{ct}{c\tau}}/(c\tau)$. The symbol ``$\otimes$'' denotes a
convolution which is with respect to the decay time resolution function
defined as
\begin{widetext}
\begin{eqnarray}
G(ct,\sigma_{ct} | f_{s_{ct1}}, s_{ct1}, s_{ct2}) = 
f_{s_{ct1}} \frac{1}{\sqrt{2\pi}s_{ct1}\sigma_{ct} }e^{-\frac{c^2t^2}{2s_{ct1}^2\sigma_{ct}^2}} + (1-f_{s_{ct1}}) \frac{1}{\sqrt{2\pi}s_{ct2}\sigma_{ct} }e^{-\frac{c^2t^2}{2s_{ct2}^2\sigma_{ct}^2}}, 
\label{eqn:sigctsmear2}
\end{eqnarray}
\end{widetext}
where $f_{s_{ct1}}$ is the fraction of the first resolution Gaussian.
From the distribution of decay time uncertainties, we find the average
of the decay time resolution function at $\sigma_{ct}\sim30~\mu$m, with
a root-mean-square deviation of about $12~\mu$m~\cite{Ref:tagged_cdf_old}.


As we are using candidate-by-candidate expected decay time
uncertainties, which are not distributed identically for the signal and
background events, it is necessary to include a PDF for the separate
uncertainty distributions~\cite{Ref:punzi}. The PDF describing the decay
time uncertainty $P_s(\sigma_{ct})$ is constructed from normalized Gamma
distributions
\begin{eqnarray}
\Gamma_{\textrm{norm}}(x) \equiv \frac{x^a e^{-x/b}}{b^{a+1}\Gamma(a+1)},
\label{eqn:punzi}
\end{eqnarray}
where $a$ and $b$ define the mean and width of the distribution.  Each
function has different values of $a$ and $b$. We find these values from
a separate lifetime-only fit before running the full angular analysis
and fix these parameters within uncertainties in the full likelihood
fit.
We handle the background lifetime resolution PDF
$P_b(\sigma_{ct})$ in the same way as the signal distribution, using a
sum of three normalized Gamma distributions $\Gamma_{\textrm{norm}}$.  

The portion of the likelihood function $P_{s}(\theta_T, \phi_T,
\psi_T, ct|\sigma_{ct}, \xi, \mathcal{D}_p)$ related to tagging can be
written as follows:
\begin{widetext}
\begin{eqnarray}
P_{s}(ct, \psi_T, \theta_T, \phi_T | \sigma_{ct}, \mathcal{D}_{SS},
\mathcal{D}_{OS}, \xi_{SS}, \xi_{OS}) &=& 
\frac{1 + \xi_{SS}s_{SS}\mathcal{D}_{SS}}{1 + |\xi_{SS}|}\,
  \frac{1 + \xi_{OS}s_{OS}\mathcal{D}_{OS}}{1 + |\xi_{OS}|}\,\rho_B(ct, \psi_T, \theta_T, \phi_T) \otimes G(ct,\sigma_{ct})
\nonumber \\ 
&+& \frac{1 - \xi_{SS}s_{SS}\mathcal{D}_{SS}}{1 + |\xi_{SS}|}\,
  \frac{1- \xi_{OS}s_{OS}\mathcal{D}_{OS}}{1 +
    |\xi_{OS}|}\,\rho_{\bar{B}}(ct, \psi_T, \theta_T, \phi_T) \otimes
  G(ct,\sigma_{ct}), 
\label{tdep}
\end{eqnarray}
\end{widetext}
where $s_{SS}$ and $s_{OS}$ are the SSKT and OST dilution scale factors.
Since the two flavor taggers search for tagging information (tracks,
jets) in complementary regions in space, we treat them as
independent tags. We have verified that the tagging decisions
and predicted dilutions are indeed independent.

The probability of a particular combined tag decision is dependent on
the efficiency of each of the taggers
\begin{eqnarray}
P(\xi) = \left\{ \begin{array}{cl} 
(1 - \varepsilon_{SS})(1 - \varepsilon_{OS}) & (\xi_{SS} = 0, \ \xi_{OS} = 0) \\
\varepsilon_{SS}(1 - \varepsilon_{OS}) & (\xi_{SS} = \pm 1, \ \xi_{OS} = 0) \\
(1 - \varepsilon_{SS})\varepsilon_{OS} & (\xi_{SS}=0, \ \xi_{OS} = \pm 1) \\
\varepsilon_{SS}\varepsilon_{OS} & (\xi_{SS} = \pm 1, \ \xi_{OS} = \pm 1).\ \
\end{array} \right.
\label{combtagprob} 
\end{eqnarray}
The predicted dilution distributions are different for signal and
background events. Therefore the likelihood function in Eq.~(\ref{lf})
contains the corresponding dilution PDFs separately for signal,
$P_s({\cal D}_p)$, and for background events, $P_b({\cal D}_p)$.  These
dilution distributions are measured with data.  In the case of the
signal, they are the sideband-subtracted distributions taken from the
$\Bs$ invariant mass signal region and stored as histograms.  The
histograms are normalized to represent probability densities for the
dilution. 

Knowledge of the fractions of positively and negatively charged events
is sufficient to describe any tagging asymmetry present in the
background. Therefore, the probability density $P_{b}(\xi)$ is equal
to the fraction of events with positive tags for $\xi=+1$, and equal
to the fraction of events with negative tags for $\xi=-1$. The
background dilution distributions are handled analogously to the
signal dilution.  The dilution distributions are taken from the $\Bs$
invariant mass sideband region, normalized to form a probability
density, and again stored as histograms.


The background proper decay time PDF, $P_b(ct|\sigma_{ct})$ is
parametrized as a prompt peak modeled by a $\delta$-function plus two
positive exponentials and one negative exponential.  This function is
convolved with the same resolution function as the signal decay time
dependence. In our parameterization, the prompt peak models the majority
of the combinatorial background events, which are expected to have no
significant lifetime, the negative exponentials defined for $t>0$
account for a small fraction of longer lived background such as other
$B$~hadron decays, and the positive exponential defined for $t<0$ takes
into account events with a mis-reconstructed vertex.
The background decay time PDF reads
\begin{widetext}
\begin{eqnarray}
P_{b}\left(ct|\sigma_{ct}\right)& = & \left \{ f_g\,\delta(ct) + 
\left(1-f_g\right)\left[f_{++}\frac{e^{-\frac{ct}{\lambda_{++}}}}{\lambda_{++}} + 
(1-f_{++})\left ( f_{-}\frac{e^{\frac{ct}{\lambda_{-}}}}{\lambda_{-}} + 
(1-f_{-})\frac{e^{-\frac{ct}{\lambda_{+}}}}{\lambda_{+}}
\right)\right]\right\} \nonumber \\
&\otimes&  G(ct,\sigma_{ct} | f_{s_{ct1}}, s_{ct1}, s_{ct2}),
\label{eqn:bkg_decay_time}
\end{eqnarray}
\end{widetext}
where $f_g$ is the fraction of the prompt background, $\lambda_{++}$,
$\lambda_{+}$ and $\lambda_{-}$ are the effective lifetimes of the
background events distributed according to the long and short lived
positive exponential as well as the negative exponential, respectively,
while $f_{++}$ and $f_{-}$ are their corresponding fractions.

\section{\boldmath{$\Bs$}~Mean  Lifetime, Decay Width Difference and Polarization Fractions }
\label{sec:smresult}

We use the likelihood function presented in the previous section to
extract measurements of the physics parameters of interest.  Before
applying the unbinned maximum likelihood fit on data, we perform an
extensive set of tests of the fitting procedure using simulated
pseudoexperiments. We observe that, with the current statistics, the
maximized likelihood function returns biased results for the parameters
of interest. Moreover, the likelihood function
shows non-Gaussian behavior with respect to the \betas and \deltaG
parameters. For these reasons
we employ frequentist techniques to determine confidence level (C.L.) regions in
the $\betas$-$\deltaG$ plane as described in Sec.~\ref{sec:statistics}. 


However, we find that the likelihood function constructed according to
the standard model expectation, in which the $CP$-violation parameter
$\betas$ is fixed to a value very close to zero ($\betas = 0.02$),
returns un-biased results for the mean $\Bs$~lifetime $\tau(\Bs)$, the
decay width difference \deltaG, the polarization fractions
$|A_{\parallel}(0)|^2$ and $|A_0(0)|^2$ as well as the strong phase
$\delta_{\perp}$.  We also observe that the likelihood function is
Gaussian with respect to all these parameters. Under these favorable
circumstances we provide point estimates with Gaussian uncertainties for
the following quantities:
\begin{eqnarray}
c\tau(\Bs) &=&458.6 \pm 7.6\,\textrm{(stat)} \pm 3.6\,\textrm{(syst)}~\mu\textrm{m}, \nonumber\\
\deltaG &=&0.075 \pm 0.035\,\textrm{(stat)} \pm 0.006\,\textrm{(syst)}~\ps, \nonumber\\
|A_{\parallel}(0)|^2 &=&0.231 \pm 0.014\,\textrm{(stat)} \pm 0.015\,\textrm{(syst)}, \nonumber\\
|A_0(0)|^2    &=&0.524 \pm 0.013\,\textrm{(stat)} \pm 0.015\,\textrm{(syst)}, \nonumber\\
\delta_{\perp} &=&2.95 \pm 0.64\,\textrm{(stat)} \pm 0.07\,\textrm{(syst)}~\textrm{rad}.
\label{eqn:pe_tag}
\end{eqnarray}

The systematic uncertainties on the measured quantities are discussed in
detail in Sec.~\ref{sec:systematics} and given above for completeness.
We are unable to also quote a result for $\delta_\parallel$ since the fit
prefers a value of $\delta_\parallel$ at the boundary of $\pi$ resulting
in a non-Gaussian likelihood shape around the minimum.

The results in Eq.~\ref{eqn:pe_tag} show good agreement with previous
measurements~\cite{Ref:tagged_cdf_old,Ref:tagged_d0}. The \Bs~mean
lifetime can be calculated as $\tau(\Bs) = 1.529 \pm
0.025\,\textrm{(stat)} \pm 0.012\,\textrm{(syst)}$~ps, which is the most
precise single measurement of this quantity. It can be compared to the
most recent measurement from the D0~collaboration using a data sample
based on 8~fb$^{-1}$ of integrated luminosity~\cite{Abazov:2011ry}
quoting $\tau(\Bs) = 1.443^{+0.038}_{-0.035}$~ps and to the Particle
Data Group (PDG) average of $\tau(\Bs) = 1.472 ^{+0.024}_{-0.026}$
ps~\cite{Ref:PDG}. The \deltaG value is of comparable precision to the
current world average of
$\deltaG=0.062^{+0.034}_{-0.037}~\ps$~\cite{Ref:PDG}.  Our central value
is somewhat smaller than the most recent measurement of
$\deltaG=0.163^{+0.065}_{-0.064}~\ps$ from the
D0~collaboration~\cite{Abazov:2011ry} but compares well to the PDG
average as well as
the SM prediction $\deltaG = 0.090 \pm 0.024~\ps$~\cite{Ref:lenz}. 

In addition to comparing the fit results with predictions and other
measurements, three cross-checks are performed using alternative
versions of the fit.  First we use the likelihood without flavor tagging
to check for any bias which could be introduced by the tagging. The fit
without flavor tagging does not have sensitivity to
$\delta_{\perp}$. This quantity is thus omitted from
Table~\ref{tab:smresults_compare} for fits without flavor tagging.  The
second and third checks are done without the $S$-wave component included
in the likelihood fit, once with flavor tagging included and once
without.  These checks are made to determine whether the $S$-wave
component has a significant effect on the fit results and to provide a
direct comparison with previous CDF results~\cite{Ref:tagged_cdf_old}
which did not account for the $S$-wave component.
Table~\ref{tab:smresults_compare} shows the results of these
cross-checks, that demonstrate good agreement between different
versions of the fit and our main results which include both flavor
tagging and the $S$-wave component.

\begin{table*}[btp]
  \caption{Results of alternative fits to cross-check the main SM results.
    All uncertainties quoted are statistical only. 
    As shown in Ref.~\cite{Aaltonen:2007gf} the untagged analysis is
    not sensitive to the strong phase $\delta_{\perp}$.
    \label{tab:smresults_compare}}  
\begin{ruledtabular} 
\begin{tabular}{lccc} 
Parameter & Un-tagged (with $S$-wave)\  & Tagged (without $S$-wave)\  &
Un-tagged (without $S$-wave)\    \\ 
\hline 
$c\tau(\Bs)\ \ [\mu {\rm m}]$          
& $456.9 \pm 8.0$ & $459.1 \pm 7.7$ & $457.2  \pm 7.9$ \\
$\deltaG$\ \ [ps$^{-1}$]            
& $0.069 \pm 0.030$ & $0.073 \pm 0.030$ & $0.070 \pm 0.040$ \\
$|A_{\parallel}(0)|^2$    
& $0.232 \pm 0.032$ & $0.232 \pm 0.014$ & $0.233 \pm 0.016$ \\
$|A_{0}(0)|^2$           
& $0.521 \pm 0.013$ & $0.523 \pm 0.012$ & $0.520 \pm 0.013$ \\
$\delta_{\perp}$ [rad]     
&	$-$	   & $ 2.8  \pm 0.6$ &		$-$		\\
\end{tabular}
\end{ruledtabular}
\end{table*}

\begin{figure*}[tbp]
\centerline{
\includegraphics[width=0.49\textwidth]{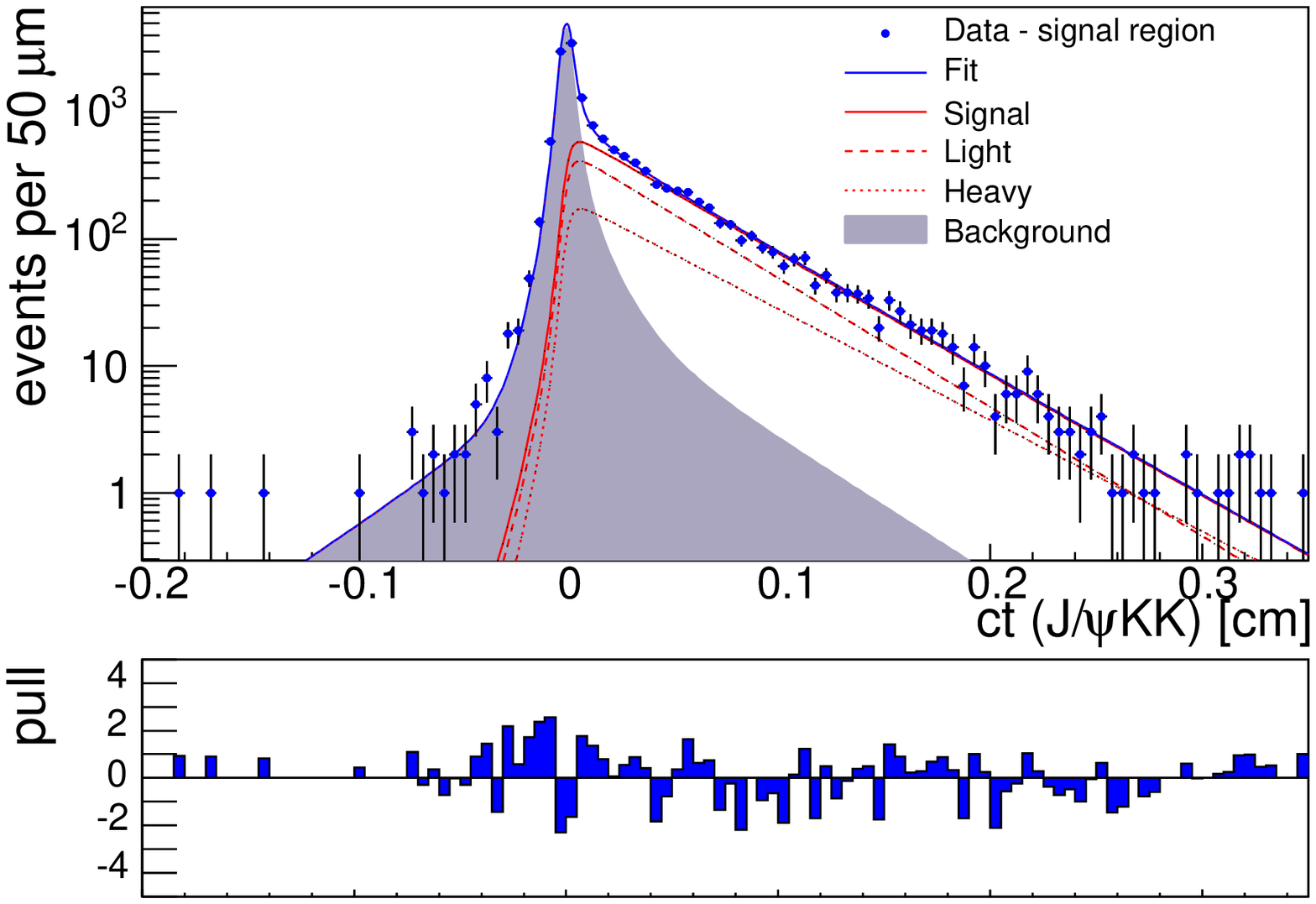}\
\includegraphics[width=0.49\textwidth]{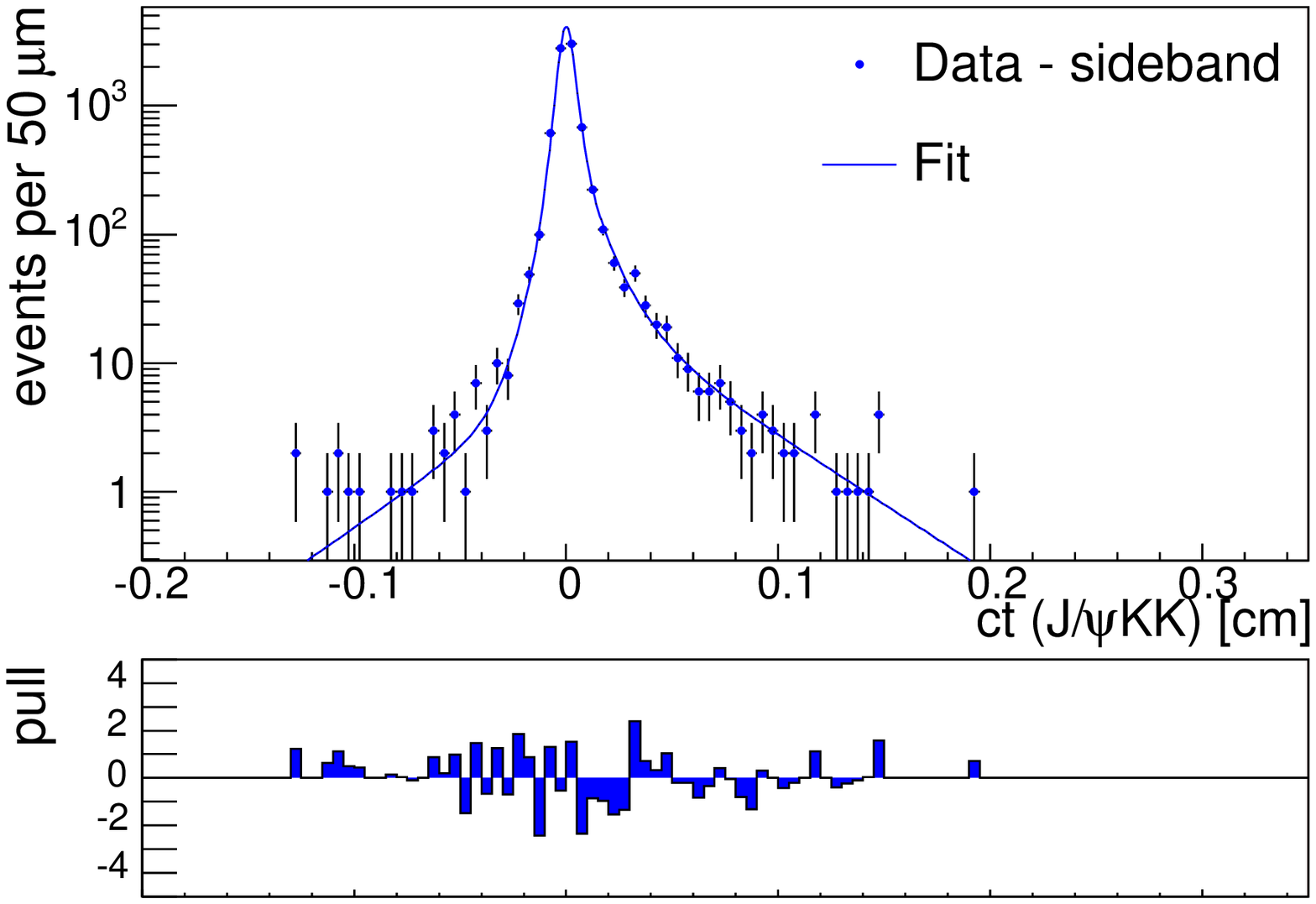}
}
\caption{(color online).
  Proper decay time fit projections for the $B^{0}_{s}$ signal
  (left) and background (right) regions. The dashed distributions
  labeled as ``Light'' and ``Heavy'' indicate the contribution of the
  $B^{L}_{s}$ and $B^{H}_{s}$, respectively. The pull distributions at
  the bottom show the difference between data and fit value normalized to
  the data uncertainty.
\label{ctau}}
\end{figure*}
\begin{figure*}[tbp]
\centerline{
\includegraphics[width=0.49\textwidth]{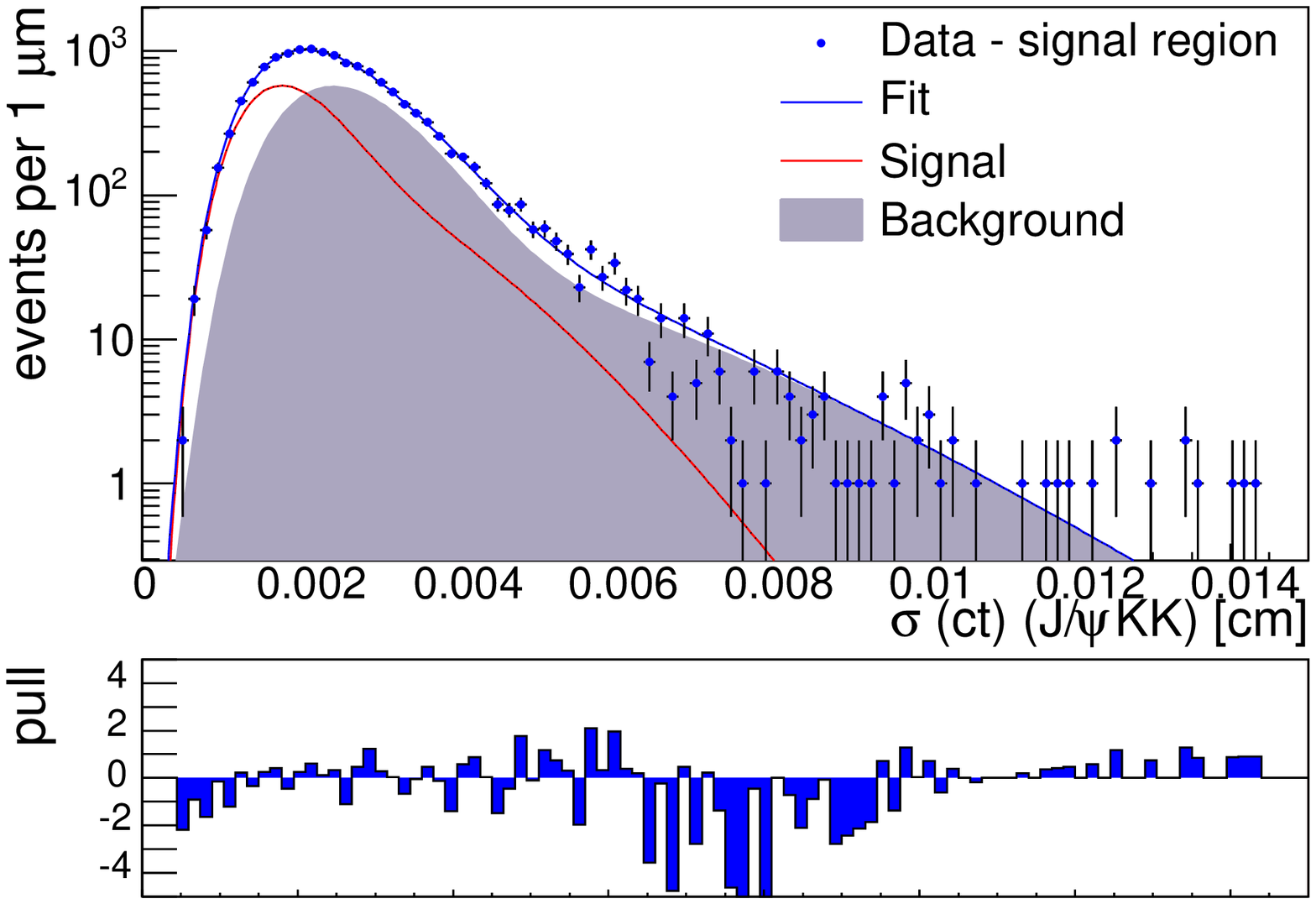}\
\includegraphics[width=0.49\textwidth]{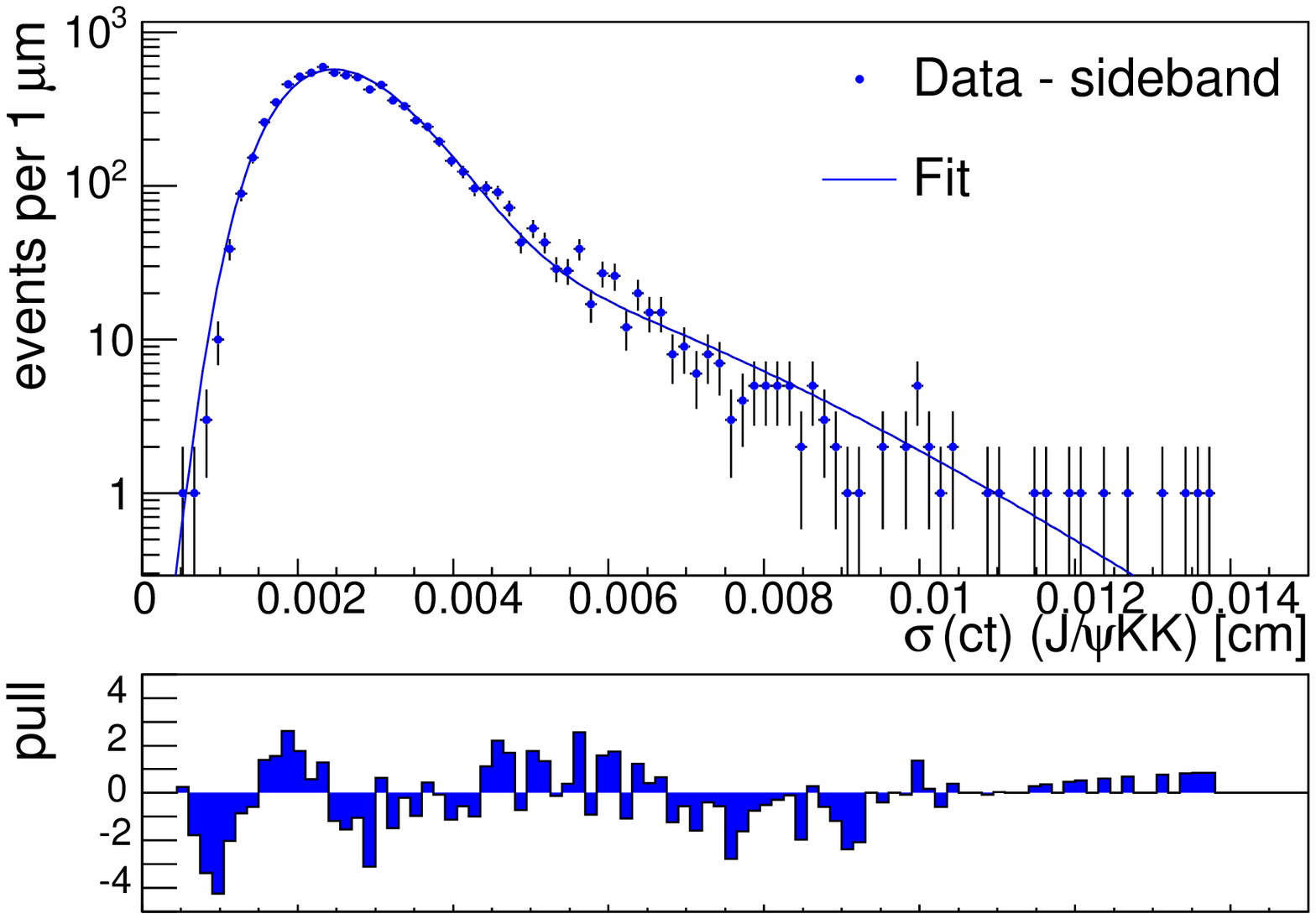}
}
\caption{(color online).
  Proper decay time uncertainty fit projections for the
  \Bs~signal (left) and background (right) regions. The pull
  distributions at the bottom show the difference between data and fit
  value normalized to the data uncertainty.
\label{ctauerr}}
\end{figure*}
\begin{figure*}[tbp]
\includegraphics[width=0.95\textwidth]{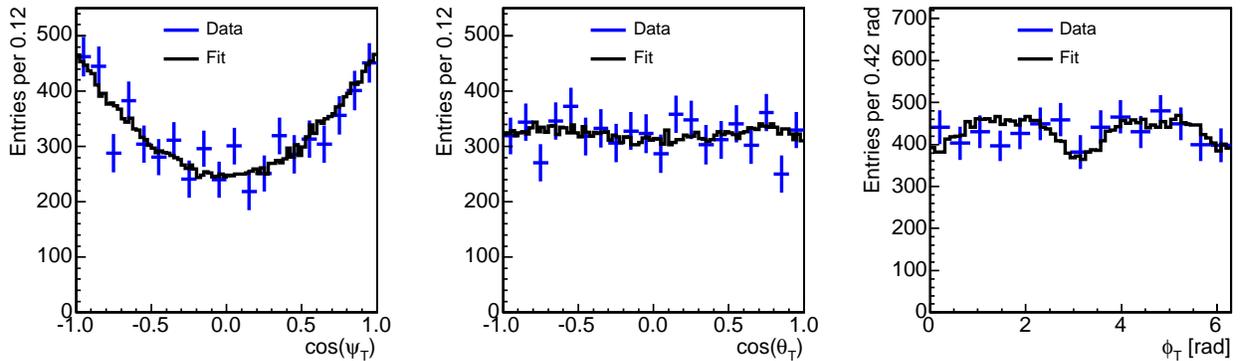}
\caption{Fit projections for transversity angles for sideband-subtracted signal.
\label{angular_projections_signal}}
\end{figure*}
\begin{figure*}[tbp]
\includegraphics[width=0.95\textwidth]{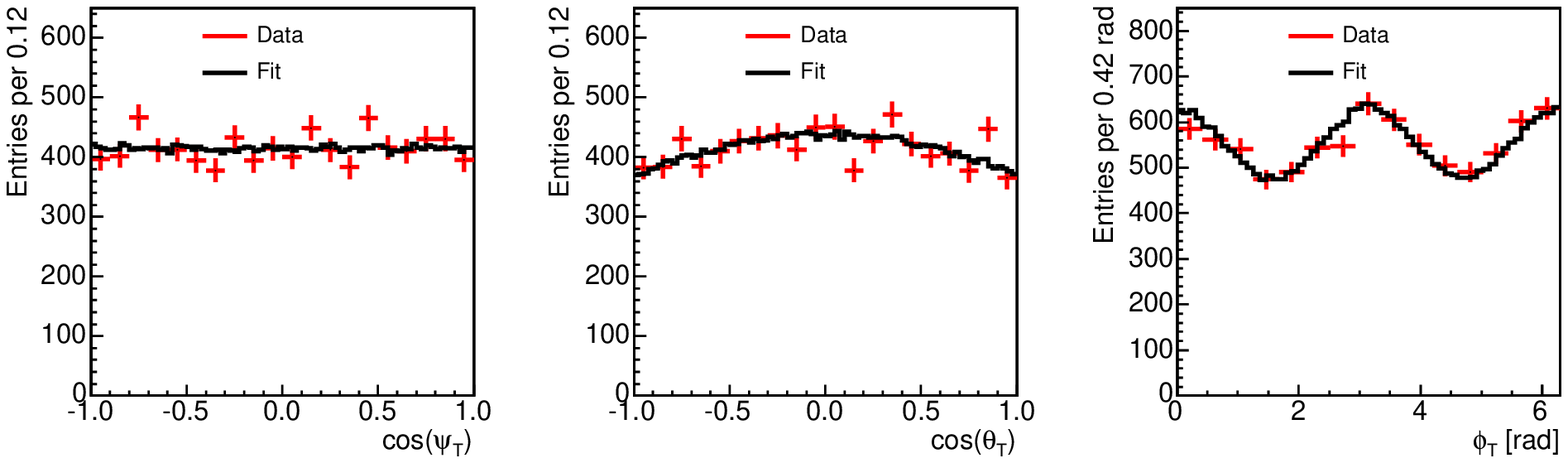}
\caption{Fit projections for transversity angles in background region.
\label{angular_projections_bkg}}
\end{figure*}

Since the unbinned maximum likelihood method does not readily provide a
goodness of fit estimator, we present fit projections onto the data to
support the quality of the fit. The likelihood function, in which all
parameters are fixed to their best-fit values, is overlaid on top of
data distributions. Such projections are performed for both signal and
background events and separately in the subspaces of the \Bs~decay time,
decay time expected uncertainty and transversity angles.  The fit
projections for the proper decay time and proper decay time
uncertainties are shown in Figs.~\ref{ctau} and \ref{ctauerr}.
Fit projections for the transversity angles $\cos\theta_T$, $\phi_T$,
and $\cos\psi_T$ from the sideband-subtracted signal region and the
background (sideband) region are shown in
Figs.~\ref{angular_projections_signal}
and~\ref{angular_projections_bkg}, respectively.  The good agreement
between the data and fit projections validates our parameterization of
both the signal and background distributions plus their uncertainties.

\subsection{Systematic uncertainties}
\label{sec:systematics}


To assess systematic uncertainties on quantities of interest other than
\betas, namely \deltaG, the \Bs~mean lifetime, the polarization
fractions, and the strong phase $\delta_{\perp}$, we set the
\CP-violating phase to its standard model expectation
$\betas=0.02$. This choice in addition improves the statistical behavior
of the likelihood function by eliminating biases on these parameters.

Systematic uncertainties are assigned by considering several effects
that are not accounted for in the likelihood
fit~\cite{Ref:Thesis_Oakes}. Such effects include potential
mis-parameterization in the fit model, impact of particular assumptions
in the fit model, and physical effects which are not well known or fully
incorporated into the model. To estimate the size of the systematic
uncertainties, we generate two sets of
pseudoexperiments by extracting random numbers distributed according
to our PDF: one set with each of the considered systematic
variations and another set of default pseudoexperiments. Each pair of
modified and not modified pseudoexperiments are generated with the same
random seeds.  The unbinned likelihood function is maximized over the
modified pseudoexperiments as well as over the corresponding default
ones.  For each systematic effect, the associated uncertainty is the
difference between the mean of the best fit value for the
pseudoexperiments with the systematic alteration included, and the
equivalent mean value for the reference set of pseudoexperiments
generated with the default model.
The individual systematic uncertainties are summed in quadrature and
presented in Table~\ref{tab:pe_syst_totals} to give the total
contribution to the uncertainties for each parameter which are due to
sources of systematic uncertainty.

\begin{table*}[tbp]
\caption{Summary of systematic uncertainties assigned to the five physics
  quantities discussed in Sec.~\ref{sec:smresult}.
\label{tab:pe_syst_totals}}
\begin{ruledtabular} 
\begin{tabular}{lccccc} 
Source of systematic effect & \ \deltaG [\ps]\ & \ $c\tau(\Bs)$ [$\mu$m]\ & 
\ $|A_{\parallel}(0)|^{2}$\ & \ $|A_{0}(0)|^{2}$\ & \ $\delta_{\perp}$ [rad]  \\  
\hline
	Signal efficiency                       & & & & & \\
	\hspace{1cm}Parameterization            & 0.0024         & 0.96  &       0.0076               &   0.008           &0.016  \\ 
	\hspace{1cm}MC re-weighting              & 0.0008         & 0.94  &       0.0129               &   0.0129          &0.022  \\ 

        Signal mass model                        & 0.0013        & 0.26  &       0.0009              &   0.0011         &0.009  \\ 
        Background mass model                    & 0.0009        & 1.4   &       0.0004              &   0.0005         &0.004   \\ 
        Resolution model                         & 0.0004        & 0.69  &       0.0002              &   0.0003         &0.022  \\ 
        Background lifetime model                & 0.0036        & 2.0   &       0.0007              &   0.0011         &0.058  \\ 
        Background angular distribution          & & & & & \\
	\hspace{1cm} Parameterization            & 0.0002        & 0.02  &       0.0001              &   0.0001         &0.001\\  
	\hspace{1cm} $\sigma_{ct}$ correlation   & 0.0002        & 0.14  &       0.0007              &   0.0007         &0.006\\  
	\hspace{1cm} Non-factorization           & 0.0001        & 0.06  &       0.0004              &   0.0004         &0.003\\  

        $B^{0}\rightarrow J/\psi K^{*}$ cross-feed & 0.0014        &  0.24 &       0.0007              &   0.0010          &0.006 \\
        SVX alignment                            & 0.0006        &  2.0  &       0.0001              &   0.0001          & 0.020\\  
        Mass resolution                          & 0.0001        &  0.58 &       0.0004              &   0.0004          & 0.002\\ 
        $\sigma_{ct}$ modeling                   & 0.0012        & 0.17  &       0.0005              &   0.0007          & 0.013  \\
        Pull bias                                & 0.0028        &       &       0.0013              &   0.0021          &       \\
        \hline
        Totals                               & 0.006 &       3.6     &       0.015               &   0.015           &  0.07 \\
\end{tabular}
\end{ruledtabular} 
\end{table*}


One source of systematic uncertainty is the modeling of the angular
efficiency of the detector described in Sec.~\ref{sec:fit}. We
model the detector efficiency with a linear combination of Legendre
polynomials and spherical harmonics as described in detail in
Ref.~\cite{Ref:beta_s_JHEP}.  The expansion coefficients of these
functions are obtained by fitting a three-dimensional efficiency
distribution obtained using simulated events. This simulated sample is
re-weighted to match the $p_{T}$ distributions observed in data.  If
the modeling is inaccurate, or the $p_{T}$ re-weighting incorrect, a
systematic uncertainty could be introduced.  We test these effects
separately.
The former effect is investigated by using the default fit model on
pseudoexperiments, generated with angular efficiencies taken directly
from the background angular distributions rather than the default
parameterization. The latter effect is investigated by generating
pseudoexperiments with non-reweighted MC events as input for the
angular efficiencies. The second test is a rather extreme case, but
shows only a small systematic effect.


The next systematic uncertainty that we consider is the modeling of the
signal $\Bs$ mass distribution, which is fitted by default with a single
Gaussian distribution. If a double Gaussian model is used, the fit
quality would be comparable and we evaluate a systematic uncertainty
from the comparison of the two fit models. To test the size of a
potential systematic effect, we generate pseudoexperiments with a
double-Gaussian signal mass model, extracted from data, and fit with the
usual single-Gaussian parameterization.


Similarly, the model used for the mass distribution of combinatorial
background events can be changed to a second order polynomial with
comparable fit quality.  To study this effect, we generate
pseudoexperiments with a second-order polynomial background model
instead of the default first order polynomial and fit with the default
straight line. The results of both tests and the corresponding
systematic uncertainties are listed in Table~\ref{tab:pe_syst_totals}.


A particularly important effect to consider for the lifetime
measurement is the lifetime resolution model.  In our standard fit we
model the detector resolution by convolving each lifetime component
with a two-Gaussian resolution function. To test the effect of a
mis-parameterization, we generate pseudoexperiments with a
three-Gaussian resolution model, extracted from data, and fit with the
default two-Gaussian model.


As well as the lifetime resolution, the modeling of the various
components of the background lifetime can systematically affect the
measured $\Bs$ lifetime. To check the effect of any inaccuracy in our
background lifetime model as described in Section~\ref{sec:fit}, we
generate pseudoexperiments with the decay time of the background events
taken from histograms of the $\Bs$ mass sidebands and fit with the
default model.

We consider three possible sources of systematic uncertainty related to
the transversity angles of the background events: mis-modeling of the
parameterization described in Section~\ref{sec:fit}, ignoring the
observed small correlations between the three angles, and correlations
between the angles and the expected proper decay time uncertainty,
$\sigma_{ct}$.  The effect of these sources of uncertainty is checked
using the actual data distributions from the mass sidebands to generate
pseudoexperiments and test the difference between our model and the
true distributions. For the parameterization check, we simply use the
data background angular distributions in the generation of the
pseudoexperiments before fitting with the default model. To check the
effect of neglecting the small correlations between the angles, we
generate pseudoexperiments where two of the background angles are
sampled randomly from the data distributions, and the third one from a
two-dimensional histogram according to the sampled value of the second
angle. To check the effect of ignoring correlations between the
transversity angles and $\sigma_{ct}$, we sample the $\phi_T$ angle
distribution, found to have the largest correlation with $\sigma_{ct}$,
using a two-dimensional histogram of $\phi_T$ versus $\sigma_{ct}$ in
order to generate the pseudoexperiments.  The effect of ignoring these
very small correlations results in an almost negligible systematic
uncertainty on the measurements (see Table~\ref{tab:pe_syst_totals}).

In the default fit, we do not account for contamination from $B^0
\rightarrow J/\psi K^{*0}$ events mis-reconstructed as \BsJpsiPhi~decays
($B^0$ cross-feed). A small fraction of these events lies in the
$\Bs$~mass signal region. The first step in identifying the size of the
systematic effect 
is to estimate the size of this
contribution by using measured production fractions of the $\Bs$ and
$B^{0}$ mesons, their relative decay rates to $\Jpsi\phi$ and $\Jpsi
K^{*0}$, respectively, and the probability for each type of event to pass
our final selection criteria when reconstructed under the \BsJpsiPhi~hypothesis.
Both the production fractions and the branching fractions are taken from
Ref.~\cite{Ref:PDG}. We estimate the efficiencies using simulation, with
both \BsJpsiPhi and $B^0\rightarrow \Jpsi K^{*0}$ modes reconstructed as
\BsJpsiPhi~decay. The fraction $f$ of $B^{0}$ cross-feed events in the
\Bs sample is calculated as
\begin{eqnarray}
&& f(B^{0} ~{\textrm{in \Bs sample}}) =  \nonumber \\
&& \hspace*{1.0cm} = \frac{f(\bar b \rightarrow B^{0})\, 
{\mathcal B}(B^0\rightarrow J/\psi K^{*0})\,\epsilon(B^{0})}
{f(\bar b \rightarrow \Bs)\,{\mathcal B}(\BsJpsiPhi)\,\epsilon(\Bs)}.
\label{eqn:crossfeed}
\end{eqnarray}
Using Eq.~(\ref{eqn:crossfeed}), we find that the fraction of $B^{0}$
cross-feed into the signal sample of this analysis is
$(1.6\pm0.6)\%$. To make a conservative estimate of the systematic
uncertainty that this effect will add to the measurement of the
parameters of interest, we generate pseudoexperiments with a fraction
of 2.2\% $B^{0}$ cross-feed, and fit with the default model which does
not account for this component.  The cross-feed component is generated
using values of the $B^0$ lifetime, decay width and transversity
amplitudes from the CDF angular analysis of $B^0 \rightarrow J/\psi
K^{*0}$~\cite{Ref:Thesis_Anikeev}.

A systematic uncertainty can be introduced by the assumption that the
silicon detector is perfectly aligned, when it could actually be
mis-aligned by bowing of the detector layers of up to 50~$\mu$m.  A
study on the effect of the limited knowledge of the CDF\,II silicon detector
alignment concluded that a conservative estimation of the systematic
uncertainty on the decay length $c\tau$ in CDF lifetime measurements is
given by a 2 $\mu$m systematic uncertainty on
$c\tau$~\cite{Ref:Alignment1,*Ref:Alignment2}. This study was done by
fully reconstructing both data and simulation under different silicon
alignment assumptions, including shifts of $\pm 50~\mu$m in all silicon
detector components.  The lifetime was fitted in several $B \rightarrow
\Jpsi X$ channels, and the worst shift was taken as the systematic
uncertainty on the lifetime due to the assumption of perfect silicon
alignment.

We use the value of 2~$\mu$m systematic uncertainty on $c\tau(\Bs)$ to
also assess secondary effects on the other parameters of interest.  Due
to correlations between the \Bs~lifetime and the other physics
parameters, it is expected that an additional uncertainty on the
lifetime measurement will also cause uncertainties in the measurement of
the other parameters. To quantify the effect on the other parameters, we
generate pseudoexperiments in which the decay time in each event is
randomly shifted $\pm2~\mu$m and fit in the usual manner to allow for
comparisons between the input and fitted values of the parameters of
interest.

In the fit model, we treat the mass resolution identically for signal
and background events.  The effect of any inaccuracy in this
assumption can be tested by generating pseudoexperiments with mass
uncertainty distributions modeled by histograms of $\Bs$ sideband data
for background events and sideband-subtracted signal region data for
signal events separately, and then fitted with the default model.

Finally, we consider the effect of mis-parameterization of the
$\sigma_{ct}$ distributions. To account for a possible
effect, we fit with the default model to pseudoexperiments generated
with the uncertainty distributions taken from data histograms rather
than the model described in Section~\ref{sec:fit}. This systematic
check also accounts for any effect caused by small observed
correlations between $\sigma_{ct}$ and the invariant mass by sampling
the background uncertainties from separate upper and lower sideband
histograms according to the generated~$\Bs$~mass.

The total systematic uncertainty assigned to \deltaG is 0.058~\ps,
while 3.6~$\mu$m is assigned to the measurement of $c\tau(\Bs)$.  Both
$|A_{\parallel}(0)|^{2}$ and $|A_{0}(0)|^{2}$ are assigned an
uncertainty of 0.015.  Finally, $\delta_{\perp}$ is assigned a 0.07~rad
uncertainty.

\section{Frequentist Analysis of  \boldmath{$\betas$} and
\boldmath{$\deltaG$}}
\label{sec:statistics}


As anticipated in Sec.~\ref{sec:smresult}, due to the pathological 
behavior of the likelihood function with respect to $\betas$ and $\deltaG$,
we perform a frequentist analysis to determine confidence regions for 
these parameters. 
In addition, we perform a cross-check in which we determine credible
intervals for $\betas$ and $\deltaG$ using Bayesian techniques described
in Sec.~\ref{06_CDF}. 
In the main analysis we use profile-likelihood ratio
ordering~\cite{Ref:FeldmanCousins} to determine the confidence level
region in the \betas-\deltaG space. The coverage of the confidence
region against deviations of the nuisance parameters from their measured
values is confirmed by explicitly checking the effect of variations of
the nuisance parameters on the profile-likelihood shape. Following our
previous publication~\cite{Ref:tagged_cdf_old}, the confidence
regions are corrected to guarantee a coverage of the true value with
at least the nominal confidence level. 
We thus choose this method to quote the main results of this paper.

The likelihood function has several symmetries that are discussed in detail in
Ref.~\cite{Ref:beta_s_JHEP}. In absence of the $S$-wave component in
$\Bs \rightarrow \Jpsi K^+K^-$ decays, the likelihood function is
symmetric under the simultaneous transformations: $\betas \rightarrow
\frac{\pi}{2} - \betas$, $\deltaG \rightarrow - \deltaG$,
$\delta_{\perp} \rightarrow \pi -\delta_{\perp}$, and
$\delta_{\parallel} \rightarrow -\delta_{\parallel}$.  An approximate
symmetry is also present under the above simultaneous transformations of
only \betas and \deltaG .  The approximate symmetry produces a
local minimum in the $\betas$-$\deltaG$ space in addition to the global
minimum. We account for this effect by performing the likelihood scan
with $\delta_{\parallel}$ being started in the fit separately in the
range [0, $\pi$] and then in the range [$\pi$, 2$\pi$]. At each point in the
$\betas$-$\deltaG$ plane, we choose the deeper of the two $-2\log\mathcal{L}$
likelihood values (absolute minimum) as evaluated for the different
$\delta_{\parallel}$ ranges. This procedure guarantees that we use the
global, not the local minimum, at each point.

Once we have minimized the likelihood function on data with respect to
the $\betas$ and $\deltaG$ parameters, we proceed with determining the
68\% and 95\% confidence regions. Constructing correct and informative
confidence regions from highly multi-dimensional likelihoods is
challenging and, as in our case, evaluating the full 35-dimensional
confidence space is computationally prohibitive. To construct a
proper coverage adjustment, which ensures that the quoted 68\% (95\%)
confidence levels do indeed contain the true values of \betas and
\deltaG at least 68\% (95\%) of the time, the choice of the ordering
algorithm is important. 
We choose the profile-likelihood ratio ordering
method~\cite{Ref:FeldmanCousins} described below. The obtained
profile-likelihood ratio is then used as a $\chi^2$ variable to derive
confidence regions in the two-dimensional space of
\betas-\deltaG. However, simulations show that the observed
profile-likelihood ratio deviates from a true $\chi^2$~distribution. In
particular, the resulting confidence regions contain true values of the
parameters of interest with lower probability than the nominal
confidence level and, in addition, the profile-likelihood ratio appears
to depend on the true values of the nuisance parameters, which are
unknown. We therefore use a large number of pseudoexperiments to derive
the actual profile-likelihood ratio distribution relevant for our
data. The effect of systematic uncertainties is accounted for by
randomly sampling a limited number of points in the space of all
nuisance parameters and using the most conservative of the resulting
profile-likelihood ratio distributions to calculate the final confidence
region. In the following, the coverage adjustment procedure is described
in detail.


\subsection{Coverage adjustment}

To construct coverage adjusted confidence level regions for each point
in the \betas-\deltaG~plane, we start with calculating a $p$-value,
which, given a certain hypothesis, describes the probability to observe
data as discrepant or more discrepant than the data observed in our
experiment.  The set of all points with a $p$-value larger than $1-x$
forms the $x\,\%$~C.L.~region. In particular, the set of points with
$p$-value larger than $1 - 0.95 = 0.05$ outlines the $95\%$ confidence
region.

Since a main goal of our analysis is to determine the compatibility of
our data with the standard model expectation for \betas, we start by
calculating the SM $p$-value. We generate pseudoexperiments at the
standard model expected point in the $\betas$-$\deltaG$ plane
($\betas=0.02$, $\deltaG=0.090$~\ps). When generating the
pseudoexperiments, we use the best fit values for all nuisance
parameters as observed in our data while $\betas$ and $\deltaG$ are
fixed to the SM expected values.  The likelihood function corresponding
to each pseudoexperiment is first maximized with all parameters
floating, and then maximized a second time with $\betas$ and $\deltaG$
fixed to their SM values while the remaining fit parameters (nuisance
parameters) are independently floating.  We then form twice the negative
difference between the logarithms of the likelihood values obtained in
each of the two steps to obtain a profile-likelihood ratio value
$-2\Delta \log\mathcal{L}$.  The profile-likelihood ratio distribution
from 1000 pseudoexperiments is used to obtain the standard model
$p$-value, which is the fraction of pseudoexperiments with $-2\Delta
\log\mathcal{L}$ larger than the corresponding quantity observed in
data.

\begin{figure}[tbp]                 
\includegraphics[width=0.49\textwidth]{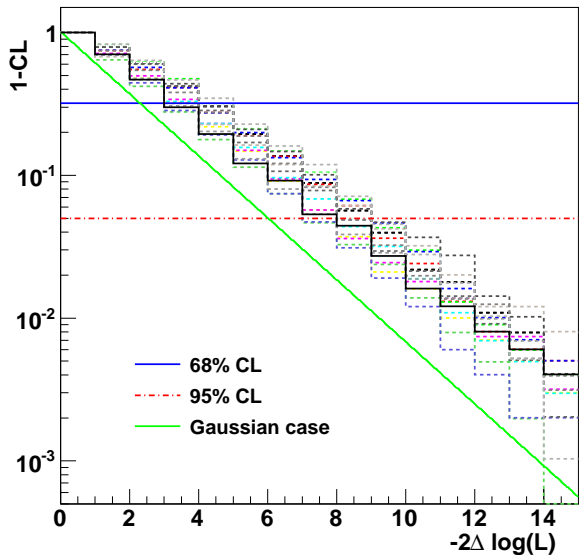}
\caption{(color online).
  Mapping of $p$-value ($1-\mathrm{C.L.}$) as a function of twice
  the negative difference of log-likelihoods ($-2\Delta\log\mathcal{L}$) as
  evaluated in pseudoexperiments. 
  The ideal dependence is a $\chi^2$~distribution with two degrees of freedom 
  as shown by the solid (green) line. The actual observed mapping for our 
  data is shown by the black histogram, while the corresponding 
  distributions for the alternative ensembles are displayed by the
  colored, dashed histograms. 	
\label{2D_coverage_final}}
\end{figure}

We construct the cumulative distribution of $-2\Delta \log\mathcal{L}$
to obtain a mapping between the $p$-value = $1-\mathrm{C.L.}$ and
$-2\Delta \log\mathcal{L}$, as shown in Fig.~\ref{2D_coverage_final} by
the solid black histogram which has been interpolated. In an ideal
situation, when the likelihood function is Gaussian with respect to
\betas and \deltaG, this dependence should be a $\chi^2$ distribution
with two degrees of freedom as indicated by the green line.  It is
evident from Fig.~\ref{2D_coverage_final} that, at least with our
current data sample size, we are not in an asymptotic, Gaussian regime.
To test the dependence of the obtained mapping on the chosen SM point
for \betas and \deltaG, we construct similar maps between the confidence
level and $-2\Delta\log\mathcal{L}$ for other random points in the
$\betas$-$\deltaG$ plane and find very similar dependencies.
Consequently, we consider the mapping determined at the SM point to
apply for all points in the $\betas$-$\deltaG$ plane.

To obtain confidence regions in \betas and \deltaG, we determine
profile-likelihood ratios for a grid on the \betas-\deltaG plane.  In a
Gaussian regime, the points with $p$-value = 0.05, corresponding to a
confidence level of 95\%, are identified by the intersection of the
two-dimensional profile-likelihood function and a horizontal plane which
is 5.99 units above the global minimum. The value 5.99 is the point on
the $-2\Delta \log\mathcal{L}$ axis where the $\chi^2$ distribution with
two degrees of freedom (green line) intersects the $1 - 0.95 = 0.05$
level (red dashed line) in Fig.~\ref{2D_coverage_final}. The
68\%~C.L.~is correspondingly obtained by the top horizontal (blue)
line. The intersection between the $0.05$ level and the actual mapping
(black histogram) is at $-2\Delta \log\mathcal{L} = 7.34$ which means
that the 95\% confidence region is obtained by taking the intersection
of the two-dimensional profile-likelihood function and a horizontal
plane which is 7.34 units above the global minimum. In this case
we find the standard model $p$-value for \betas to be 0.27. Clearly,
this procedure leads to confidence regions larger than in the ideal,
Gaussian case.


In order to guarantee additional coverage over a conservative range of
possible values of nuisance parameters, sixteen alternative ensembles
are generated.  As we do not know the true values for these nuisance
parameters, we compute the coverage over a wide range of possible values
but always within their physically allowed
range~\cite{Ref:5sigma_stat_reference}.  In particular, each alternative
ensemble is produced by generating pseudoexperiments with nuisance
parameters randomized uniformly within $\pm 5\,\sigma$ of their best fit
values as obtained from maximizing the likelihood function on data. In
these pseudoexperiments, the parameters $\betas$ and $\deltaG$ are again
fixed to their standard model expectation. We choose a random variation
of $\pm5\,\sigma$ over the nuisance parameters because we aim to cover
the space of nuisance parameters with a C.L.~much larger than the
anticipated C.L.~for our final result.  Exceptions to this approach are
the strong phases which are generated only within the range from zero to
$2\pi$ and the dilution scale factors which are generated so that the
dilution is always between zero and one.  The other exception to
applying a $\pm 5\,\sigma$ range is the phase $\delta_{SW}$, which is
generated flat between 0 and $2\pi$.  Since the $S$-wave fraction
$f_{SW}$ is consistent with zero as discussed in
Sec.~\ref{sec:nonsmresults}, we lack sensitivity to the associated phase
and choose to vary it over its full range possible.


To determine this additional coverage adjustment, we again generate 1000
pseudoexperiments for each of the sixteen alternative ensembles. The
same profile-likelihood ratio procedure is performed on
each ensemble, and the broadest and thus most conservative $p$-value
is taken to form the final confidence regions. The colored, dashed lines
in Fig.~\ref{2D_coverage_final} show the resulting mappings between
$1-\mathrm{C.L.}$ and $-2\Delta \ln{\cal L}$ for each of the sixteen
alternative ensembles. We use the $p$-value of the most conservative
ensemble to determine the corresponding 68\% and 95\% confidence regions
for our data. The intersection between the $0.05$ ($0.32$) confidence
level ($1-\mathrm{C.L.}$) and the most conservative mapping is at
$-2\Delta \log\mathcal{L} = 8.79\ (4.19)$. This means that the 95\%
(68\%) confidence region with guaranteed coverage is obtained by taking
the intersection of the two-dimensional profile-likelihood ratio and a
horizontal plane which is 8.79 (4.19) units above the global minimum.
Note that this procedure of randomly sampling a limited number of points in
the space of all nuisance parameters and using the most conservative of
the resulting profile-likelihood ratio distributions automatically
accounts for the effect of systematic uncertainties.


A similar coverage adjustment procedure is carried out to determine
individual confidence intervals for $\betas$ and $\deltaG$ separately.
When determining the $\betas$ confidence interval, $\deltaG$ is
randomized in the pseudoexperiment generation and treated analogously
with the other nuisance parameters.
We again generate 1000 pseudoexperiments per alternative ensemble for
the final coverage adjustment in the one-dimensional case.  Using a
similar approach as in the two-dimensional case, we obtain again
mappings of $1-\mathrm{C.L.}$ versus $-2\Delta\log\mathcal{L}$ for
alternative ensembles with randomized nuisance parameters.



\subsection{Results using frequentist approach} 
\label{sec:nonsmresults}

\begin{figure}[tbp]
\includegraphics[width=0.48\textwidth]{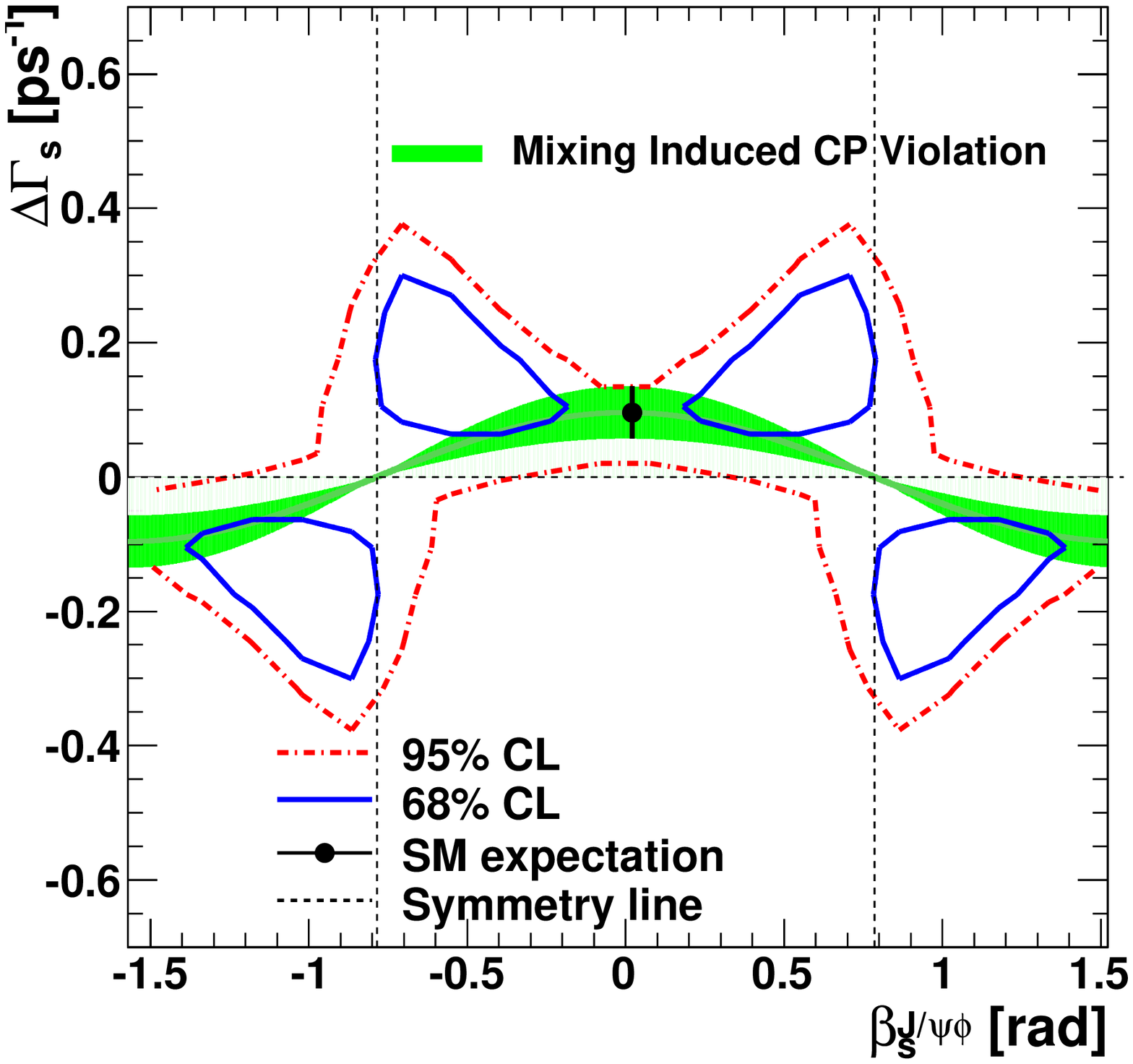}
\caption{(color online).
  Confidence regions in $\betas$-$\deltaG$ plane for the fit
  without application of flavor tagging.  The solid (blue) and dot-dashed
  (red) contours show the 68\% and 95\% confidence regions,
  respectively. The dotted lines are the symmetry axes corresponding to
  the profiled likelihood invariance under $\betas \rightarrow
  \frac{\pi}{2} - \betas$ and $\deltaG \rightarrow - \deltaG$.  In
  addition, the likelihood is invariant under $\betas \rightarrow -
  \betas$. The shaded (green) band is the theoretical prediction of
  mixing-induced \CP~violation.
\label{betas_deltaG_untagged}}
\end{figure}

We present frequentist $\betas$-$\deltaG$ confidence regions and
$p$-values obtained according to the procedure described in
Sec.~\ref{sec:statistics} above.  The $\betas$-$\deltaG$ confidence
regions without the application of flavor tagging are shown in
Fig.~\ref{betas_deltaG_untagged}.  The SM prediction is indicated by the
black marker, and the 68\% and 95\% C.L.~regions are shown as solid
(blue) and dot-dashed (red) contours, respectively. We find the $p$-value
for \betas to agree with the standard model prediction to be 0.10. The
shaded (green) band is the theoretical prediction of mixing-induced
\CP~violation. As discussed above,
in the absence of an $S$-wave component, the likelihood function is
symmetric under the simultaneous transformations $\betas \rightarrow
\frac{\pi}{2} - \betas$, $\deltaG \rightarrow - \deltaG$,
$\delta_{\perp} \rightarrow \pi -\delta_{\perp}$ and $\delta_{\parallel}
\rightarrow -\delta_{\parallel}$.  In addition, if no flavor tagging
information is used, an additional symmetry is present in the likelihood
$\betas \rightarrow -\betas$.  As a consequence of these symmetries, the
likelihood function has four global maxima as can be seen in
Fig.~\ref{betas_deltaG_untagged}.

\begin{figure}[tbp]
\includegraphics[width=0.48\textwidth]{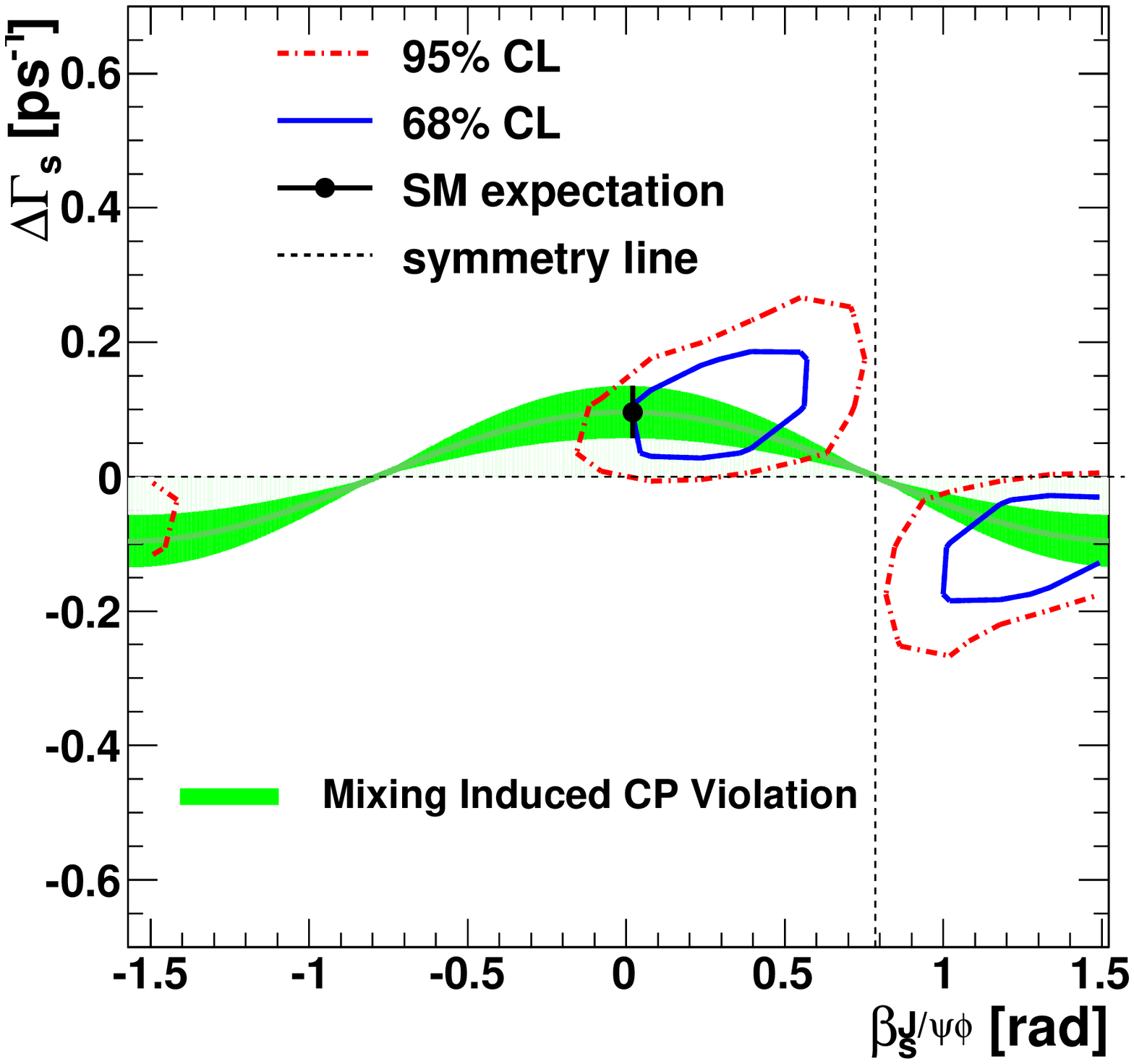}
\caption{(color online).
  Confidence regions in $\betas$-$\deltaG$ plane for the fit
  including flavor tagging information. The solid (blue) and dot-dashed
  (red) contours show the 68\% and 95\% confidence regions,
  respectively. The dotted lines are the symmetry axes corresponding to
  the profiled likelihood invariance under $\betas \rightarrow
  \frac{\pi}{2} - \betas$ and $\deltaG \rightarrow - \deltaG$.  The
  shaded (green) band is the theoretical prediction of mixing-induced
  \CP~violation.  
\label{betas_deltaG}}
\end{figure}

Once the flavor tagging information is added to the analysis, the
$\betas \rightarrow -\betas$ symmetry is removed and the likelihood
function has only two global maxima corresponding to the likelihood
invariance under $\betas \rightarrow \frac{\pi}{2} - \betas$ and
$\deltaG \rightarrow - \deltaG$.  The $\betas$-$\deltaG$ confidence
regions for the flavor tagged analysis, after coverage adjustment, are
shown in Fig.~\ref{betas_deltaG}. Our sensitivity to \betas and \deltaG
has substantially improved compared to our previously published
measurement~\cite{Ref:tagged_cdf_old}, as evidenced by
the decrease in size of the confidence region. The result is also
more consistent with the standard model prediction. 

\begin{figure*}[tbp]
\centerline{
\includegraphics[width=0.48\textwidth]{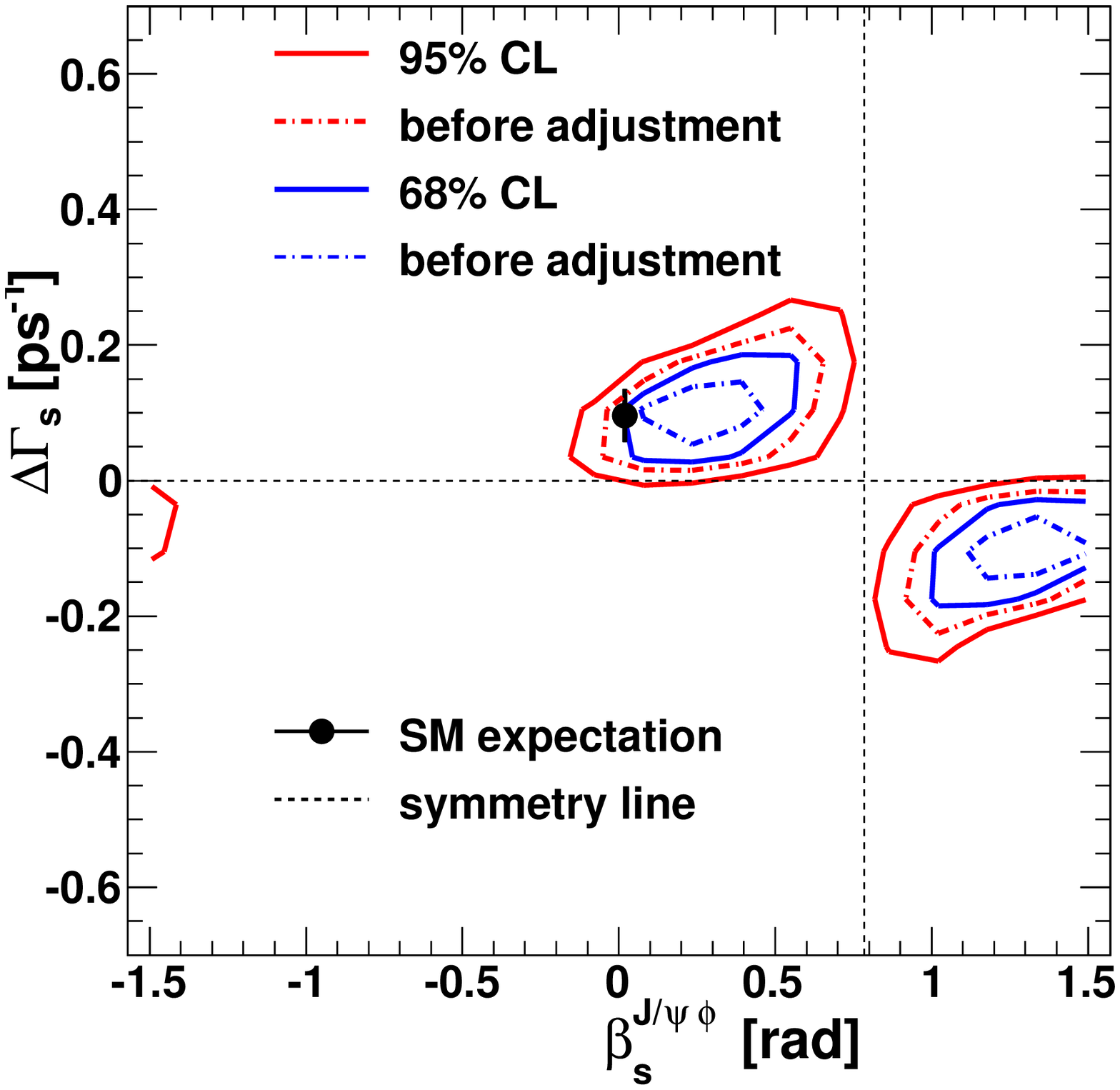}\ \ \
\includegraphics[width=0.5\textwidth]{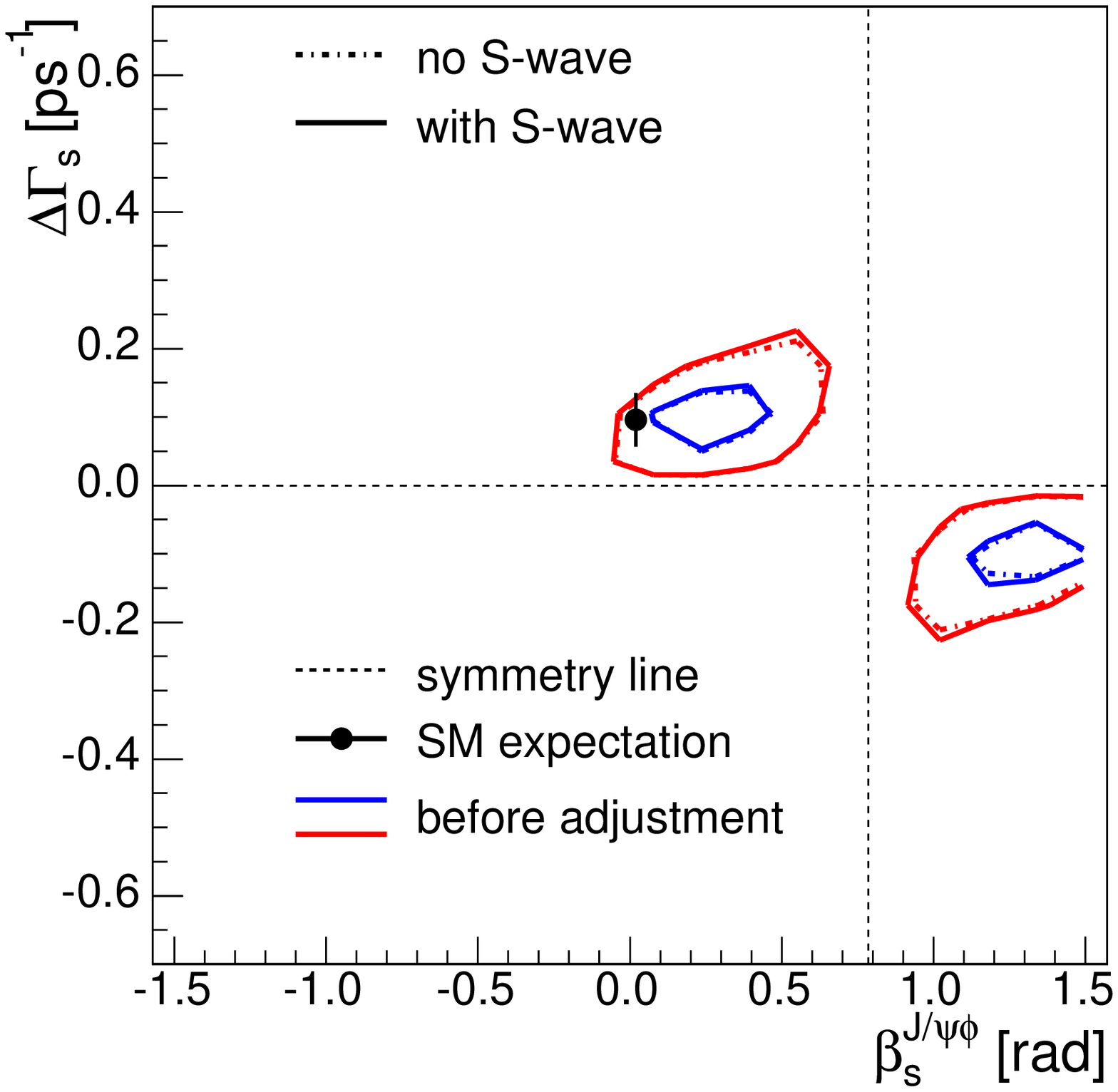}
}
\caption{(color online).
  Left: Confidence regions in $\betas$-$\deltaG$ plane for the fit
  including flavor tagging information before (dashed) and after (solid)
  performing the
  coverage adjustment. Right: Comparison of including (solid) and not
  including (dashed) the $S$-wave contribution in the likelihood fit.
\label{betas_checks}}
\end{figure*}

To illustrate the effect of the coverage adjustment, the left-hand side of
Fig.~\ref{betas_checks} compares the 68\% and 95\% C.L.~contours 
after coverage adjustment with the corresponding contours before the
coverage adjustment procedure. A small increase in the
size of the contours can be seen. As a further cross check, we also
performed the same fit setting the $S$-wave fraction to zero as shown on
the right-hand side of Fig.~\ref{betas_checks}. The contour regions
corresponding to a profile-likelihood ratio variation of
$-2\Delta\log\mathcal{L}=2.30$ (blue) and 
$-2\Delta\log\mathcal{L}=5.99$ (red) 
are compared when
including (solid) and not including (dashed) the $S$-wave fraction in
the likelihood fit. The contours are almost identical.

\begin{figure*}[tbp]
\centerline{
\includegraphics[width=0.49\textwidth]{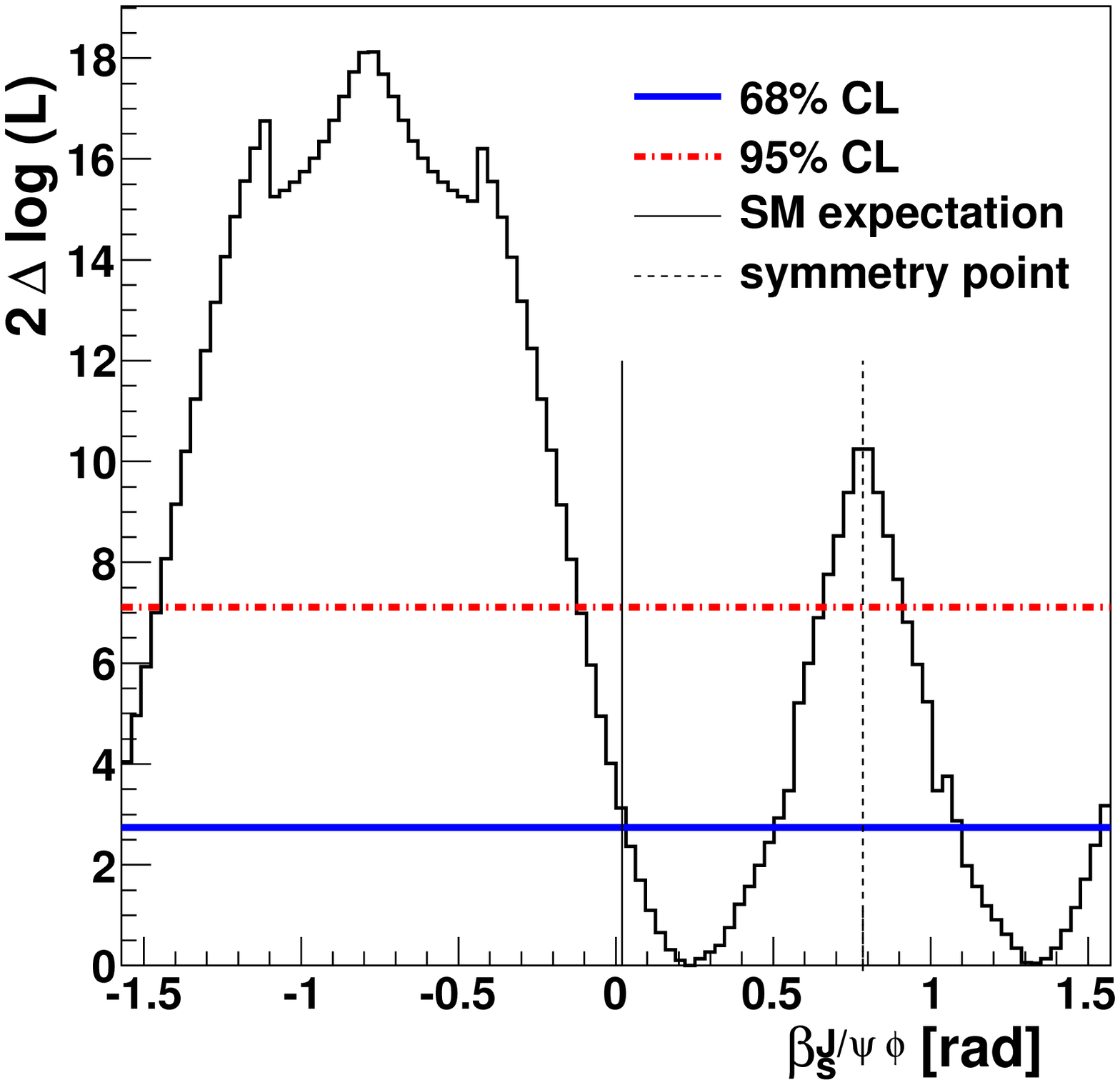}\
\includegraphics[width=0.49\textwidth]{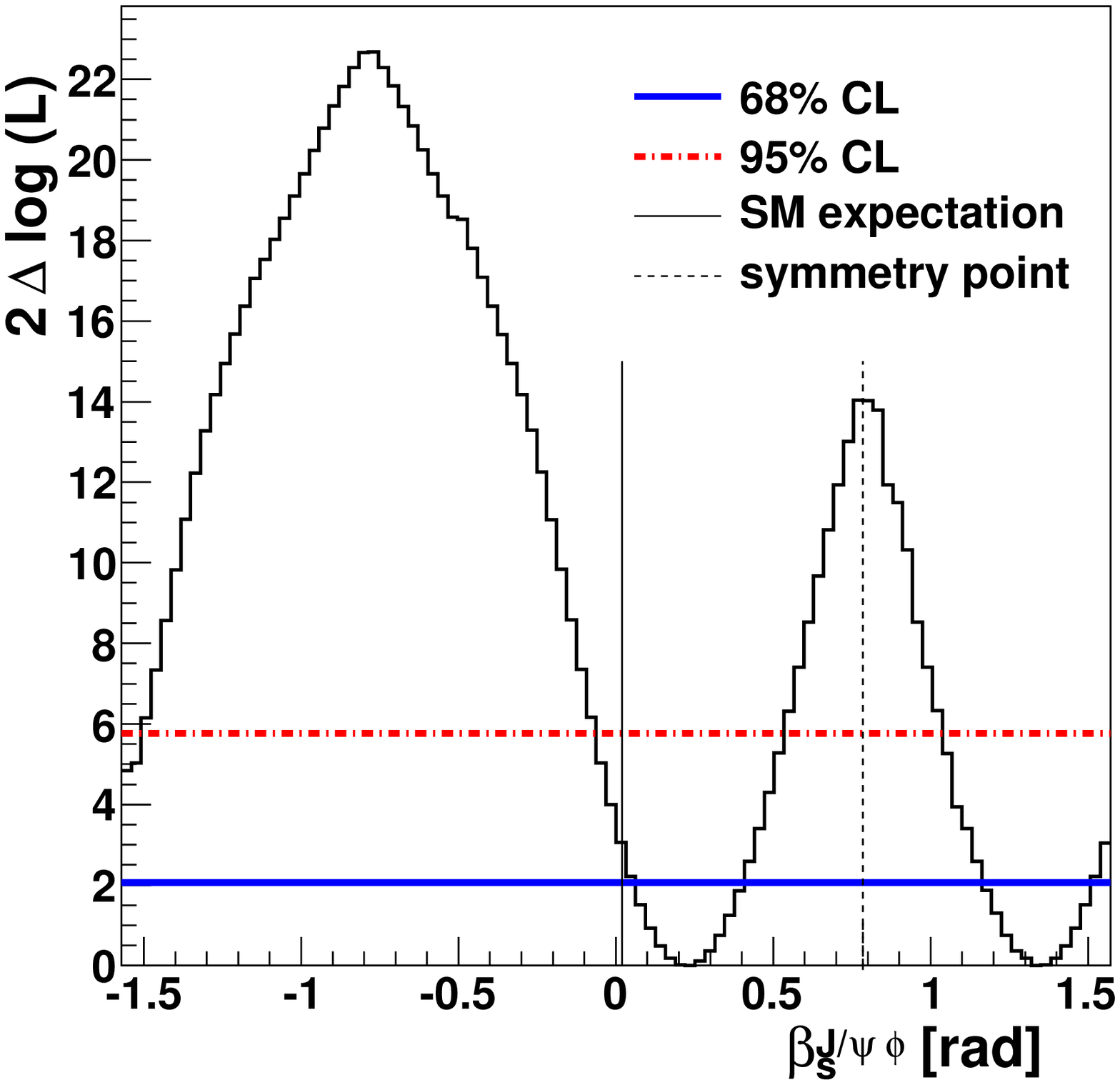}
}
\caption{Likelihood scan in \betas with no constraint (left) and with
  \deltaG constrained to the SM prediction (right). After coverage adjustment
  the solid (blue) and dot-dashed (red) horizontal lines indicate the
  68\% (95\%) C.L.~range above the global minimum.
\label{betas}}
\end{figure*}

The one-dimensional likelihood scan in the quantity \betas after coverage
adjustment is shown in Fig.~\ref{betas} on the left-hand side.
In a Gaussian scenario the 68\% (95\%) C.L.~range is between the points
of intersection of the profile-likelihood scan curve and a horizontal
line which is one unit (four units) above the global minimum. In our
case after coverage adjustment the solid (blue) and dot-dashed (red)
horizontal lines which indicate the 68\% and 95\%~C.L.~ranges are at
2.74 and 7.11 units above the global minimum, respectively.  We obtain
\begin{widetext}
\begin{eqnarray*}
\betas & \in & [0.02, 0.52] \cup 
[1.08,1.55]~\textrm{at 68\% confidence level,}\\
& \in & [-\pi/2, -1.46] \cup [-0.11, 0.65] \cup 
[0.91, \pi/2]~\textrm{at 95\% confidence level.} 
\end{eqnarray*}
\end{widetext}
We find the standard model $p$-value for \betas to be 0.30 corresponding
to about one Gaussian standard deviation from the SM expectation as is
also evidenced in Fig.~\ref{betas_deltaG}.
In comparison with the recent measurement of \betas from the
D0~collaboration using a data sample based on 8~fb$^{-1}$ of integrated
luminosity~\cite{Abazov:2011ry}, we find a similar region to constrain
\betas at the 68\% C.L. and obtain a similar $p$-value for comparison
with the SM expectation. However, our result constrains \betas to a
narrower region at the 95\% confidence level.

\begin{figure}[tbp]
\includegraphics[width=0.48\textwidth]{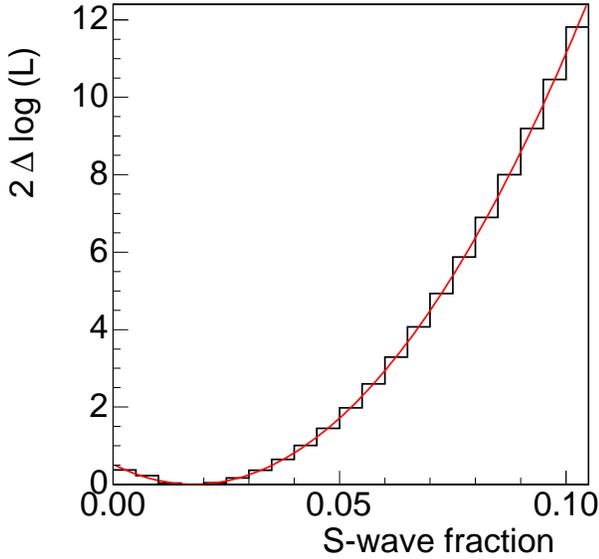}
\caption{Likelihood scan of the $S$-wave fraction with a quadratic fit
  overlaid indicating the parabolic shape of the likelihood around the minimum.
\label{swave-scan}}
\end{figure}

In addition, we quote a confidence interval for the $S$-wave fraction after
performing a likelihood scan for $f_{SW}$ as shown in
Fig.~\ref{swave-scan}. We also show a quadratic fit
overlaid indicating the parabolic shape of the likelihood around the
minimum which we integrate to calculate upper limits on
the $S$-wave fraction.
The upper limit on the $S$-wave fraction over the mass interval $1.009 <
m(K^+K^-)<1.028~\gevcc$ corresponding to the selected $K^+K^-$ signal
region is 4\% of the total signal at the 68\% confidence level, and
$f_{SW}<6\%$ at 95\%~C.L.  Since the analysis is limited to events in a
narrow $K^+K^-$ mass range around the $\phi$~signal, the observed
$S$-wave fraction is small and its effect on the observables quoted in
this analysis is minor.  We verified with pseudoexperiments that a
sizeable amount of $S$-wave would affect the measured value of \betas.
In contrast to our result, the recent D0~publication~\cite{Abazov:2011ry} 
quotes a sizeable fraction of $17.3\pm3.6\%$ for the $S$-wave fraction
over almost the same $K^+K^-$~mass range.  We also perform a likelihood
scan to determine the associated $S$-wave phase, but, as expected from
simulated experiments, we find that we are not sensitive to $\delta_{SW}$
with the current data sample size.

Finally, we perform a flavor~tagged analysis with \deltaG Gaussian
constrained to the 
theoretical prediction of 
$2\,|\Gamma^s_{12}| = (0.090 \pm 0.024)$~ps$^{-1}$~\cite{Ref:lenz}.  
Under this constraint, \betas is found in the range $[0.05, 0.40] \cup
[1.17, 1.49]$ at the 68\% confidence level, and within $[-\pi/2, -1.51]
\cup [-0.07, 0.54] \cup [1.03, \pi/2]$ at 95\% C.L. as shown in
Fig.~\ref{betas} on the right-hand side. The $p$-value for the SM
expected value of \betas~from this constrained fit is 0.21,
corresponding to a deviation from the SM expectation of $1.3\,\sigma$
significance. We note that the likelihood scans in Fig.~\ref{betas}
exhibit small deviations from the symmetry in \betas~that is expected
according to our discussion above. The reason is given by the small
$S$-wave fraction that our likelihood fit finds as well as the choice of
binning and numerical precision in determining the displayed
$-2\Delta\log\mathcal{L}$ values.


\section {Results on \boldmath{\betas} and \boldmath{\deltaG} in a
  Bayesian Approach}  
\label {06_CDF}

In addition to the frequentist results shown in the previous section,
we use a Bayesian analysis to provide cross-checks on the determination of the  
physics parameters.
We use Bayesian inference via integration of the
posterior density obtained from the
likelihood function described in Sec.~\ref{sec:fit} over the nuisance
parameters and over those physics parameters in which we are
not presently interested. 

The starting point for this Bayesian approach is the likelihood
function, ${\cal L} ({\vec x}\,|\,{\vec\theta},{\vec \nu}\,)$, where
${\vec x}$ are the experimental measurements including the
$\Bs$~candidate decay time and invariant mass, the transversity angles
and tagging information, while 
$\vec{\mu} = ({\vec \theta},{\vec \nu}\,)$ 
is a vector distinguishing the physics parameters ${\vec \theta}$ 
described in Table~\ref{tab:physicsParameters} from the
remaining nuisance parameters ${\vec \nu}$ in the fit describing
features such as background shapes, tagging performance, and detector
resolution (see Sec.~\ref{sec:fit}). In our analysis the dimensionality
of $\vec\theta$ and $\vec\nu$ is 11 and 24, respectively.

Within the Bayesian approach to statistical inference, Bayes' theorem
defines the posterior probability density given the observed data set
${\vec x}$
\begin{eqnarray}
p({\vec \theta}\,|\,{\vec x}) = 
\frac{p({\vec x}\,|\,{\vec \theta}\,)\,p({\vec \theta}\,)}
{\int p({\vec x}\,|\,{\vec \theta}\,)\,p({\vec \theta}\,)\ d^N\theta}\, ,
\label{eqn:BayesND}
\end{eqnarray}
where $p({\vec x}\,|\,{\vec \theta}\,)$ is the likelihood
function ${\cal L}(\vec x\,|\,{\vec \theta}\,)$ and $p({\vec \theta}\,)$ is the
prior probability density for ${\vec \theta}$, which describes the
knowledge about parameters ${\vec \theta}$ that we assume prior to our
measurement.  The projection of the $N$-dimensional posterior density
onto $M$ parameters of physical interest corresponds to an integration
over all the other $N-M$ parameters, where the limits of integration
cover all possible values for the $N-M$~parameters. This projection
gives a new posterior density $p(\vec \alpha\,|\,{\vec x})$ for
parameters $\alpha_i \in {\vec \theta}$. For a single variable $\alpha$
and a uniform prior density this expression reduces to
\begin{eqnarray}
p({\alpha}\,|\,{\vec x}) = 
\int p({\vec \theta}\,|\,{\vec x})\,(\theta)\ d^{N-1}\theta\,.
\label{eqn:Bayes1D}
\end{eqnarray}

We compute a representation of the 35-dimensional likelihood function
including nuisance parameters, which we store in a computer-readable
data format (Ntuple). These Ntuples contain Markov chains, which have
been generated by a Markov chain Monte Carlo (MCMC)
technique~\cite{ref:MCMCBook}. Projections of the Markov chains onto
subspaces of physical parameters of interest, in particular \betas,
\deltaG, etc., are then performed. In addition, we compute credible
intervals for certain parameters and credible contours for pairs of
parameters derived from these projections, as discussed in more detail
below.

From a Markov chain projection 
one may easily draw conclusions about specific values of
parameters such as \betas and \deltaG
with a view toward propagating that information into global fits
of, e.g., CKM parameters and incorporating certain prior information,
about, e.g., mixing-induced \CP~violation and the
values of the strong phases that appear in \BsJpsiPhi~decays.
By projecting a Markov chain onto a subspace of parameters of
dimension $M$, i.e., making a histogram or a scatter-plot from
the Ntuples, one is in fact performing a numerical integration of the
posterior density, over the other $N-M$ parameters.  The normalization
factor, i.e., the denominator in Eq.~(\ref{eqn:BayesND}), is easily
identified as the total number of points of the Markov Chain.

We can define a credible interval [$\alpha_L,\alpha^U$] for the
parameter $\alpha$ with probability content $\beta$ through
\begin{eqnarray}
\beta = \int_{\alpha_L}^{\alpha^U}p(\alpha\,|\,{\vec x}) \ d\alpha
= P(\alpha_L < \alpha < \alpha^U).
\end{eqnarray}
%
The credible interval [$\alpha_L,\alpha^U$] contains a fraction $\beta$
of the posterior density about $\alpha$ but it is not unique. However, we
can build a ``shortest interval'' using the straightforward algorithm of
maximum probability ordering by accepting into the interval the
largest values of the PDF $p(\alpha\,|\,{\vec x})$.  Using the same
algorithm as in the one-dimensional case, we build credible contours
in the \betas-\deltaG plane. A credible interval (or contour) does
not necessarily cover the true value of a parameter (or parameters) with
any given frequentist probability. On the other hand, regarding the
technical aspect it allows
for the combination of experimental results and theoretical inputs
in a straightforward manner. In addition, the Bayesian technique
can be trivially modified to incorporate other conditions on, for
example, 
mixing-induced \CP violation or constraints on the strong phase angles
through non-uniform prior probability densities (see
Sec.~\ref{08_Mixing} and Sec.~\ref{08A_MixingAndSP}).

To verify convergence of the Markov chain, we generate sixteen
independent chains.  A burn-in phase of approximately 10\,000
steps is identified.  We discard the first 250\,000 states in each chain
and keep the following one million states. This means the probability
densities shown below are based on sixteen million states.

\subsection{Results using Bayesian approach} 

\begin{figure*}[tbp]
\centerline{
\includegraphics[width=0.49\textwidth]{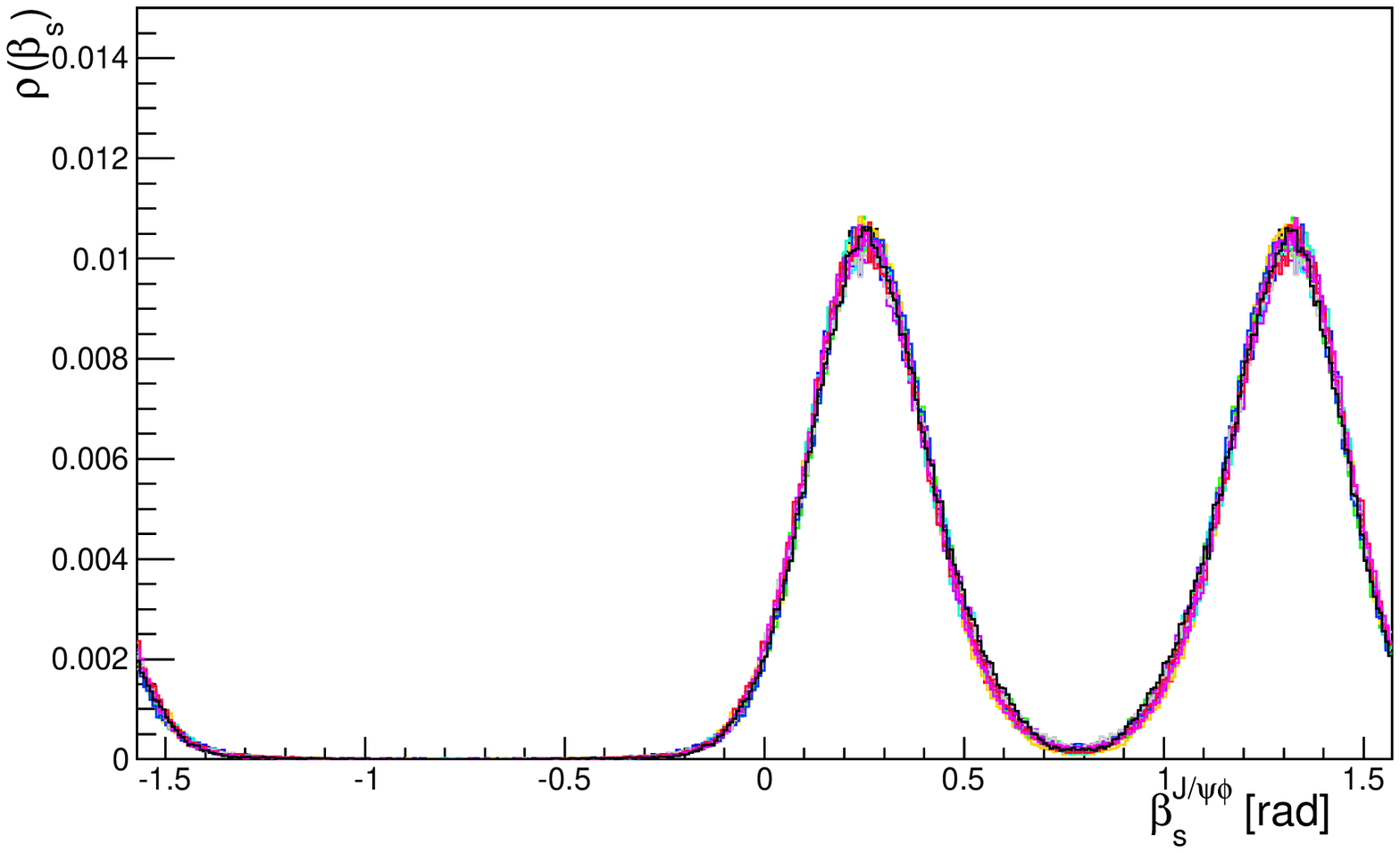}
\includegraphics[width=0.49\textwidth]{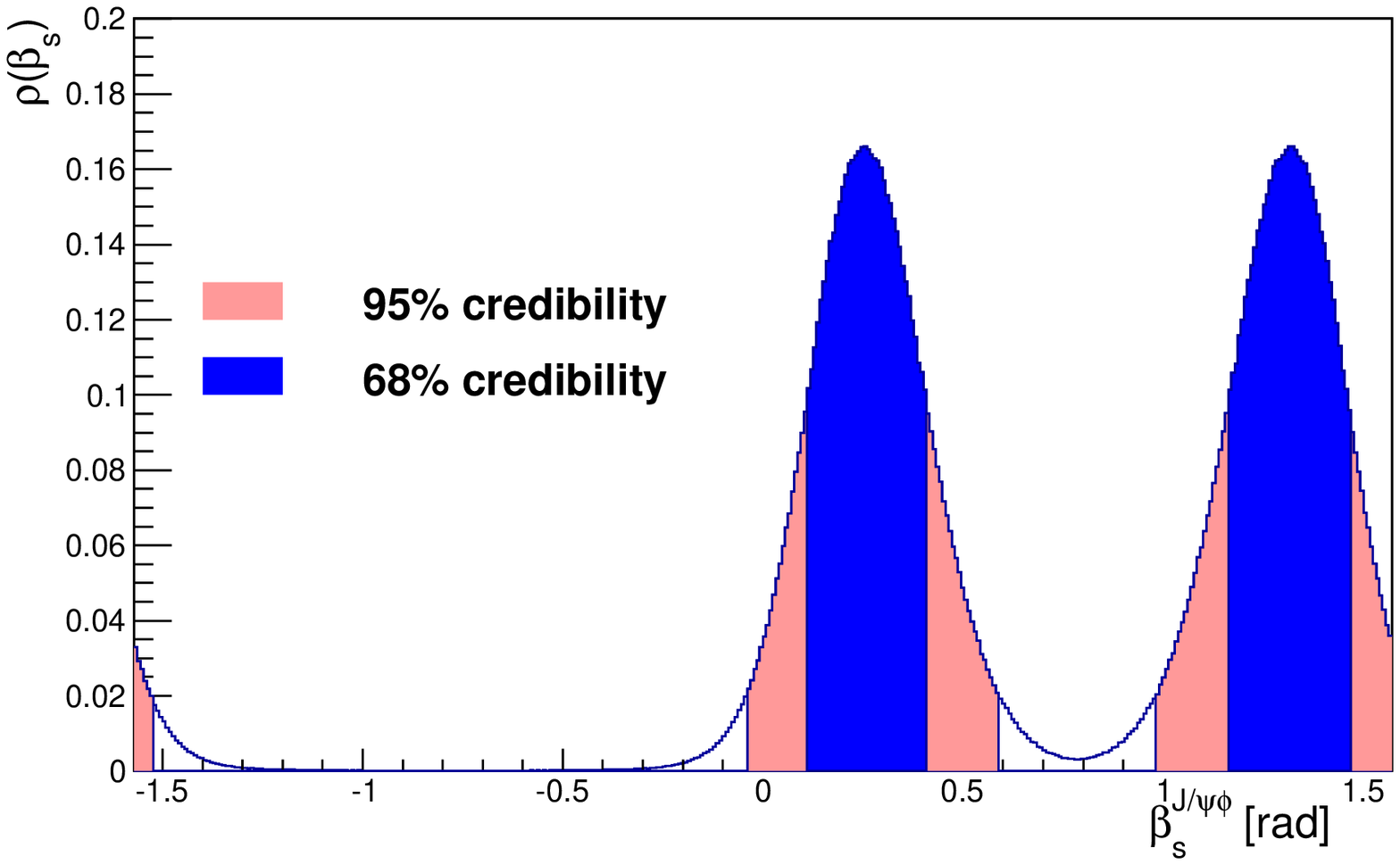}}
\centerline{
\includegraphics[width=0.49\textwidth]{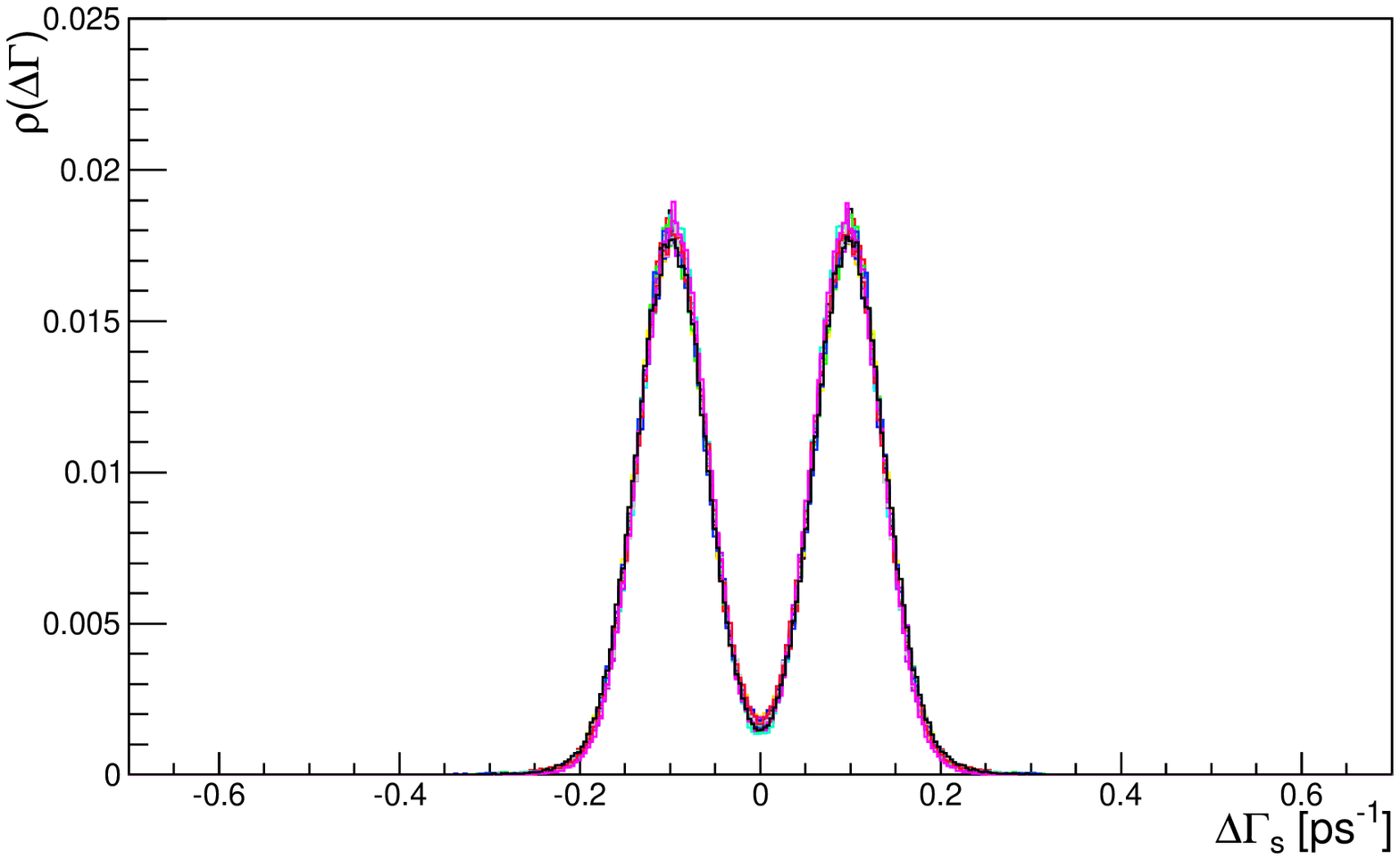}
\includegraphics[width=0.49\textwidth]{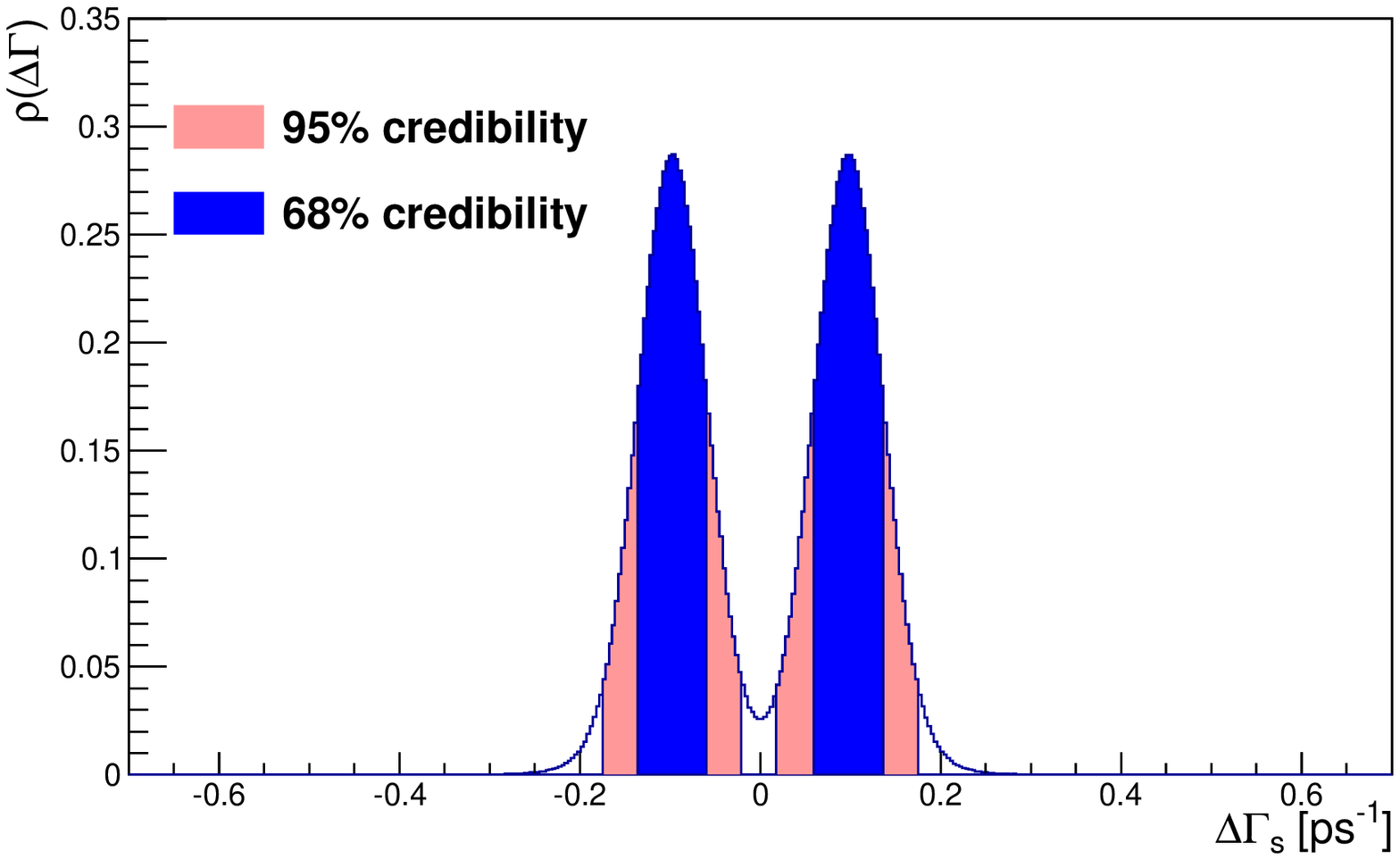}}
\caption{(color online).
Bayesian posterior densities for the variables \betas (top) and
\deltaG (bottom). The left plots show projections of sixteen independent Markov
Chains, while the right two plots show the posterior densities with
68\% and 95\% credible intervals in dark-solid (blue) and light-solid
(red) areas, respectively.
\label{fig:COMBO-S}}
\end{figure*}

The projections of the sixteen chains onto the variables \betas and
\deltaG are displayed in Fig.~\ref{fig:COMBO-S} together with the sum of
all sixteen chains. The close agreement between the sixteen independent
chains on the left-hand side of Fig.~\ref{fig:COMBO-S} also demonstrates
the convergence of the computation.
Using this Bayesian analysis of the data, we obtain 
\begin{widetext}
\begin{eqnarray*}
\betas &\in& \left [ 0.11, 0.41 \right] \cup \left [ 1.16, 1.47 \right]~\textrm{as 68\% credible interval,} \\
       &\in& \left [ -0.04, 0.59 \right] \cup  \left [0.98, 1.62 \right]~\textrm{as 95\% credible interval,} \\
\deltaG &\in& \left[-0.14~\ps, -0.06~\ps \right] \cup 
\left[0.06~\ps, 0.14~\ps\right]~\textrm{as 68\% credible interval,}\\
        &\in& \left [-0.18~\ps, -0.02~\ps \right] \cup 
\left[0.02~\ps, 0.18~\ps\right]~\textrm{as 95\% credible interval}.
\end{eqnarray*}
\end{widetext}

\begin{figure}[tbp]
\includegraphics[width=0.49\textwidth]{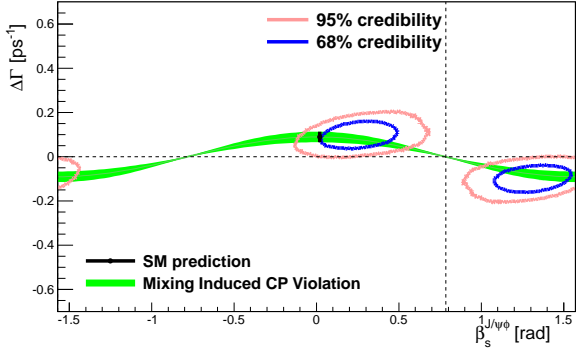}
\caption{(color online).
Joint posterior probability density in the 
\betas-\deltaG~plane for the combined analysis. The dark-solid (blue)
and light-solid (red) contours show the 68\% and 95\% credible regions,
respectively. 
The narrow band (green) is the theoretical prediction
of mixing-induced \CP~violation.
\label{fig:COMBO-bsDG-True}}
\end{figure}

The joint posterior probability densities and credible contours in the
\betas-\deltaG~plane are displayed in
Fig.~\ref{fig:COMBO-bsDG-True}. The narrow band shown in the figure is
the theoretical prediction of mixing-induced \CP violation using
$2|\Gamma^s_{12}| = (0.090 \pm 0.024)~\ps$~\cite{Ref:lenz,ref:multipleHiggs}
overlapping with our result.

Furthermore, we calculate
\begin{eqnarray}
\int_{{\cal A}_{\beta_s}} p(\betas)\ d\betas = 0.119,
\end{eqnarray}
where ${\cal A}_{\beta_s} = \{\betas ; p(\betas) < p(\beta_s^{SM})\}$,
indicating that, as the contour is expanded to enclose larger
credibility, the standard model prediction $\beta_s^{SM}$ is first
included within the enclosed region at a credibility of 88.1\%.

\begin{figure}[tbph]
\includegraphics[width=0.5\textwidth]{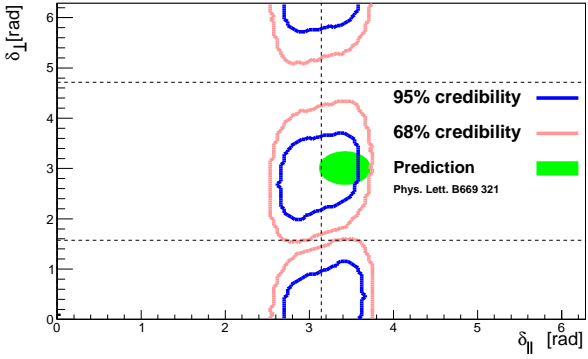}
\caption{(color online).
  Posterior density in the strong phases $\delta_{\perp}$ versus
  $\delta_{\parallel}$ overlaid with the prediction that the phases in
  \BsJpsiPhi match those in $B^0\rightarrow J/\psi K^{0*}$ decays to
  within 10$\degree$~\cite{ref:Rosner}. The dark-solid (blue) and
  light-solid (red) contours show the 68\% and 95\% credible regions,
  respectively.  The width of the light-shaded (green) ellipse includes
  the 10$\degree$ theoretical uncertainty added in quadrature with the
  experimental uncertainties on $\delta_{\parallel}$ and
  $\delta_{\perp}$ from $B^0\rightarrow J/\psi K^{0*}$.
\label{fig:PhasesII}}
\end{figure}

We also examine the posterior density in the variables
$\delta_{\parallel}$ versus $\delta_{\perp}$ as shown in
Fig.~\ref{fig:PhasesII}.  
It is predicted that the phases in \BsJpsiPhi match those in
the equivalent decay $B^0\rightarrow J/\psi K^{*0}$ to within
10$\degree$~\cite{ref:Rosner}.
The measured values of these phases in the
$B^0 \rightarrow J/\psi K^{*0}$ decay
($\delta_{\parallel}=-2.93\pm0.08\,\textrm{(stat)}
\pm0.04\,\textrm{(syst)}~\textrm{rad}$ and 
$\delta_{\perp}=2.91\pm0.05\,\textrm{(stat)}
\pm0.03\,\textrm{(syst)}~\textrm{rad}$~\cite{ref:KStarBabar}) are overlaid
in form of a green ellipse in Fig.~\ref{fig:PhasesII}. 
The width of this ellipse
includes the 10$\degree$ theoretical uncertainty added in quadrature with
the experimental uncertainties on $\delta_{\parallel}$ and
$\delta_{\perp}$. For one mode of the probability density good
agreement between the \Bs and $B^0$~system is observed as predicted in
Ref.~\cite{ref:Rosner}. 

\subsection {Constrained results} 
\label {08_Mixing} 

Figure~\ref{fig:COMBO-bsDG-True} shows that our measurement in the
\betas-\deltaG~plane is consistent with the hypothesis of mixing-induced
\CP violation as well as with the hypothesis that the
measured \CP~violation originates from the standard model. We apply the
hypothesis of mixing-induced \CP violation, together with the
theoretical calculation of $\Gamma^s_{12}$, to our data in form of a
prior density during the computation of the MCMC. We carry out this
calculation by simply re-weighting the Markov chains using a new
flat prior density derived from the theoretical calculation which
gives $2\,|\Gamma^s_{12}| = (0.090 \pm 0.024)~\ps$~\cite{Ref:lenz}.

\begin{figure*}[tbph]
\centerline{
\includegraphics[width=0.5\textwidth]{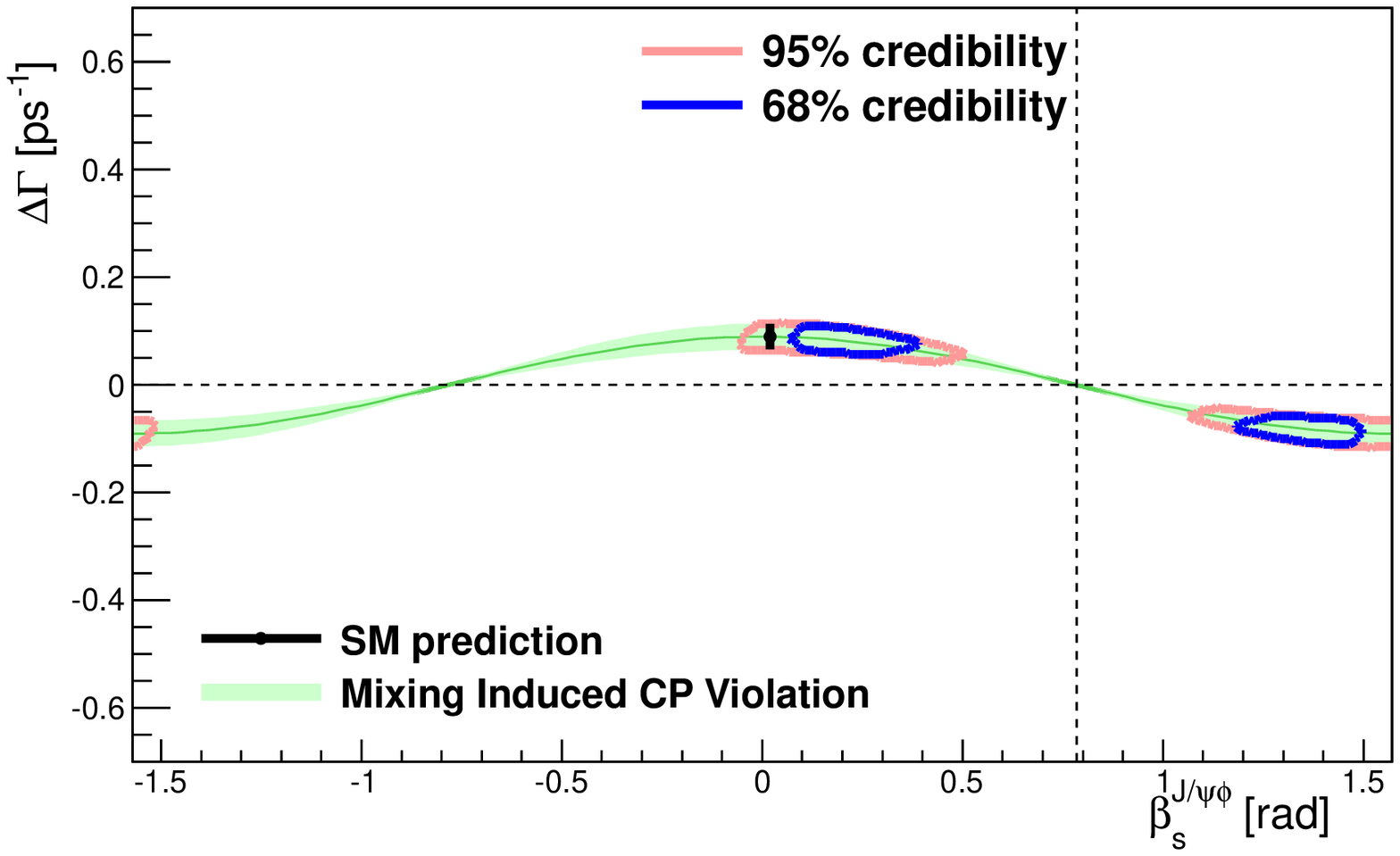}
\includegraphics[width=0.5\textwidth]{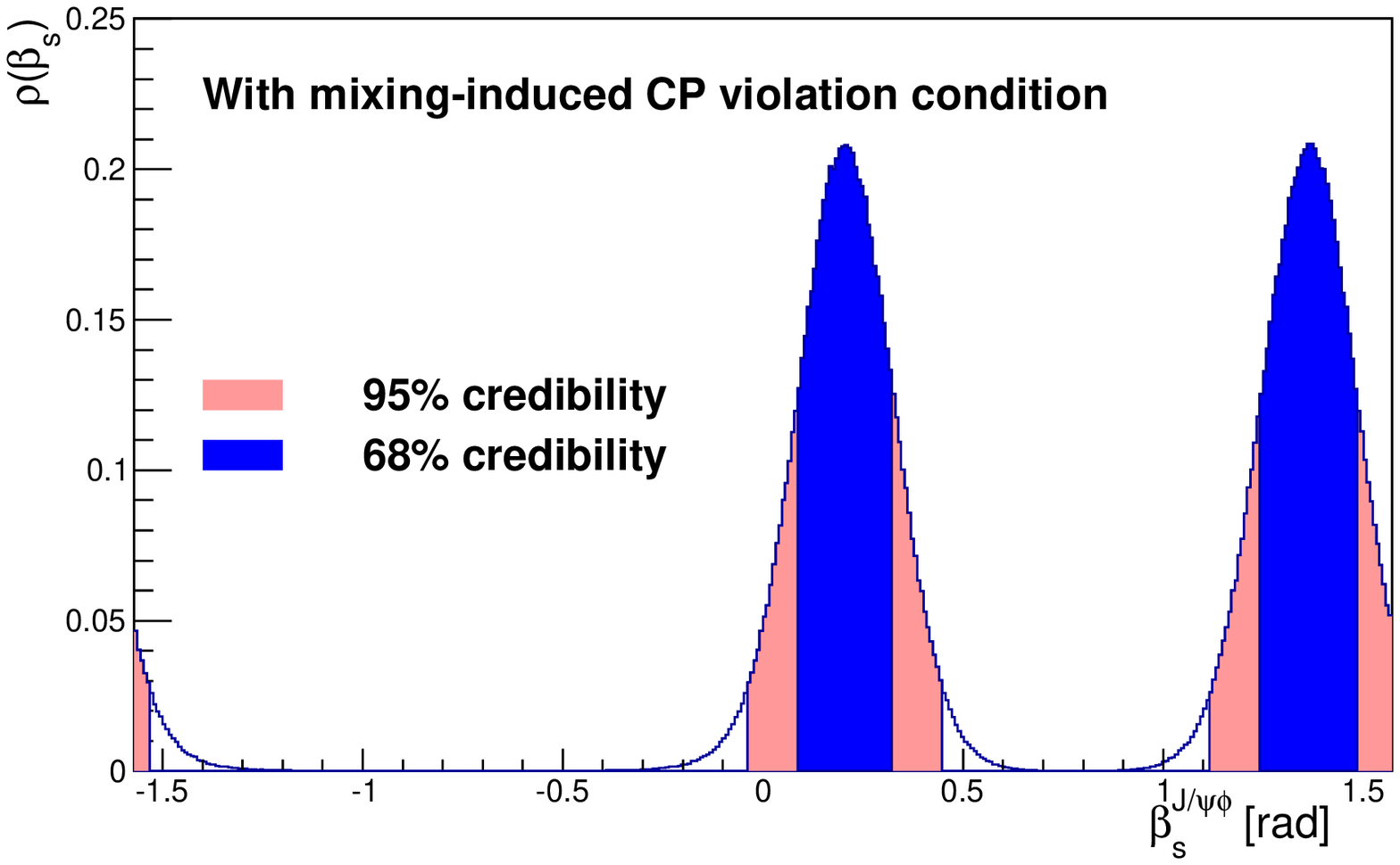}}
\caption{(color online).
  Two-dimensional posterior densities for the variables \betas
  and \deltaG (left), and one-dimensional posterior density for the
  variable \betas (right) including a prior density for mixing-induced
  \CP violation. The dark-solid (blue) and light-solid (red) contours
  show the 68\% and 95\% credible regions, respectively. 
\label{fig:MI-bs}}
\end{figure*}

The 68\% credible interval on the \CP-violating quantity \betas is
$\betas \in \left [ 0.09 , 0.32 \right ] \cup \left [ 1.24 , 1.48
\right]$.  The posterior density in \betas alone is shown in
Fig.~\ref{fig:MI-bs} on the left-hand side. Again, we calculate the
quantity 
\begin{eqnarray}
\int_{{\cal A}_{\beta_s}} p(\betas)\ d\betas = 0.131,
\end{eqnarray}
where ${\cal A}_{\beta_s} = \{\betas ; p(\betas) < p(\beta_s^{SM})\}$.
The SM is first included at a credibility of 86.9\%.

\label {08A_MixingAndSP} 

\begin{figure*}[tbp]
\centerline{
\includegraphics[width=0.5\textwidth] {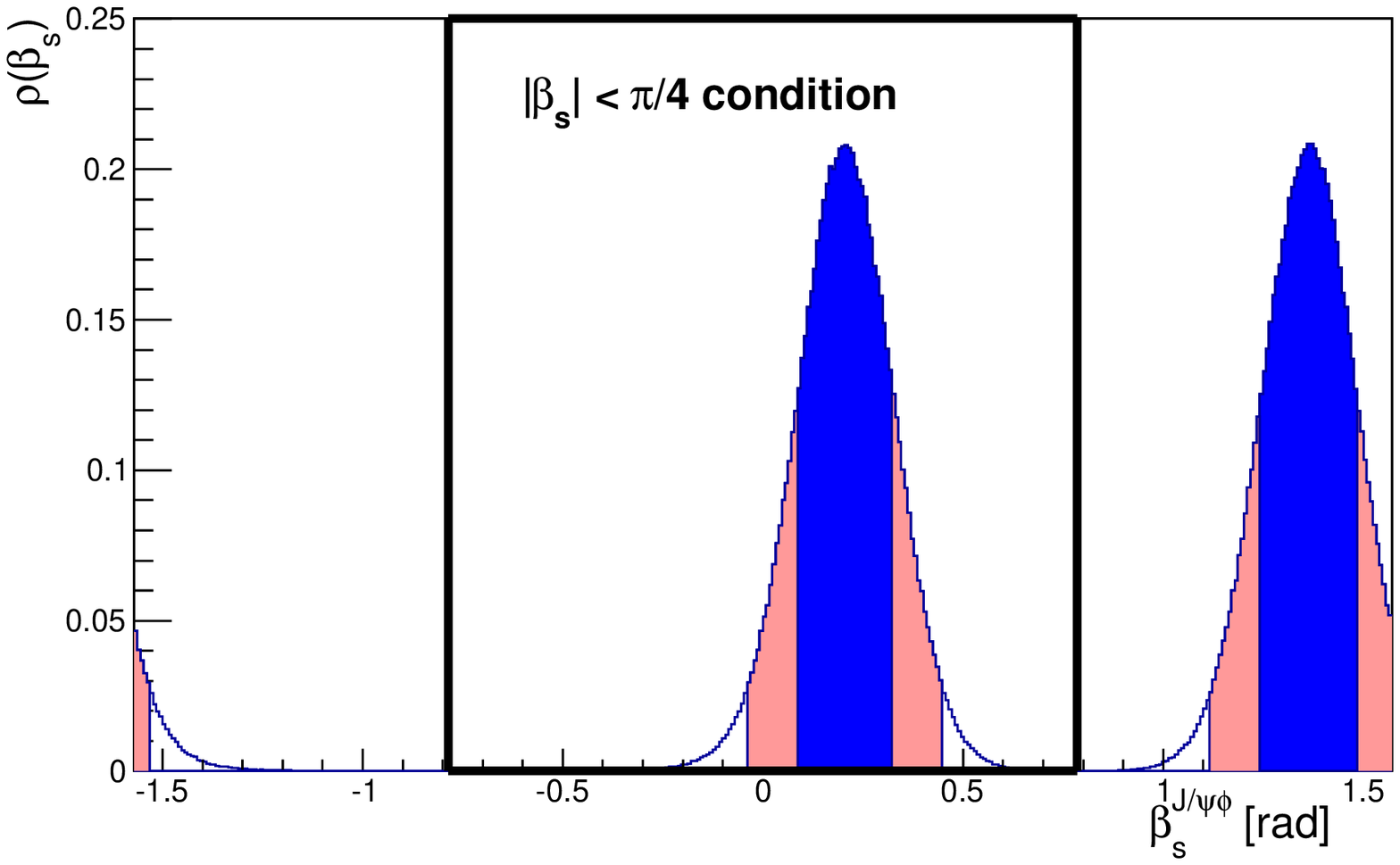}
\includegraphics[width=0.5\textwidth] {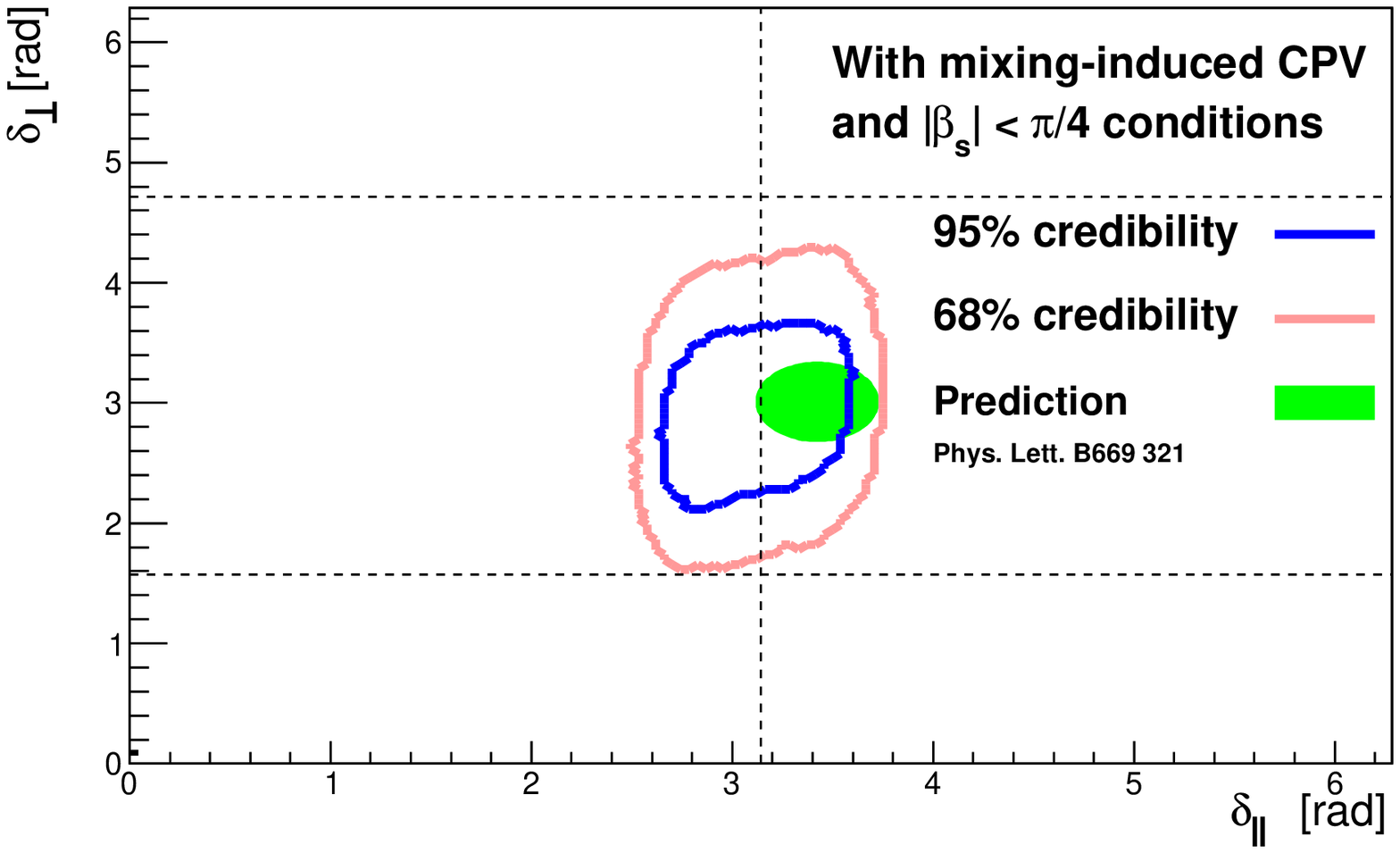} }
\centerline{
\includegraphics[width=0.5\textwidth] {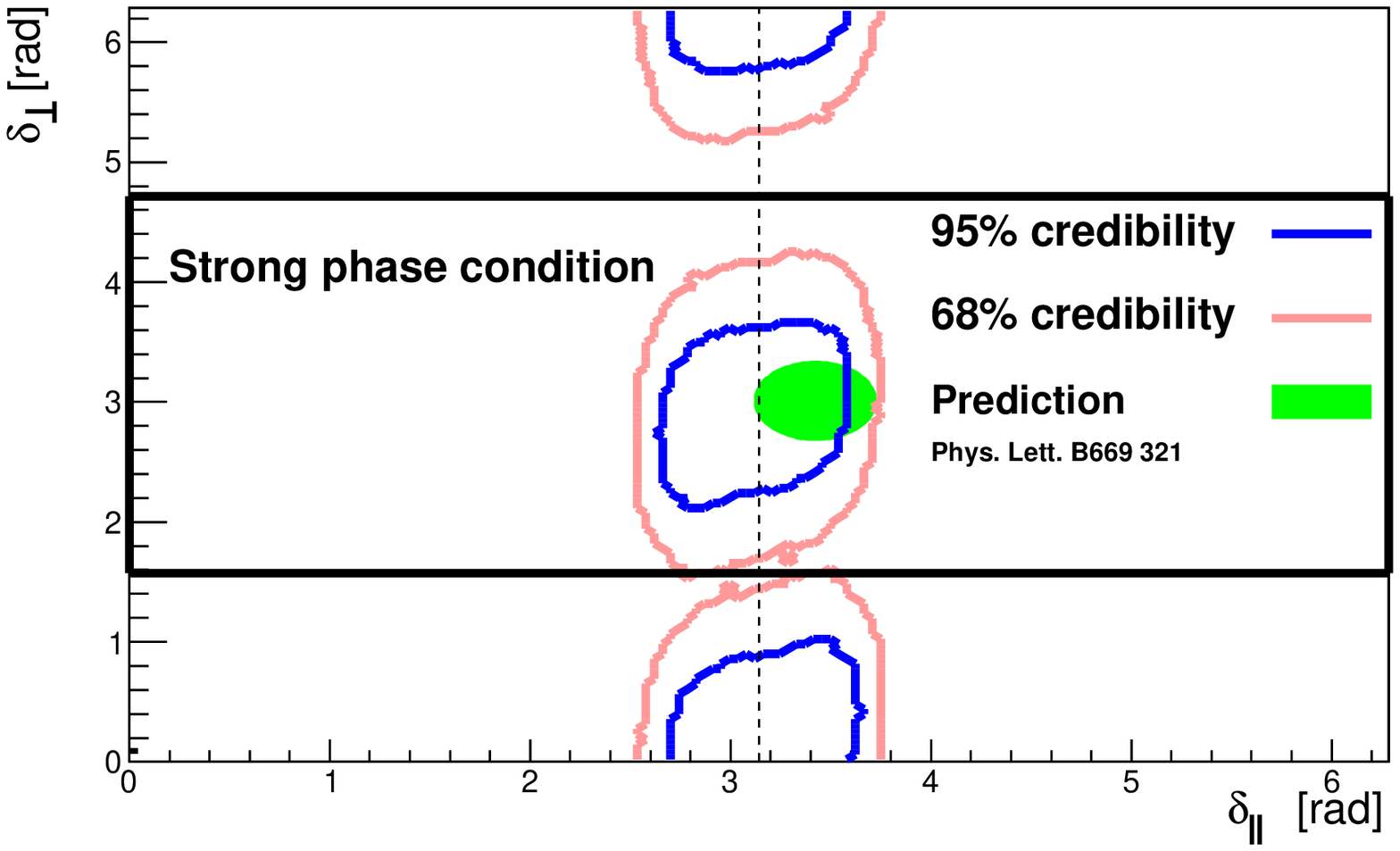}
\includegraphics[width=0.5\textwidth] {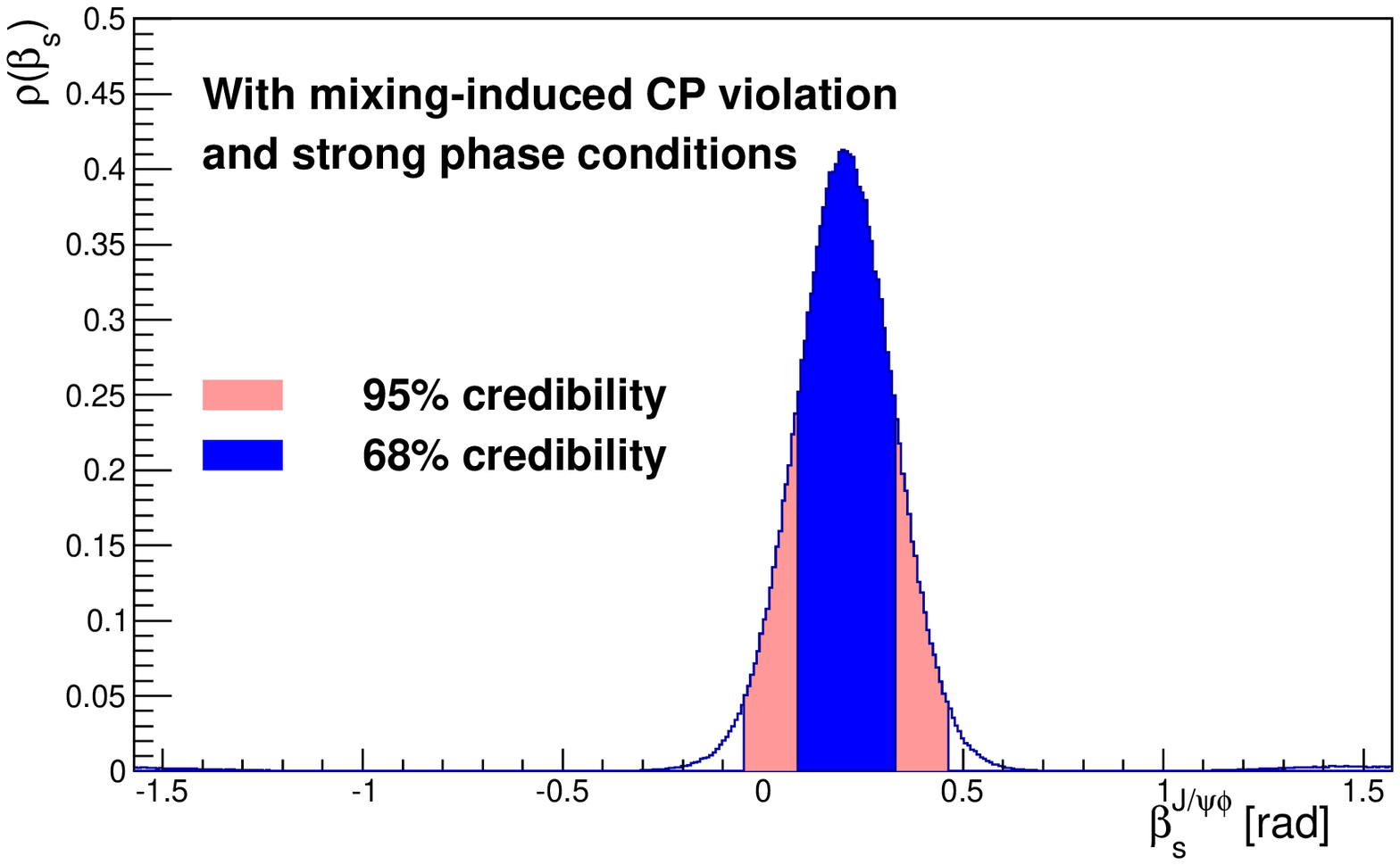} }
\caption{(color online).
  Conditional posterior densities for $\delta_\perp$ versus
  $\delta_{\parallel}$ and for $\betas$.  The extra conditions applied
  to \betas (top row) and $\delta_\perp$ (bottom row) are shown on the
  left, and the resulting conditional probabilities are displayed on the
  right. The theoretical prediction of Ref.~\cite{ref:Rosner} is
  indicated as green ellipse in the bottom left plot.  All plots in this
  figure are subject to the constraint of mixing-induced \CP~violation.
  The dark-solid (blue) and light-solid (red) contours
  show the 68\% and 95\% credible regions, respectively.
\label{fig:flatPhasePrior}}
\end{figure*}

As can be seen in 
Fig.~\ref{fig:PhasesII}, the theoretical predictions
of Ref.~\cite{ref:Rosner} are consistent with one of the modes
of the probability density in the $\delta_\perp$ versus $\delta_\parallel$ plane
but not with the other mode.  In the following, conditional posterior
densities are used to show that the favored mode of probability from
Fig.~\ref{fig:flatPhasePrior} corresponds to the solution  $\betas \in
\left [ 0.09 , 0.32 \right ]$.

The sequence of plots in Fig.~\ref{fig:flatPhasePrior} illustrates this
statement. The figures in the top row show the conditional posterior
density in the parameters $\delta_{\parallel}$ and $\delta_{\perp}$ 
after imposing the requirement $|\betas| < \pi/4$.  It can be seen
that this condition completely eliminates one of the modes of the
probability density in the $\delta_{\parallel}$ versus $\delta_{\perp}$
parameter space.  The plots in the bottom row of
Fig.~\ref{fig:flatPhasePrior} show that, conversely, the condition
$\pi/2 < \delta_{\perp} < 3\pi/2$ eliminates one mode of the probability
projected onto \betas, while there is virtually no impact on the other
mode with $|\betas| < \pi/4$.

These conditional probabilities allow us to visualize what is happening
in the larger space of all four parameters (\betas, \deltaG,
$\delta_{\parallel}$ and $\delta_{\perp}$), and to identify the
prediction of Ref.~\cite{ref:Rosner} with only one of the solutions for
$\betas$, namely $\betas \in \left [ 0.09 , 0.32 \right]$ 
at the 68\% credible interval. This result confirms the early finding in
our previous publication~\cite{Ref:tagged_cdf_old} which indicated that
the solution centered in $0\le\betas\le\pi/4$ and $\deltaG > 0$
corresponds to $\cos(\delta_{\perp})< 0$ and
$\cos(\delta_{\perp}-\delta_\parallel)>0$, while the opposite is true for
the solution centered in $\pi/4\le\betas\le\pi/2$ and $\deltaG < 0$.

\subsection {Sensitivity analysis}
It is a well-known fact that Bayesian results depend on the chosen prior
densities; a flat prior in a given metric might, in general,
not be flat in another metric. To study such effects, we
carried out a sensitivity analysis in order to characterize the degree to
which the Bayesian results of this section depends upon the chosen
input priors.  The sensitivity analysis was performed by weighting the
probability density with the Jacobian of the transformation from the
default parameterization to the desired parameterization.  Using this
technique we checked the variation of the Bayesian result with respect
to the following six variations. First, the prior is taken flat in
$\sin{2\betas}$ rather than flat in \betas; second, the prior is taken
flat in $\cos{\delta_{\perp}}$, and third, the prior is taken flat in
$\cos{\delta_{\parallel}}$. Afterwards, all three conditions are applied 
together at the same time. Fifth, the prior is taken flat in the
amplitudes $A_{\parallel}(0)$ and $A_{\perp}(0)$ rather than in their squares
and finally, the mixing-induced \CP~violation constraint is taken as a
Gaussian rather than flat constraint.
The effect of changing the priors on the 68\%~credibility intervals on
\betas is
summarized in Table~\ref{table:sensitivity}. Modest changes are
observed for the unconstrained result and the result with
$|\Gamma^s_{12}|$ constrained to $2\,|\Gamma^s_{12}| = (0.090 \pm
0.024)~\ps$~\cite{Ref:lenz}.
Only the effect on the first \betas~credibility interval
is shown, since the second interval can be trivially derived from
the numbers in Table~\ref{table:sensitivity}.

\begin{table}[tbp]
\caption {Summary of the sensitivity study. The 68\% credibility
interval on \betas is given for the unconstrained
result and when $2\,|\Gamma^s_{12}|$ is constrained to its SM prediction.
\label{table:sensitivity}}
\begin{ruledtabular}
\begin{tabular}{lcc} 
Variation                         &   Constrained\                   & \  Unconstrained         \\  \hline
Default                           &       [0.09,0.32]                &        [0.11,0.41]      \\
Flat $\sin{2\betas}$              &       [0.08,0.31]                &        [0.09,0.37]      \\
Flat $\cos{\delta_{\perp}}$       &       [0.09,0.33]                &        [0.10,0.43]      \\
Flat $\cos{\delta_{\parallel}}$   &       [0.09,0.32]                &        [0.11,0.41]      \\
Previous three together           &       [0.07,0.31]                &        [0.09,0.39]      \\
Flat in amplitudes                &       [0.09,0.32]                &        [0.11,0.41]      \\
Gaussian mix.-ind. \CP~viol. \          &       [0.09,0.34]                &                         \\ 
\end{tabular}
\end{ruledtabular}
\end{table}


\section{Summary}
\label{sec:conclusions}

In summary, we have presented a measurement of \CP~violation in
\BsJpsiPhi decays using ~6500 signal events from a data sample with
5.2~\fb integrated luminosity collected with the CDF\,II detector
operating at the Tevatron $p\bar p$~collider.  We find the \CP-violating
phase \betas to be within the range $\betas \in [0.02, 0.52] \cup [1.08,
1.55]$ at 68\%~confidence level and within the interval $[-\pi/2, -1.46]
\cup [-0.11, 0.65] \cup [0.91, \pi/2]$ at 95\%~C.L. where the coverage
property of the quoted interval is guaranteed using a frequentist
statistical analysis. Assuming the standard model expectation for
\betas, the probability to observe a fluctuation as large as in our data
or larger is given by a $p$-value of 0.30 corresponding to about one
Gaussian standard deviation. This result shows less of a discrepancy
with the SM expectation than our previously published result using
1.3~fb$^{-1}$ of integrated luminosity~\cite{Ref:tagged_cdf_old}.  In
comparison with the recent measurement of \betas from the
D0~collaboration using a data sample based on 8~fb$^{-1}$ of integrated
luminosity~\cite{Abazov:2011ry}, our result agrees within uncertainty
but constrains \betas to a narrower region.  With \deltaG constrained to
its SM prediction, we find \betas in the range $[0.05, 0.40] \cup [1.17,
1.49]$ at the 68\%~C.L.  The measurement of the \CP-violating phase
\betas is still statistics-limited. It will improve with the final 2011
CDF dataset approximately doubling the current integrated luminosity.

This analysis also incorporates the possibility of contributions to the
$\Bs \to J/\psi K^+K^-$~final state in the region of the
$\phi$~resonance from $\Bs\rightarrow \Jpsi f_{0}$ and $\Bs\rightarrow
\Jpsi K^{+}K^{-}$ (non-resonant) decays.  We measure the $S$-wave
contribution over the mass interval $1.009 < m(K^+K^-)<1.028~\gevcc$
corresponding to the selected $K^+K^-$ signal region to be less than 6\%
(4\%) at the 95\% (68\%) confidence level.  We do not confirm a sizeable
fraction of $17.3\pm3.6\%$ for the $S$-wave fraction as quoted over
almost the same $K^+K^-$~mass range in a recent
D0~publication~\cite{Abazov:2011ry}.

Assuming the standard model prediction for the \CP-violating
phase~\betas, we measure several other parameters describing the
\Bs~system. These include the \Bs mean lifetime $\tau(\Bs)$, the decay
width difference \deltaG between the heavy and light \Bs~mass
eigenstates, the transversity amplitudes $|A_{\parallel}(0)|^2$ and
$|A_{0}(0)|^2$, as well as the strong phase $\delta_{\perp}$. The
measurements for $\tau(\Bs) = 1.529 \pm 0.025\,\textrm{(stat)} \pm
0.012\,\textrm{(syst)}$~ps and $\deltaG = 0.075 \pm 0.035\,\textrm{(stat)}
\pm 0.006\,\textrm{(syst)}~\ps$ are the most precise measurements of these
quantities using a single decay mode. They are also in good agreement
with the PDG world averages~\cite{Ref:PDG}. The measurements of the
transversity amplitudes are consistent with previous measurements in the
\BsJpsiPhi system.
Finally, we report an alternative Bayesian analysis based on Markov
chain integration that gives results consistent with the frequentist
approach. 




\begin{acknowledgments}
  We thank the Fermilab staff and the technical staffs of the
  participating institutions for their vital contributions. This work
  was supported by the U.S. Department of Energy and National Science
  Foundation; the Italian Istituto Nazionale di Fisica Nucleare; the
  Ministry of Education, Culture, Sports, Science and Technology of
  Japan; the Natural Sciences and Engineering Research Council of
  Canada; the National Science Council of the Republic of China; the
  Swiss National Science Foundation; the A.P. Sloan Foundation; the
  Bundesministerium f\"ur Bildung und Forschung, Germany; the Korean
  World Class University Program, the National Research Foundation of
  Korea; the Science and Technology Facilities Council and the Royal
  Society, UK; the Russian Foundation for Basic Research; the Ministerio
  de Ciencia e Innovaci\'{o}n, and Programa Consolider-Ingenio 2010,
  Spain; the Slovak R\&D Agency; the Academy of Finland; and the
  Australian Research Council (ARC).
\end{acknowledgments}

\bibliography{paper_twocol}

\end{document}